\documentclass[screen, manuscript, nonacm]{acmart}

\usepackage[utf8]{inputenc}

\usepackage{bm} % bold math fonts, e.g. $\bm \Sigma$ will give a bold-face $\Sigma$

\usepackage[colorinlistoftodos]{todonotes}

\usepackage[small,nohug,heads=vee]{diagrams}
\diagramstyle[labelstyle=\scriptstyle]

\usepackage[ruled, noend]{algorithm2e}
\usepackage{enumitem}
\usepackage{quiver}
\usepackage{subcaption}

\usepackage{microtype}

\newcommand{\rai}{Relational\underline{AI}}

\newcommand{\norm}[1]{\|#1\|}
\newcommand{\set}[1]{\{#1\}}                    % Set (as in \set{1,2,3}).
\newcommand{\bag}[1]{\{\hspace{-1mm}\{#1\}\hspace{-1mm}\}}                    % bag (as in \bag{1,2,3}).
\newcommand{\setof}[2]{\{{#1}\mid{#2}\}}        % Set (as in \setof{x}{x>0}).
\newcommand{\bagof}[2]{\{\hspace{-1mm}\{{#1}\mid{#2}\}\hspace{-1mm}\}}        % Set (as in \setof{x}{x>0}).
\newcommand{\degree}{\text{\sf deg}}

\newcommand{\lfp}{\text{\sf lfp}}

\newcommand{\calB}{\mathcal B}
\newcommand{\calC}{\mathcal C}

\newcommand{\calP}{\mathcal P}

\newcommand{\calL}{\mathcal L}

\newcommand{\calT}{\mathcal T}

\theoremstyle{plain}                   % default
\newtheorem{thm}{Theorem}[section]
\newtheorem{lmm}[thm]{Lemma}
\newtheorem{prop}[thm]{Proposition}
\newtheorem{cor}[thm]{Corollary}
\newtheorem{defn}[thm]{Definition}

\theoremstyle{definition}              % Examples and all

\newtheorem{ex}[thm]{Example}

\newcommand{\defeq}{\stackrel{\text{def}}{=}}

\newcommand{\mult}{\cdot}

\newcommand{\B}{\mathbb B} % the Booleans
\newcommand{\N}{\mathbb N} % the natural numbers
\newcommand{\R}{\mathbb R} % the real numbers
\newcommand{\cd}{\text{ :- }}

\newcommand{\name}{\text{\sf datalog}^\circ}
\newcommand{\Name}{\text{\sf Datalog}^\circ}
\newcommand{\trop}{\text{\sf Trop}}

\newcommand{\ground}{\textsf{GA}}
\newcommand{\arity}{\textsf{arity}}
\newcommand{\adom}{\textsf{ADom}}
\newcommand{\inst}{\textsf{Inst}}

\title{Convergence of Datalog over (Pre-) Semirings}
\author{Mahmoud Abo Khamis}
\affiliation{
  \institution{\rai{}}
  \city{Berkeley}
  \state{CA}
  \country{USA}
}
\author{Hung Q. Ngo}
\affiliation{
  \institution{\rai{}}
  \city{Berkeley}
  \state{CA}
  \country{USA}
}
\author{Reinhard Pichler}
\affiliation{
  \institution{TU Wien}
  \city{Vienna}
  \country{Austria}}
\author{Dan Suciu}
\affiliation{
  \institution{University of Washington}
  \city{Seattle}
  \state{WA}
  \country{USA}
}
\author{Yisu Remy Wang}
\affiliation{
  \institution{University of Washington}
  \city{Seattle}
  \state{WA}
  \country{USA}
}

\begin{abstract}
  Recursive queries have been traditionally studied in the framework
  of datalog, a language that restricts recursion to monotone queries
  over sets, which is guaranteed to converge in polynomial time in the
  size of the input.  But modern big data systems require recursive
  computations beyond the Boolean space.  In this paper we study the
  convergence of datalog when it is interpreted over an arbitrary
  semiring.  We consider an ordered semiring, define the semantics of
  a datalog program as a least fixpoint in this semiring, and study
  the number of steps required to reach that fixpoint, if ever.  We
  identify algebraic properties of the semiring that correspond to
  certain convergence properties of datalog programs.  Finally, we
  describe a class of ordered semirings on which one can use the
  semi-na\"ive evaluation algorithm on any datalog program.
\end{abstract}

\begin{document}

\maketitle

% \fancyhead{}

\section{Introduction}
\label{sec:intro}

For fifty years, the relational data model has been the main choice for
representing, modeling, and processing data.  Its main query
language, SQL, is found today in a wide range of applications and
devices, from smart phones, to database servers, to distributed
cloud-based clusters.  The reason for its success is the {\em data
    independence principle}, which separates the declarative model from
the physical implementation~\cite{DBLP:journals/cacm/Codd70}, and
enables advanced implementation techniques, such as cost-based
optimizations, indices, materialized views, incremental view
maintenance, parallel evaluation, and many others, while keeping the
same simple, declarative interface to the data unchanged.

But scientists today often need to perform tasks that require
iteration over the data.
Gradient descent, clustering, page-rank, network centrality, inference
in knowledge graphs are some examples of common tasks in data science
that require iteration.  While SQL has introduced recursion since 1999
(through Common Table Expressions, CTE), it has many cumbersome
limitations and is little used in practice~\cite{frankmcsherry-2022}.

The need to support recursion in a declarative language led to the
development of {\em datalog} in the mid 80s~\cite{DBLP:conf/pods/Vianu21}.
Datalog adds recursion to the relational query language, yet enjoys several elegant
properties: it has a simple, declarative semantics; its na\"ive
bottom-up evaluation algorithm always terminates; and it admits a few
powerful optimizations, such as semi-na\"ive evaluation and magic set
rewriting.  Datalog has been extensively studied in the literature;
see~\cite{DBLP:journals/ftdb/GreenHLZ13} for a survey
and~\cite{DBLP:books/mc/18/MaierTKW18,DBLP:conf/pods/Vianu21} for historical notes.

However, datalog is not the answer to modern needs, because it only
supports monotone queries over sets.  Most tasks in data science today
require the interleaving of recursion and aggregate computation.
Aggregates are not monotone under set inclusion, and therefore they
are not supported by the framework of pure datalog.  Neither SQL'99
nor popular open-source datalog systems like
Souffl\'e~\cite{DBLP:conf/cav/JordanSS16} allow recursive queries to
have aggregates.  While several proposals have been made to extend
datalog with
aggregation~\cite{DBLP:conf/pods/GangulyGZ91,DBLP:conf/pods/RossS92,DBLP:journals/jcss/GangulyGZ95,DBLP:journals/vldb/MazuranSZ13,DBLP:conf/icde/ShkapskyYZ15,DBLP:conf/sigmod/ShkapskyYICCZ16,DBLP:conf/amw/ZanioloYDI16,DBLP:journals/tplp/ZanioloYDSCI17,DBLP:conf/amw/ZanioloYIDSC18,DBLP:journals/tplp/CondieDISYZ18,DBLP:conf/sigmod/0001WMSYDZ19,DBLP:journals/corr/abs-1910-08888,DBLP:journals/corr/abs-1909-08249,DBLP:journals/debu/ZanioloD0LL021},
these extensions are at odds with the elegant properties of datalog
and have not been adopted by either datalog systems or SQL engines.

In this paper we propose a foundation for a query language that
supports both recursion and aggregation.  Our proposal is based on the
concept of $K$-relations, introduced in a seminal
paper~\cite{DBLP:conf/pods/GreenKT07}.  In a $K$-relation, tuples are
mapped to a fixed semiring. Standard relations (sets) are
$\B$-relations where tuples are mapped to the Boolean semiring $\B$,
relations with duplicates (bags) are $\N$-relations, sparse tensors
are $\R$-relations, and so on.  Queries over $K$-relations are the
familiar relational queries, where the operations $\wedge, \vee$ are
replaced by the operations $\otimes, \oplus$ in the semiring;
importantly, an existential quantifier $\exists$ becomes an
$\oplus$-aggregate operator.
% relations with duplicates (also called bags), where each tuple is
% mapped to the semiring $\N$, indicating its multiplicity, sparse
% tensors, where each tuple is mapped to $\R$, and many other types of
% collections.
$K$-relations are a very powerful abstraction, because they open up
the possibility of adapting query processing and optimization
techniques to other domains~\cite{DBLP:conf/pods/KhamisNR16}.

Our first contribution is to introduce an extension of datalog to
$K$-relations.  We call the language $\name$, where the superscript
$o$ represents a (semi)-ring.  $\Name$ has a declarative semantics
based on the least fixpoint, and supports both recursion and
aggregates.  We illustrate throughout this paper its utility through
several examples that are typical for recursive data processing.  In
order to define the least fixpoint semantics of $\name$, the semiring
needs to be partially ordered.  For this purpose, we introduce an
algebraic structure called a {\em Partially Ordered Pre-Semiring (POPS)\/},
which generalizes the more familiar naturally ordered semirings.  This
generalization is necessary for some applications.  For example, the
bill-of-material program (Example~\ref{ex:sum1:sum2}) is naturally
expressed over the lifted reals, $\R_\bot$, which is a POPS that is
not naturally ordered.

Like datalog, $\name$ can be evaluated using the {\em na\"ive algorithm},
by repeatedly applying all rules of the program, until there is no
more change.  However, unlike datalog, a $\name$ program may diverge.
Our second contribution consists of a full characterization of the
POPS that guarantee that every $\name$ program terminates.  More
precisely, we show that termination is guaranteed iff the POPS enjoys
a certain algebraic property called {\em
    stability}~\cite{semiring_book}.  The result is based on an analysis
of the fixpoint of a vector-valued polynomial function over a semiring, which is of
independent interest.  With some abuse, we will say in this paper that
a $\name$ program {\em converges}, if the na\"ive algorithm terminates
in a finite number of steps; we do not consider ``convergence in the
limit'', for example in $\omega$-continuous
semirings~\cite{DBLP:conf/pods/GreenKT07,DBLP:journals/jacm/EsparzaKL10}.

Finally, we describe how the {\em semi-na\"ive algorithm} can be
extended to $\name$, under certain restrictions on the POPS.  This
should be viewed as an illustration of the potential for applying
advanced optimizations to $\name$: in a companion
paper~\cite{DBLP:conf/sigmod/WangK0PS22}, we introduced a simple, yet
powerful optimization technique for $\name$, and showed, among other
things, that magic set rewriting can be obtained using several
applications of that rule.

At its essence, a $\name$ program consists of solving a fixpoint
equation in a semiring, which is a problem that was studied in a variety
of areas, like automata theory, program analysis, and many
others~\cite{MR1470001,DBLP:conf/popl/CousotC77,MR1728440,MR1059930,
  DBLP:conf/lics/HopkinsK99, DBLP:journals/tcs/Lehmann77,
  semiring_book,MR609751}.  The existence of the fixpoint is usually
ensured by requiring the semiring to be $\omega$-continuous. For
example, Green et al.~\cite{DBLP:conf/pods/GreenKT07} studied the
provenance of datalog on $K$-relations, while Esparza et
al.~\cite{DBLP:journals/jacm/EsparzaKL10} studied dataflow equations,
in both cases requiring the semiring to be $\omega$-continuous.  This
guarantees that the least fixpoint exists, even if the na\"ive algorithm
diverges.

In addition to the (semi-)na\"ive method, which is a first-order method for solving
fixpoint equations, there is a second-order method called the {\em Newton's method},
discovered relatively recently \cite{DBLP:journals/jacm/EsparzaKL10,DBLP:conf/lics/HopkinsK99}
in the program analysis literature.
It was shown that Newton's method requires a smaller number of iterations than the
na\"ive algorithm.\footnote{The na\"ive algorithm is called Kleene iteration
  method in~\cite{DBLP:journals/jacm/EsparzaKL10}.} In the particular
case of a commutative and idempotent semiring, Newton's method always converges, while the
na\"ive algorithm may diverge~\cite{DBLP:journals/jacm/EsparzaKL10}.
However, every iteration of Newton's method requires solving a least fixpoint solution
to an inner-recursion, which is expensive.
Here, we are seeing the exact same phenomenon when comparing gradient descent vs.
Newton's method in continuous optimization: while Newton's method may converge in fewer
steps, every step is more expensive, and requires the materialization of a large intermediate
result (the Hessian matrix).
In continuous optimization, Newton's method is rarely used for large-scale problems for
this reason~\cite{DBLP:books/sp/NocedalW99}.
For $\name$, it remains unclear whether Newton's method is more efficient in practice than
the na\"ive algorithm.  One experimental evaluation~\cite{DBLP:conf/popl/RepsTP16} has
found that it is not. In contrast, the na\"ive algorithm, and its
extension to the semi-na\"ive algorithm, is simple, and implemented by
all datalog systems today.  In this paper we focus only on the
convergence of the na\"ive algorithm, and do not consider Newton's
method.

A preliminary version of this paper appeared
in~\cite{DBLP:conf/pods/Khamis0PSW22}.  The convergence theorem in
stable semirings, Theorem~\ref{th:non-unformly:stable}, is new.  The
convergence semiring in $p$-stable semirings,
Theorem~\ref{th:uniform:stable}, is improved and extended.  We have
also included many proofs omitted
from~\cite{DBLP:conf/pods/Khamis0PSW22} and added several examples.

\subsection{Overview of the Results}

\label{subsec:overview}

We present here a high-level overview of the main results in this
paper.  We start by recalling the syntax of a (traditional) datalog
program $\Pi$, which consists of a set of rules of the form:
\begin{align}
  R_0(\bm X_0) & \cd R_1(\bm X_1) \wedge \cdots \wedge R_m(\bm X_m) \label{eq:datalog}
\end{align}
where $R_0, \ldots, R_m$ are relation names (not necessarily distinct)
and each $\bm X_i$ is a tuple of variables and/or constants.  The atom
$R_0(\bm X_0)$ is called the head, and the conjunction
$R_1(\bm X_1) \wedge \cdots \wedge R_m(\bm X_m)$ is called the body.
A relation name that occurs in the head of some rule in $\Pi$ is
called an {\em intensional database predicate} or IDB, otherwise it is
called an {\em extensional database predicate} or EDB.  The EDBs form
the input database, which is denoted $I$, and the IDBs represent the
output instance computed by $\Pi$, which we denote by $J$.  The finite
set of all constants occurring in $I$ is called the {\em active
    domain} of $I$, and denoted $\adom(I)$.
The textbook example of a datalog program is one that computes the
transitive closure of a graph defined by the edge relation $E(X,Y)$:
\begin{align*}
  T(X,Y) & \cd E(X,Y)               \\
  T(X,Y) & \cd T(X,Z) \wedge E(Z,Y)
\end{align*}
Here $E$ is an EDB predicate, and $T$ is an IDB predicate.  Multiple
rules with the same head are interpreted as a disjunction, and in this
paper we prefer to combine them in a single rule, with an explicit
disjunction.  In particular, the program above becomes:
\begin{align}
  T(X,Y) & \cd E(X,Y) \vee \exists_Z (T(X,Z) \wedge E(Z,Y)) \label{eq:datalog:intro}
\end{align}
The semantics of a datalog program is the least fixpoint of the
function defined by its rules, and can be computed by the {\em na\"ive
    evaluation algorithm}, as follows: initialize all IDBs to the empty
set, then repeatedly apply all rules, updating the state of the IDBs,
until there is no more change to the IDBs.  The algorithm always
terminates in time polynomial in $\adom(I)$.

We introduce $\name$, a language that generalizes datalog to relations
over a {\em partially ordered, commutative pre-semiring}, or POPS.  A
POPS, $\bm P$, is an algebraic structure with operations
$\oplus, \otimes$ on a domain $P$, which is also a partially ordered set with an order
relation $\sqsubseteq$ and a smallest element $\bot$. (Formal
definition is in Sec.~\ref{sec:pops}).  For example, the Boolean
semiring $\B$ forms a POPS where $(\oplus,\otimes) = (\vee, \wedge)$, over domain
$\{0,1\}$, and the order relation is $0 \leq 1$.  For another example, the {\em
    tropical semiring}, $\trop$, consists of the real numbers $\R$, the
operations are $\min, +$, and the order relation $x \sqsubseteq y$ is
the reverse of the usual order $x \geq y$.  If
we restrict the tropical semiring to the non-negative reals and infinity, $\R_+ \cup \infty$, then
we denote the POPS by $\trop^+$.  For a fixed POPS $\bm P$, a
$\bm P$-relation of arity $k$ is a function that maps all $k$-tuples
over some domain to values in $\bm P$; for example, a standard
relation is a $\B$-relation, where $\B$ is the Boolean semiring, while
a relation with duplicates (a bag) is an $\N$-relation.  A $\name$
program consists of a set of rules, e.g.
like~\eqref{eq:datalog:intro}, where all relations are
$\bm P$-relations, and the operators $\wedge, \vee, \exists_Z$ are
replaced by $\otimes, \oplus, \bigoplus_Z$, and its semantics is
defined as its least fixpoint, when it exists, or undefined otherwise.
We will illustrate the utility of $\name$ through multiple examples in
this paper, and give here only a preview.

\begin{ex} \label{ex:intro} The {\em all-pairs shortest paths} (APSP) problem is to compute the
  shortest path length $T(X,Y)$ between all pairs $X,Y$ of vertices in
  the graph, given that $E(X,Y)=$ the length of the edge $(X,Y)$ in
  the graph.  We assume that both $E$ and $T$ are $\trop^+$-relations.
  The APSP problem can be expressed very compactly in $\name$ as
  \begin{align}
    T(X,Y) & \cd E(X,Y) \oplus \bigoplus_Z T(X,Z) \otimes E(Z,Y), \label{eqn:linear:tc}
  \end{align}
  where $(\oplus,\otimes) = (\min, +)$ are the ``addition'' and
  ``multiplication'' in $\trop^+$.  If we instantiate these operations
  in~\eqref{eqn:linear:tc}, the program becomes:
  \begin{align}
    T(X,Y) & \cd \min\left(E(X,Y), \min_Z (T(X,Z) + E(Z,Y))\right), \label{eqn:apsp}
  \end{align}
  If we use a different POPS in~\eqref{eqn:linear:tc}, then the same
  $\name$ program will express similar problems, in exactly the same
  way. For example, if the relations $E, T$ are $\B$-relations, then
  the program~\eqref{eqn:linear:tc} becomes the {\sf transitive
      closure} program~\eqref{eq:datalog:intro}; alternatively, if both
  $E, T$ are $\trop^+_p$-relations, where $\trop^+_p$ is a semiring
  defined in Sec.~\ref{sec:pops}, then the program computes the {\em
      top $p+1$-shortest-paths}.
\end{ex}

The least fixpoint of $\name$, if exists, can be computed by the na\"ive
algorithm, quite similar to datalog: initialize all IDBs to $\bot$,
then repeatedly apply all rules of the $\name$ program, obtaining an
increasing chain of IDB instances,
$J^{(0)} \sqsubseteq J^{(1)} \sqsubseteq J^{(2)} \sqsubseteq \cdots$
% where $\sqsubseteq$ is the partial order on the POPS, extended
% component-wise to IDB instances $J$
When $J^{(t)} = J^{(t+1)}$, then the algorithm stops and returns
$J^{(t)}$; in that case we say that the $\name$ program converges in
$t$ steps, or we just say that it converges; otherwise we say that it
diverges.  Traditional datalog always converges, but this is no longer
the case for $\name$ programs.  There are five possibilities,
depending on the choice of the POPS $\bm P$:
%
%
%
% The first question in this paper is whether every $\name$ program
% terminates.  Unlike datalog, this is no longer guaranteed for $\name$,
% because the value of an atom $u=R(\bm a)$ can potentially increase
% forever $\bot \sqsubset u_1 \sqsubset u_2 \sqsubset \cdots$ Relations
% over semirings were first introduced by Green et
% al.~\cite{DBLP:conf/pods/GreenKT07}, who also extended datalog to
% semirings.  They required the semiring to be {\em
%   $\omega$-continuous}, meaning that every increasing sequence
% $(u_n)_{n \geq 0}$ has a limit, and proved that every datalog program
% has a least fixpoint.  However, datalog systems implementations run
% the na\"ive or semi-na\"ive algorithm, and $\omega$-continuity is not
% sufficient to ensure termination.  This may lead to cases where a
% datalog program has a well defined semantics, but the system cannot
% compute it. Instead, we study when the na\"ive algorithm is guaranted to
% terminate; in that case it will also return the least fixpoint.  Given
% a POPS $P$, there are four possibilities:
%
\begin{enumerate}[label=(\roman*)]
  \item \label{item:converge:1} For some $\name$ programs, $\bigvee_t J^{(t)}$ is not the least fixpoint.
  \item \label{item:converge:2} Every $\name$ program has the least fixpoint $\bigvee_t J^{(t)}$, but may not necessarily converge.
  \item \label{item:converge:3} Every $\name$ program converges.
  \item \label{item:converge:4} Every $\name$ program converges in a
        number of steps that depends only on $|\adom(I)|$.
  \item \label{item:converge:5} Every $\name$ program converges in a
        number of steps that is polynomial in $|\adom(I)|$.
\end{enumerate}
In this paper, we will only consider
data-complexity~\cite{DBLP:conf/stoc/Vardi82}, where the $\name$
program is assumed to be fixed, and the input consists only of the EDB
instance $I$.

We study algebraic properties of the POPS $\bm P$ that ensure that we
are in one of the cases~\ref{item:converge:3}-\ref{item:converge:5};
we do not address cases~\ref{item:converge:1}-\ref{item:converge:2} in
this paper.  We give next a necessary and sufficient condition for
each of the cases~\ref{item:converge:3} and~\ref{item:converge:4}, and
give a sufficient condition for case~\ref{item:converge:5}.  For any
POPS $\bm P$, the set
$\bm P \oplus \bot \defeq \setof{u\oplus \bot}{u \in P}$ is a
semiring (Proposition~\ref{prop:s:plus:bot}); our characterization is based entirely on a certain
property, called {\em stability}, of the semiring $\bm P \oplus \bot$,
which we describe here.

Given a semiring $\bm S$ and $u \in S$, denote by
$u^{(p)} := 1 \oplus u \oplus u^2 \oplus \cdots \oplus u^{p}$, where
$u^{i} := u \otimes u \otimes \cdots \otimes u$ ($i$ times).  We say
that $u$ is {\em $p$-stable} if $u^{(p)}=u^{(p+1)}$; we say that the
semiring $\bm S$ is {\em $p$-stable} if every element $u \in S$ is $p$-stable, and
we say that $\bm S$ is {\em stable} if every element $u$ is stable for some
$p$ that may depend on $u$.
A $\name$ program is {\em linear} if every rule has at most one IDB predicate in the body.
We prove:

\begin{thm} \label{th:main:intro} Given a POPS $\bm P$, the following hold.
  \begin{itemize}
    \item Every $\name$ program converges iff the semiring $\bm P\oplus \bot$ is stable.
    \item Every program converges in a number of steps that depends only
          on $|\adom(I)|$ iff $\bm P \oplus \bot$ is $p$-stable for some
          $p$.  More precisely, every $\name$ program converges in
          $\sum_{i=1}^{N}(p+2)^i$ steps, where $N$ is the number of ground
          tuples consisting of IDB predicates and constants from $\adom(I)$.
          Furthermore, if the program is linear, then it converges in $\sum_{i=1}^N(p+1)^i$ steps.
    \item If $\bm P\oplus \bot$ is $0$-stable, then every $\name$
          program converges in $N$ steps; in particular, the program runs in
          polynomial time in the size of the input database.
  \end{itemize}
\end{thm}

In a nutshell, the theorem says that convergence of $\name$ is
intimately related to the notion of stability.  The proof, provided in
Sec.~\ref{sec:complexity}, consists of an analysis of the infinite
powerseries resulting from unfolding the fixpoint definition; for
the proof of the first item we also use Parikh's
theorem~\cite{MR209093}.  As mentioned earlier, most prior work on
fixpoint equations assumes an $\omega$-continuous semiring; when
convergence is desired, one usually offers the Ascending Chain
Condition, ACC (see Sec.~\ref{sec:lfp}) as a sufficient condition for
convergence.  Our theorem implies that ACC is only a sufficient, but
not a necessary condition for convergence, for example $\trop^+$ is
$0$-stable, and therefore every $\name$ program converges on
$\trop^+$, yet it does not satisfy the ACC condition.  A somewhat
related result is proven by Luttenberger and
Schlund~\cite{DBLP:journals/iandc/LuttenbergerS16} who showed that, if
$1$ is $p$-stable, then Newton's method requires at most
$N + \log\log p$ iterations.  As mentioned earlier, each step of
Newton's method requires the computation of another least fixpoint,
hence that result does not inform us on the convergence of the na\"ive
algorithm.

Next, we introduce an extension of the semi-na\"ive evaluation
algorithm to $\name$.  It is known that the na\"ive evaluation
algorithm is inefficient in practice, because at each iteration it
repeats all the computations that it has done at the previous
iterations.  Most datalog systems implement an improvement called the
  {\em semi-na\"ive} evaluation, which keeps track of the delta between
the successive states of the IDBs and applies the datalog rules as a
function of these deltas to obtain the new iteration's deltas.
Semi-na\"ive evaluation is one of the major optimization techniques
for evaluating datalog, however, it is defined only for programs that
are monotone under set inclusion, and the systems that implement it
enforce monotonicity, preventing the use of aggregation in
recursion. Our second result consists of showing how to adapt the
semi-na\"ive algorithm to $\name$, under certain restrictions of the
POPS $\bm P$, thus, enabling the semi-na\"ive algorithm to be applied
to programs with aggregation in recursion.

Let's illustrate the semi-na\"ive algorithm on the APSP program in
Example~\ref{ex:intro}.  For that, let's first review the standard
semi-na\"ive algorithm for pure datalog on the transitive closure
program~\eqref{eq:datalog:intro}.  While the na\"ive algorithm starts
with $T^{(0)}(X,Y)=\emptyset$ and repeats
$T^{(t+1)}(X,Y) = E(X,Y) \vee \bigvee_Z T^{(t)}(X,Z) \wedge E(Z,Y)$
for $t=0,1,2,\ldots$, the semi-na\"ive algorithm starts by
initializing $T^{(1)}(X,Y) = \delta^{(0)}(X,Y) = E(X,Y)$, then
performs the following steps for $t=1,2,3,\ldots$:
\begin{align}
  \delta^{(t)}(X,Y) & = \left(\bigvee_z \delta^{(t-1)}(X,Z) \wedge E(Z,Y) \right) \setminus T^{(t)}(X,Y) \label{eqn:deltat} \\
  T^{(t+1)}(X,Y)    & = T^{(t)}(X,Y) \cup \delta^{(t)}(X,Y).\nonumber
\end{align}
%
% Furthermore, starting from iteration $t=2$ onwards, we can
% simplify~\eqref{eqn:deltat} further by {\em removing} the base-case
% $E(X,Y)$ because we know $E(X,Y) \subseteq T^{(t-1)}(X,Y)$ for
% $t\geq 2$:
% %
% \begin{align}
%   \delta^{(t)}(X,Y) &= \left(\bigvee_Z \delta^{(t-1)}(X,Z) \wedge E(Z,Y) \right) \setminus T^{(t-1)}(X,Y).
%     \label{eqn:base:case:removal}
% \end{align}
% %
Thus, the semi-na\"ive algorithm avoids re-discovering at iteration
$t$ tuples already discovered at iterations $1,2,\ldots, t-1$.

In order to apply the same principle to the APSP
program~\eqref{eqn:apsp}, we need to define an appropriate ``minus''
$\ominus$ operator on the tropical semiring.  This operator is:
\begin{align}
  v\ominus u= &
  \begin{cases} v      & \mbox{if } v < u    \\
              \infty & \mbox{if } v \geq u
  \end{cases} \label{eqn:trop:minus}
\end{align}
The semi-na\"ive algorithm for the APSP problem in~\eqref{eqn:linear:tc}
is a mirror of that in~\eqref{eqn:deltat}:
\begin{align}
  \delta^{(t)}(X,Y) & = (\min_Z \delta^{(t-1)}(X,Z) + E(Z,Y)) \ominus T^{(t)}(X,Y), \label{eq:semi-naive:trop} \\
  T^{(t+1)}(X,Y)    & = \min(T^{(t)}(X,Y), \delta^{(t)}(X,Y))\nonumber
\end{align}
where the difference operator $\ominus$ is defined
in~\eqref{eqn:trop:minus}. The reason why this algorithm is more
efficient than the na\"ive algorithm is the fact that, in general, a
database system needs to store only the tuples that are ``present'' in
a relation, i.e. those whose value is $\neq \bot$.  In our example,
only those tuples $\delta^{(t)}(X,Y)$ whose value is $\neq \infty$
need to be stored.  The $\ominus$ operator
in~\eqref{eq:semi-naive:trop} checks if the new value
$\min_Z \delta^{(t-1)}(X,Z) + E(Z,Y)$ is strictly less than the old
value $T^{(t)}(X,Y)$: if not, then it returns $\infty$, signaling that
the value of $(X,Y)$ in both $\delta^{(t)}(X,Y)$ and $T^{(t)}(X,Y)$
does not need to be updated.  As a consequence, only those tuples
$T^{(t+1)}(X,Y)$ need to be processed at step $t$ where the value has
strictly decreased from the previous step.

Finally, the last topic we address in this paper is a possible way of
introducing negation in $\name$.  We will show that by interpreting
$\name$ over a particular POPS called THREE
(Sec.~\ref{subsec:three:pops}) and by appropriately defining a
function \texttt{not}, $\name$ can express datalog queries with
negation under Fitting's three-valued
semantics~\cite{DBLP:journals/jlp/Fitting85a}.  The crux in Fitting's
approach is to apply to the three-valued logic
(with truth values $\{0, 1,\bot\}$, where $\bot$ means ``undefined'')
the
  {\em knowledge order} $\leq_k$ with $\bot \leq_k 0$ and
$\bot \leq_k 1$. Then, the function \texttt{not} defined as
$\mathtt{not}(0) = 1$, $\mathtt{not}(1) = 0$, and
$\mathtt{not}(\bot) = \bot$ is monotone w.r.t.\ $\leq_k$, and one can
apply the usual least fixpoint semantics to datalog programs with
negation (and, likewise, to $\name$ programs with the function
\texttt{not}).  Hence, in cases when Fitting's 3-valued semantics
coincides with the well-founded semantics, so does $\name$ equipped
with the function \texttt{not} when interpreted over the POPS THREE.

\subsection{Paper organization}

We define POPS in Sec.~\ref{sec:pops} and give several examples.
In Sec.~\ref{sec:lfp} we consider the least fixpoint of monotone
functions over posets, and prove an upper bound on the number of
iterations needed to compute the least fixpoint.  We define $\name$
formally in Sec.~\ref{sec:datalogo}, and give several examples.
The convergence results described in Theorem~\ref{th:main:intro} are
stated formally and proven in Sec.~\ref{sec:complexity}.
Sec.~\ref{sec:semi:naive} presents a generalization of semi-na\"ive
evaluation to $\name$.  We discuss how $\name$ can express datalog
queries with negation using 3-valued logic in
Sec.~\ref{sec:fitting}.  Finally, we conclude in
Sec.~\ref{sec:conclusions}.

\section{Partially Ordered Pre-semirings}
\label{sec:pops}

In this section, we review the basic algebraic notions of (pre-)semirings,
$P$-relations, and sum-product queries.
We also introduce an extension called partially ordered pre-semiring (POPS).

\subsection{(Pre-)Semirings and POPS}

\begin{defn}[(Pre-)semiring]
  A {\em pre-semiring}~\cite{semiring_book} is a tuple
  $\bm S = (S, \oplus, \otimes, 0, 1)$ where $\oplus$ and $\otimes$ are
  binary operators on $S$ for which $(S, \oplus, 0)$ is a commutative
  monoid, $(S, \otimes, 1)$ is a monoid, and $\otimes$ distributes over
  $\oplus$.
  When the {\em absorption rule} $x \otimes 0 = 0$ holds
  for all $x \in S$, we call $\bm S$ a {\em semiring}.\footnote{Some
    references, e.g.~\cite{DBLP:journals/ai/KohlasW08}, define a
    semiring without absorption.}
  When $\otimes$ is commutative, then we say that the
  pre-semiring is {\em commutative}.  In this paper we only consider
  commutative pre-semirings, and we will simply refer to them as
  pre-semirings.
\end{defn}

In any (pre-)semiring $\bm S$, the relation $x \preceq_S y$ defined as
$\exists z: x \oplus z = y$, is a {\em preorder}, which means that it
is reflexive and transitive, but it is not anti-symmetric in general.
When $\preceq_S$ is anti-symmetric, then it is a partial order, and is
called the {\em natural order} on $\bm S$; in that case we say that
$\bm S$ is {\em naturally ordered}.

\begin{ex}
  Some simple examples of naturally ordered semirings are the Booleans
  $(\B \defeq \set{0,1}, \vee, \wedge, 0, 1)$, the natural numbers
  $(\N, +, \times , 0, 1)$, and the real numbers
  $(\R, +, \times, 0, 1)$.  We will refer to them simply as $\B, \N$
  and $\R$.  The natural order on $\B$ is $0 \preceq_\B 1$ (or {\sf
      false} $\preceq_\B$ {\sf true}); the natural order on $\N$ is the
  same as the familiar total order $\leq$ of numbers.  $\R$ is not
  naturally ordered, because $x \preceq_\R y$ holds for every
  $x,y \in \R$.  Another useful example is the {\em tropical semiring}
  $\trop^+ = (\R_+ \cup \{\infty\}, \min,+, \infty, 0)$, where the
  natural order $x \preceq y$ is the {\em reverse} order $x \geq y$ on
  $\R_+ \cup \{\infty\}$.
\end{ex}

A key idea we introduce in this paper is the decoupling of the partial order from the
algebraic structure of the (pre-)semiring.
The decoupling allows us to inject a partial order when the (pre-)semiring is not
naturally ordered, or when we need a {\em different} partial order than the natural order.

\begin{defn}[POPS] \label{def:pops} A {\em partially ordered pre-semiring},
  in short POPS, is a tuple
  $\bm P = (P, \oplus, \otimes, 0, 1, \sqsubseteq)$, where
  $(P, \oplus, \otimes, 0, 1)$ is a pre-semiring, $(P, \sqsubseteq)$
  is a poset, and $\oplus, \otimes$ are {\em monotone}\footnote{Monotonicity means
    $x\sqsubseteq x'$ and $y \sqsubseteq y'$ imply $x\oplus y \sqsubseteq x' \oplus y'$
    and $x \otimes y \sqsubseteq x' \otimes y'$.}
  operators under $\sqsubseteq$.
  In this paper, we will assume that every poset $(P, \sqsubseteq)$ has a minimum
  element denoted by $\bot$.
\end{defn}

A POPS satisfies the identities $\bot \oplus \bot = \bot$ and
$\bot \otimes \bot = \bot$, because, by monotonicity and the fact that
$(P,\oplus,0)$ and $(P,\otimes,1)$ are commutative monoids, we have
$\bot \oplus \bot \sqsubseteq \bot \oplus 0 = \bot$, and
$\bot \otimes \bot \sqsubseteq \bot \otimes 1 = \bot$.  We say that
the multiplicative operator $\otimes$ is {\em strict} if the identity
$x \otimes \bot = \bot$ holds for every $x \in P$.

Throughout this paper we will assume that $\otimes$ is strict,
unless otherwise stated.
One of the reasons for insisting on the strictness assumption is that
we can ``extract'' from a POPS a semiring, called the {\em core semiring} of the POPS,
as shown in the following simple proposition.

\begin{prop} \label{prop:s:plus:bot} Given an arbitrary POPS
  $\bm P=(P,\oplus,\otimes,0,1,\sqsubseteq)$.  Define the subset
  $P\oplus \bot \defeq \setof{x\oplus\bot}{x \in P} \subseteq P$.
  Then, $(P\oplus \bot, \oplus, \otimes, 0\oplus \bot, 1 \oplus \bot)$
  is a semiring.  We denote this semiring by $\bm P \oplus \bot$, and
  refer to it as the {\em core} semiring of $\bm P$.
\end{prop}
\begin{proof}
  We use the fact that $\otimes$ is strict, and check that
  $\bm P\oplus\bot$ is closed under $\oplus$ and $\otimes$:
  $(x\oplus\bot)\oplus(y\oplus\bot)=(x\oplus
    y)\oplus(\bot\oplus\bot)=(x\oplus y)\oplus\bot$, and similarly
  $(x\oplus\bot)\otimes(y\oplus\bot)=(x\otimes y)\oplus (x\otimes\bot)
    \oplus (y\otimes \bot) \oplus\bot = (x\otimes
    y)\oplus\bot\oplus\bot\oplus\bot=(x\otimes y)\oplus\bot \in \bm P
    \oplus \bot$.  Finally, it suffices to observe that $\bot= 0\oplus\bot$ is the identity
  for $\oplus$ and $1\oplus\bot$ is the identity for $\otimes$.
\end{proof}

Every naturally ordered semiring is a POPS, where $\bot = 0$ and
$\otimes$ is strict, and its core is itself, $\bm S \oplus 0 = \bm S$.
The converse does not hold: some POPS are not naturally ordered; a
simple example of a non-naturally ordered POPS is the set of {\em lifted
    reals}\footnote{The term ``lifted'' comes from the programming languages community where adding bottom to a set `lifts' the set.},
$\R_\bot \defeq (\R \cup \set{\bot}, +, *, 0, 1, \sqsubseteq)$, where
$x+\bot = x*\bot = \bot$ for all $x$, and $x \sqsubseteq y$ iff
$x = \bot$ or $x=y$.  Its core semiring is the trivial semiring
$\R_\bot + \bot = \set{\bot}$ consisting of a single element.  We
will consider similarly the lifted natural numbers, $\N_\bot$.

\subsection{Polynomials over POPS}

\label{subsec:polynomial}

Fix a POPS $\bm P = (P, \oplus, \otimes, 0, 1, \sqsubseteq)$.
We are interested in analysing behaviors of vector-valued multivariate functions on $P$
defined by composing $\oplus$ and $\otimes$.
These functions are multivariate polynomials.
Writing polynomials in $\bm P$ using the symbols $\oplus, \otimes$
is cumbersome and difficult to parse. Consequently, we replace them with $+, \cdot$ when the underlying
POPS $\bm P$ is clear from context; furthermore, we will also
abbreviate a multiplication $a\cdot b$ with $ab$.  As usual, $a^k$
denotes the product of $k$ copies of $a$, where $a^0 \defeq 1$.

Let $x_1, \ldots, x_N$ be $N$ variables.  A {\em monomial} (on $\bm P$) is an
expression of the form:
\begin{align}
  m \defeq & c \cdot x_1^{k_1}\cdot \cdots \cdot x_N^{k_N} \label{eq:def:monomial}
\end{align}
where $c \in P$ is some constant.  Its {\em degree} is
$\degree(m) \defeq k_1+\cdots+k_N$.  A (multivariate) {\em polynomial}
is a sum:
\begin{align}
  f(x_1, \ldots, x_N) \defeq & m_1 + m_2 + \cdots + m_q \label{eq:def:polynomial}
\end{align}
where each $m_i$ is a monomial.  The polynomial $f$ defines a function
$P^N \rightarrow P$ in the obvious way, and, with some abuse,
we will denote by $f$ both the polynomial and the function it defines.
Notice that $f$ is monotone in each of its arguments.

A {\em vector-valued polynomial function} on $\bm P$ is a function
$\bm f : P^N \rightarrow P^M$ whose component functions are polynomials.
In particular, the vector-valued polynomial function is a
  {\em tuple of polynomials} $\bm f = (f_1, \ldots, f_M)$ where each $f_i$ is a
polynomial in variables $x_1, \ldots, x_N$.

We note a subtlety when dealing with POPS:
when the POPS is not a semiring, then we cannot ``remove'' monomials
by setting their coefficient $c=0$, because $0$ is not absorbing.
Instead, we must ensure that they are not included in the polynomial~\eqref{eq:def:polynomial}.
For example, consider the POPS of the lifted reals, $\bm R_\bot$, and
the linear polynomial $f(x) = ax + b$.  If we set $a = 0$, we do not
obtain the constant function $g(x)=b$, because
$f(\bot) = a\bot + b = \bot + b = \bot \neq g(\bot)=b$.  We just have
to be careful to not include monomials we don't want.

\subsection{$P$-Relations}

\label{subsec:p:relations}

Fix a relational vocabulary, $\sigma = \set{R_1, \ldots, R_m}$, where
each $R_i$ is a relation name, with an associated arity.  Let $D$ be
an infinite domain of constants, for example the set of all strings
over a fixed alphabet, or the set of natural numbers.  Recall that an
instance of the relation $R_i$ is a finite subset of
$D^{\text{arity}(R_i)}$, or equivalently, a mapping
$D^{\text{arity}(R_i)} \rightarrow \B$ assigning $1$ to all tuples
present in the relation.  Following~\cite{DBLP:conf/pods/GreenKT07},
we generalize this abstraction from $\B$ to an arbitrary POPS $\bm P$.

Given a relation name $R_i \in \sigma$, a {\em ground atom} of $R_i$
is an expression of the form $R_i(\bm u)$, where
$\bm u \in D^{\arity(R_i)}$.  Let $\ground(R_i,D)$ denote the set of
all ground atoms of $R_i$, and
$\ground(\sigma, D) \defeq \bigcup_i \ground(R_i,D)$ denote the set of
all ground atoms over the entire vocabulary $\sigma$.  The set
$\ground(\sigma, D)$ is the familiar Herbrand base in logic
programming.  Note that each ground atom is prefixed by a relation
name.

Let $\bm P$ be a POPS.  A {\em $\bm P$-instance for $\sigma$} is a function
$I : \ground(\sigma, D) \rightarrow \bm P$ with {\em finite support},
where the support is defined as the set of ground atoms that are
mapped to elements {\em other than} $\bot$.
For example, if $\bm P$ is a naturally ordered semiring, then the support of the function
$I$ is the set of ground atoms assigned to a non-zero value.
The {\em active domain} of the instance $I$,
denoted by $\adom(I)$, is the finite set $\adom(I) \subseteq D$ of all constants that occur
in the support of $I$.  We denote by $\inst(\sigma, D, \bm P)$ the set
of $\bm P$-instances over the domain $D$.  When $\sigma$ consists of a
single relation name, then we call $I$ a {\em $\bm P$-relation}.

An equivalent way to define a $\bm P$-instance is as a function
$I : \ground(\sigma, D_0) \rightarrow \bm P$, for some finite subset
$D_0 \subseteq D$; by convention, this function is extended to the
entire set $\ground(\sigma, D)$ by setting $I(a) := \bot$ for all
$a \in \ground(\sigma,D) \setminus \ground(\sigma,D_0)$.  The set
$\inst(\sigma, D_0, \bm P)$ is isomorphic to $P^N$, where
$N=|\ground(\sigma, D_0)|$, and, throughout this paper, we will
identify a $\bm P$-instance with a tuple in $P^N$.

Thus, a $\bm P$-instance involves two domains: $D$, which is called
the {\em key space}, and the POPS $\bm P$, which is called the {\em
    value space}.  For some simple illustrations, a $\B$-relation is a
standard relation where every ground tuple is either true or false,
while an $\R$-relation is a sparse tensor.

\subsection{(Sum-)Sum-Product Queries on POPS}
\label{subsec:sum-products}

In the Boolean world, conjunctive queries and unions of conjunctive
queries are building-block queries.  Analogously, in the POPS world,
we introduce the concepts of {\em sum-product queries} and {\em
    sum-sum-product queries}.  In the most simple setting, these queries
have been studied in other communities (especially AI and machine
learning as reviewed below).  In our setting, we need to introduce one
extra feature called ``conditional'', in order to cope with the fact
that $0$ is not absorptive.

Fix two disjoint vocabularies, $\sigma, \sigma_\B$; the relation names
in $\sigma$ will be interpreted over a POPS $\bm P$, while those in
$\sigma_\B$ will be interpreted over the Booleans.  Let $D$ be a
domain, and $V = \set{X_1, \ldots, X_p}$ a set of ``key variables''
whose values are over the key space $D$. They should not be
confused with variables used in polynomials, which are interpreted
over the POPS $\bm P$; we refer to them as ``value variables'' to
contrast them with
the key variables.
We use upper case for key variables, and
lower case for value variables.  A {\em $\sigma$-atom} is an expression
of the form $R_i(\bm X)$, where $R_i \in \sigma$ and $\bm X \in (V \cup D)^{\text{arity}(R_i)}$.
% (Recall that the set of all atoms of the form $R_i(X)$ for $X \in D^{\text{arity}(R_i)}$
% is the familiar {\em Herbrand Base} in logic programming.)

\begin{defn} \label{def:sum:product} A (conditional) {\em sum-product query}, or
    {\em sum-product rule} is an expression of the form
  \begin{align}
    T(X_1, \ldots, X_k) & \cd \bigoplus_{X_{k+1}, \ldots,X_p} \setof{R_1(\bm X_1) \otimes
      \cdots \otimes R_m(\bm X_m)}{\Phi(V)}  \label{eq:t:monomial}
  \end{align}
  where $T$ is a new relation name of arity $k$, each $R_j(\bm X_j)$
  is a $\sigma$-atom, and $\Phi$ is a first-order (FO) formula over
  $\sigma_\B$, whose free variables are in
  $V = \set{X_1, \ldots, X_p}$.  The LHS of $\cd$ is called the {\em
      head}, and the RHS the {\em body} of the rule.  The variables
  $X_1, \ldots, X_k$ are called {\em free variables} of the query
  (also called {\em head variables}), and $X_{k+1}, \ldots, X_p$ are
  called {\em bound variables}.
\end{defn}

{\em Without} the conditional term $\Phi$,
the problem of computing efficiently sum-products over semirings has
been extensively studied both in the database and in the AI
literature. In databases, the query optimization and evaluation
problem is a special case of sum-product computation over the
value-space of Booleans (set semantics) or natural numbers (bag
semantics). The functional aggregate queries (FAQ)
framework~\cite{DBLP:conf/pods/KhamisNR16} extends the formulation to
queries over multiple semirings. In AI, this problem was studied by
Shenoy and Schafer~\cite{DBLP:conf/uai/ShenoyS88},
Dechter~\cite{DBLP:journals/constraints/Dechter97}, Kohlas and
Wilson~\cite{DBLP:journals/ai/KohlasW08} and others.  Surveys and more
examples can be found
in~\cite{DBLP:journals/tit/AjiM00,DBLP:books/daglib/0008195}.  These
methods use a sparse representation of the $\bm P$-relations,
consisting of a collection of the tuples in their support.

The use of a conditional $\Phi$ in the sum-product is non-standard,
but it is necessary for sum-product expressions over a POPS that is
not a semiring, as we illustrate next.
% We will define its semantic shortly; before doing so, let us illustrate the motivation
% for $\Phi$'s existence with an example.

\begin{ex} \label{ex:conditional:sum:product}
  Let $E(X,Y)$ be a $\B$-relation (i.e. a standard relation),
  representing a graph.  The following sum-product expression over $\B$ computes
  all pairs of nodes connected by a path of length 2:
  \begin{align*}
    T(X,Z) & \cd \exists_Y \left(E(X,Y) \wedge E(Y,Z)\right)
  \end{align*}
  This is a standard conjunctive
  query~\cite{DBLP:books/aw/AbiteboulHV95} (where the semantics of
  quantification over $Y$ is explicitly written).  Here
  $\sigma = \set{E}$, and $\sigma_\B = \emptyset$: we do not need the
  conditional term $\Phi$ yet.

  For the second example, consider the same graph given by $E(X,Y)$,
  and let $C(X)$ be an $\R_\bot$-relation associating to each node $X$
  a real number representing a cost, or $\bot$ if the cost is unknown;
  now $\sigma = \set{C}, \sigma_\B = \set{E}$.  The following
  sum-product expression computes the total costs of all neighbors of
  $X$:
  \begin{align}
    T(X) \cd & \sum_Y \setof{C(Y)}{E(X,Y)} \label{eq:explicit:conditional}
  \end{align}
  Usually, conditionals are avoided by using an indicator function
  $1_{E(X,Y)}$, which is defined to be $1$ when $E(X,Y)$ is true and
  $0$ otherwise, and writing the rule as
  $T(X) \cd \sum_Y \left(1_{E(X,Y)}\cdot C(Y)\right)$.  But this does
  not work in $\R_\bot$, because, when $Y$ is mapped to a
  non-neighboring node which so happens to have an unknown cost (while
  all neighbors' costs are known), we have $C(Y) = \bot$. In this
  case, $1_{E(X,Y)} \cdot C(Y) = 0 \cdot \bot = \bot$, instead of
  $0$. Since $x + \bot = \bot$ in $\R_\bot$, the result is also
  $\bot$.  One may ask whether we can re-define the POPS $\R_\bot$ so
  that $\bot \cdot 0 = 0$, but we show in
  Lemma~\ref{lemma:no:extension:for:r} that this is not possible.  The
  explicit conditional in~\eqref{eq:explicit:conditional} allows us to
  restrict the range of $Y$ only to the neighbors of $X$.
\end{ex}

We now formally define the semantics of (conditional) sum-product queries.
Due to the subtlety with POPS, we need to consider an alternative approach to evaluating
the results of sum-product queries: first compute the {\em provenance polynomials} of
the query~\eqref{eq:t:monomial} to obtain the component polynomials of a vector-valued
function, then evaluate these polynomials.  The provenance polynomials, or simply
provenance, are also called lineage, or groundings in the literature~\cite{DBLP:conf/pods/GreenKT07}.

Given an input instance $I_\B \in \inst(\sigma_\B, D, \B)$,
$I \in \inst(\sigma, D, \bm P)$, and let $D_0 \subseteq D$ be the
finite set consisting of their active domains and all constants
occurring in the sum-product expression~\eqref{eq:t:monomial}.  Let
$N \defeq |\ground(\sigma, D_0)|$ and
$M \defeq |\ground(T,D_0)| = |D_0|^k$ be the number of input ground
atoms and output ground atoms respectively.

To each of the $N$ input atoms $\ground(\sigma,D_0)$ we associate a unique POPS variable
$x_1, \ldots, x_N$. (Recall that we use upper case for key variables and
lower case for value variables.)
Abusing notation, we also write $x_{R(\bm u)}$ to mean the variable associated to the ground
atom $R(\bm u)$.
A {\em valuation} is a function $\theta : V \rightarrow D_0$.
When applied to the body of the rule~\eqref{eq:t:monomial}, the valuation $\theta$ defines the
following monomial:
\begin{align}
  \theta(\text{body}) \defeq & x_{R_1(\theta(\bm X_1))}\cdot x_{R_2(\theta(\bm X_2))}\cdots x_{R_m(\theta(\bm X_m))} \label{eq:grounding:monomial}
\end{align}
The {\em provenance polynomial}~\cite{DBLP:conf/pods/GreenKT07} of the output tuple
$T(\bm a) \in \ground(T,D_0)$ is the following:
\begin{align}
  f_{T(\bm a)}(x_1, \ldots, x_N) \defeq &
  \sum_{\substack{\theta: V \rightarrow D_0: \\ \theta(X_1,\ldots,X_k)=\bm a, \\ I_\B \models
      \Phi[\theta]}} \theta(\text{body}) \label{eq:grounding:polynomial}
\end{align}
In other words, we consider only valuations $\theta$ that map the head
variables to the tuple $\bm a$ and satisfy the FO sentence $\Phi$.
There are $M$ provenance polynomials, one for each tuple in
$\ground(T,D_0)$, and they define an $M$-tuple of polynomials in $N$
variables, $\bm f$, which in turn defines a function
$\bm f : \bm P^N \rightarrow \bm P^M$.  The semantics of the
query~\eqref{eq:t:monomial} is defined as the value of this polynomial
on the input instance $I \in \inst(\sigma, D_0, \bm P)$, when viewed
as a tuple $I \in \bm P^N$.

Note that, once we have constructed the provenance polynomial, we no longer need to deal with the
conditional $\Phi$, because the grounded version does not have $\Phi$ anymore.
In most of the rest of the paper we will study properties of vector-valued
functions whose components are these provenance polynomials.

We notice that, as defined, our semantics depends on the choice of the
domain $D_0$: if we used a larger finite domain $D'_0 \supseteq D_0$,
then the provenance polynomials will include additional spurious
monomials, corresponding to the spurious grounded tuples in $D_0'$.
Traditionally, these spurious monomials are harmless, because their
value is 0.  However, in our setting, their value is $\bot$, and they
may change the result.  This is precisely the role of the conditional
$\Phi$ in~\eqref{eq:t:monomial}: to control the range of the variables
and ensure that the semantics is domain independent.  All examples in
this paper are written such that they are domain independent.

Finally, (conditional) sum-sum-product queries are defined in the natural way:

\begin{defn} \label{def:sum:sum:product}
  A (conditional) {\em sum-sum-product query} or {\em sum-sum-product rule} has the
  form:
  \begin{align}
    T(X_1, \ldots, X_k) & \cd E_1 \oplus \cdots \oplus E_q \label{eq:sum:sum:product}
  \end{align}
  where $E_1, E_2, \ldots, E_q$ are the bodies of sum-product
  expressions~\eqref{eq:t:monomial}, each with the same free variables
  $X_1, \ldots, X_k$.
\end{defn}

The provenance polynomials of a sum-sum-product query are defined as
the sum of the provenance polynomials of the expressions
$E_1, \ldots, E_q$.

% For a simple illustration, we show a modification
% of~\eqref{eq:explicit:conditional} where we include in the total sum
% $T(X)$ the cost of $X$:
% %
% \begin{align*}
%   T(X) \cd & C(X) + \sum_Y \setof{C(Y)}{E(X,Y)}
% \end{align*}

\subsection{Properties and Examples of POPS}
\label{subsec:examples:pops}

We end this section by presenting several properties of POPS and
illustrating them  with a few examples.

\subsubsection{Extending Pre-semirings to POPS}

If $\bm S$ is a pre-semiring, then we say that a POPS $\bm P$ {\em
    extends} $\bm S$ if $S \subseteq P$ ($S$ and $P$ are their domains), and the
operations $\oplus, \otimes, 0, 1$ in $\bm S$ are the same as those in $\bm P$.
We describe three procedures to extend a pre-semiring $\bm S$ to a
POPS $\bm P$, inspired by abstract interpretations in programming
languages~\cite{DBLP:conf/popl/CousotC77}.

\begin{description}
  \item[Representing Undefined] The {\em lifted POPS} is
    $\bm S_\bot = (S \cup \set{\bot}, \oplus, \otimes, 0, 1,
      \sqsubseteq)$, where $x \sqsubseteq y$ iff $x = \bot$ or $x = y$,
    and the operations $\oplus, \otimes$ are extended to $\bot$ by
    setting $x \oplus \bot = x \otimes \bot = \bot$.  Notice that
    $\bm S_\bot$ is not a semiring, because $0$ is not absorbing:
    $0 \otimes \bot \neq 0$.  Its core semiring is the trivial semiring
    $S_\bot \oplus \bot = \set{\bot}$.  Here $\bot$ represents {\em
        undefined}.

  \item[Representing Contradiction] The {\em completed POPS} is
    $\bm S_\bot^\top = (S \cup \set{\bot, \top}, \oplus, \otimes, 0, 1,
      \sqsubseteq)$, where $x \sqsubseteq y$ iff $x=\bot$, $x=y$, or
    $y=\top$ and the operations $\oplus, \otimes$ are extended to
    $\bot, \top$ as follows: $x \oplus \bot = x \otimes \bot = \bot$ for
    all $x$ (including $x=\top$), and
    $x \oplus \top = x \otimes \top = \top$ for all $x \neq \bot$.  As
    before, its core semiring is the trivial semiring
    $S_\bot^\top \oplus \bot = \set{\bot}$.  Here $\bot, \top$ represent
    undefined and contradiction respectively.  Intuitively: $\bot$ is
    the empty set $\emptyset$, each element $x \in \bm S$ is a singleton
    set consisting of one value, and $\top$ is the entire set $S$.

  \item[Representing Incomplete Values] More generally, define
    $\calP(\bm S) = (\calP(S), \oplus, \otimes, 0, 1, \subseteq)$.  It
    consists of all subsets of $\bm S$, ordered by set inclusion, where
    the operations $\oplus, \otimes$ are extended to sets, e.g.
    $A \oplus B = \setof{x\oplus y}{x \in A, y \in B}$.  Its core
    semiring is itself, $\calP(\bm S) \oplus \set{0} = \calP(\bm S)$.
    Here $\bot=\emptyset$ represents undefined, $\top=S$ represents
    contradiction, and, more generally, every set represents some degree
    of incompleteness.
\end{description}

A lifted POPS $\bm S_\bot$ is never a semiring, because
$\bot \otimes 0 = \bot$, and the reader may ask whether there exists
an alternative way to extend it to a POPS that is also a semiring,
i.e. $0 \otimes x = 0$.  For example, we can define
$\N \cup \set{\bot}$ as a semiring by setting $x + \bot = \bot$ for
all $x$, $0 \mult \bot = 0$ and $x \mult \bot = \bot$ for $x > 0$: one can
check that the semiring laws hold.  However, this is not possible in
general.  We prove:

\begin{lmm} \label{lemma:no:extension:for:r} If $\bm S$ is any POPS
  extension of $(\R, +, \mult, 0, 1)$, then $\bm S$ is not a semiring,
  i.e. it fails the absorption law $0 \mult x = 0$.
\end{lmm}
\begin{proof}
  Let $\bm S = (S, +, \mult, 0, 1, \sqsubseteq)$ be any POPS that is an
  extension of $\R$.  In particular $\R \subseteq S$ and $S$ has a
  minimal element $\bot$.  Since 0, 1 are additive and multiplicative identities,
  we have:
  \begin{align*}
    \bot+0 & = \bot & \bot \mult 1 & = \bot
  \end{align*}
  We claim that the following more general identities hold:
  \begin{align*}
    \forall x \in \R:\  \bot+x = & \bot & \forall x \in \R \setminus \set{0}: \bot \mult x = & \bot
  \end{align*}
  To prove the first identity, we use the fact that $+$ is monotone in
  $\bm S$ and $\bot$ is the smallest element, and derive
  $\bot + x \sqsubseteq (\bot + (y-x)) +x = \bot + y$ for all
  $x,y\in\R$.  This implies $\bot + x = \bot + y$ for all $x,y$ and
  the claim follows by setting $y=0$.  The proof of the second
  identity is similar: first observe that
  $\bot \mult x \sqsubseteq (\bot \mult \frac{y}{x}) \mult x=\bot
    \mult y$ hence $\bot \mult x = \bot \mult y$ for all
  $x, y \in \R \setminus \set{0}$, and the claim follows by setting
  $y=1$.

  Assuming $\bm S$ is a semiring, it satisfies the absorption law:
  $\bot \mult 0 = 0$.  We prove now that $0 = \bot$.  Choose any
  $x \in \R \setminus \set{0}$, and derive:
  \begin{align*}
    \bot = \bot + \bot =
    (\bot  \mult  x) + (\bot \mult (-x))
    = \bot \mult (x+(-x)) & = \bot  \mult  0 = 0.
  \end{align*}
  The middle identity follows from distributivity. From $0 = \bot$, we
  conclude that $0$ is the smallest element in $\bm S$.  Then, for
  every $x \in \R$, we have $x = x+0 \sqsubseteq x+(-x) = 0$, which
  implies $x = 0$, $\forall x \in \R$, which is a contradiction.
  Thus, $\bm S$ is not a semiring.
\end{proof}

\subsubsection{The POPS THREE} \label{subsec:three:pops}

Consider the following POPS:
$\texttt{THREE} \defeq (\set{\bot, 0,  1}, \vee, \wedge,  0,
  1, \leq_k)$, where:
\begin{itemize}
  \item $\vee, \wedge$ have the semantics of 3-valued
        logic~\cite{DBLP:journals/jlp/Fitting85a}.  More precisely, define
        the {\em truth ordering} $0 \leq_t \bot \leq_t 1$ and set
        $x \vee y \defeq \max_t(x,y)$, $x \wedge y \defeq \min_t(x,y)$.  We
        note that this is precisely Kleene's three-valued
        logic~\cite{DBLP:journals/logcom/Fitting91}.
  \item $\leq_k$ is the {\em knowledge order\/}, defined as
        $\bot <_k  0$ and $\bot <_k  1$.
\end{itemize}
$\texttt{THREE}$ is not the same as the lifted Booleans, $\B_\bot$,
because in the latter $0 \wedge \bot = \bot$, while in
$\texttt{THREE}$ we have $0 \wedge \bot = 0$.  Its core semiring is
$\texttt{THREE} \vee \bot = \set{\bot, 1}$, and is isomorphic to $\B$.
We will return to this POPS in Sec.~\ref{sec:fitting}.

\subsubsection{Stable Semirings}
\label{subsec:stable:semirings}

We illustrate here two examples of semirings that are {\em stable}, a
property that we define formally in Sec.~\ref{sec:complexity}.  Both
examples are adapted from~\cite[Example 7.1.4]{semiring_book}
and~\cite[Chapt.8, Sec.1.3.2]{semiring_book} respectively.  If $A$ is
a set and $p \geq 0$ a natural number, then we denote by $\calP_p(A)$
the set of subsets of $A$ of size $p$, and by $\calB_p(A)$ the set of
bags of $A$ of size $p$.  We also define
\begin{align*}
  \calP_{\texttt{fin}}(A) & \defeq \bigcup_{p\geq 0}\calP_p(A)  &
  \calB_{\texttt{fin}}(A) & \defeq \bigcup_{p\geq 0}\calB_p(A).
                          &
\end{align*}
We denote bags as in $\bag{a,a,a,b,c,c}$.  Given
$\bm x,\bm y \in \calP_{\texttt{fin}}(\R_+ \cup \infty)$, we denote by:
\begin{align*}
  \bm x \cup \bm y \defeq & \mbox{set union of $\bm x,\bm y$}     &
  \bm x + \bm y \defeq    & \setof{u+v}{u \in \bm x, v \in \bm y}
\end{align*}
Similarly, given $\bm x,\bm y \in \calB_{fin}(\R_+ \cup \infty)$, we denote
by:
\begin{align*}
  \bm x \uplus \bm y \defeq & \mbox{bag union of $\bm x,\bm y$}     &
  \bm x + \bm y \defeq      & \bagof{u+v}{u \in \bm x, v \in \bm y}
\end{align*}

\begin{ex} \label{ex:trop:p} For any bag
  $\bm x = \bag{x_0, x_1, \ldots, x_n}$, where
  $x_0\leq x_1 \leq \ldots \leq x_n$, and any $p \geq 0$, define:
  \begin{align*}
    {\min}_p(\bm x) \defeq & \bag{x_0, x_1, \ldots, x_{\min(p,n)}}
  \end{align*}
  In other words, $\min_p$ returns the smallest $p+1$ elements of the
  bag $\bm x$.  Then, for any $p \geq 0$, the following is a semiring:
  \begin{align*}
    \trop^+_p \defeq & (\calB_{p+1}(\R_+\cup\set{\infty}), \oplus_p, \otimes_p, \bm 0_p, \bm 1_p)
  \end{align*}
  where:
  \begin{align*}
    \bm x \oplus_p \bm y \defeq  & {\min}_p(\bm x \uplus \bm y)         &
    \bm 0_p \defeq               & \bag{\infty, \infty, \ldots, \infty}   \\
    \bm x \otimes_p \bm y \defeq & {\min}_p(\bm x + \bm y)              &
    \bm 1_p \defeq               & \bag{0,\infty, \ldots, \infty}
  \end{align*}
  For example, if $p=2$ then
  $\bag{3,7,9} \oplus_2 \bag{3,7,7} = \bag{3,3,7}$ and
  $\bag{3,7,9} \otimes_2 \bag{3,7,7} = \bag{6,10,10}$.
  The following identities are easily checked, for any two finite bags
  $\bm x, \bm y$:
  \begin{align}
    {\min}_p({\min}_p(\bm x) \uplus {\min}_p(\bm y))= & {\min}_p(\bm x \uplus \bm y)                     &
    {\min}_p({\min}_p(\bm x) + {\min}_p(\bm y))=      & {\min}_p(\bm x + \bm y) \label{eq:minp:identity}
  \end{align}
  This implies that, an expression in the semiring $\trop^+_p$ can be
  computed as follows.  First, convert $\oplus, \otimes$ to
  $\uplus, +$ respectively, compute the resulting bag, then apply
  $\min_p$ only once, on the final result. $\trop^+_p$ is naturally
  ordered (see Prop~\ref{prop:trop:p:stable}) and therefore its core
  semiring is itself, $\trop^+_p \oplus_p \bm 0_p = \trop^+_p$.
  When $p=0$, then $\trop^+_p = \trop^+$, which we introduced in
  Example~\ref{ex:intro}.
\end{ex}

\begin{ex} \label{ex:trop:eta} Fix a real number $\eta \geq 0$, and
  denote by $\calP_{\leq \eta}(\R_+\cup\set{\infty})$ the set of
  nonempty, finite sets $\bm x = \set{x_0, x_1, \ldots, x_p}$ where
  $\min(\bm x) \leq \max(\bm x)\leq \min(\bm x) + \eta$.  Given any
  finite set $\bm x \in \calP_{\texttt{fin}}(\R_+\cup\set{\infty})$,
  we define
  \begin{align*}
    {\min}_{\leq \eta}(\bm x) \defeq & \setof{u}{u \in \bm x, u \leq \min(\bm x) + \eta}
  \end{align*}
  In other words, $\min_{\leq \eta}$ retains from the set $\bm x$ only
  the elements at distance $\leq \eta$ from its minimum.  The
  following is a semiring:
  \begin{align*}
    \trop^+_{\leq\eta} \defeq & (\calP_{\leq \eta}(\R_+\cup\set{\infty}),\oplus_{\leq\eta},\otimes_{\leq\eta},\bm 0_{\leq\eta},\bm 1_{\leq\eta})
  \end{align*}
  where:
  \begin{align*}
    \bm x \oplus_{\leq\eta} \bm y \defeq  & {\min}_{\leq\eta}(\bm x \cup \bm y) &
    \bm 0_{\leq\eta} \defeq               & \set{\infty}                          \\
    \bm x \otimes_{\leq\eta} \bm y \defeq & {\min}_{\leq\eta}(\bm x + \bm y)    &
    \bm 1_{\leq\eta} \defeq               & \set{0}
  \end{align*}
  For example, if $\eta = 6.5$ then:
  $\set{3,7} \oplus_{\leq \eta} \set{5,9,10} = \set{3,5,7,9}$ and
  $\set{1,6} \otimes_{\leq \eta} \set{1,2,3} = \set{2,3,4,7,8}$.
  The following identities are easily checked, for any two finite
  sets $\bm x, \bm y$:
  \begin{align}
    {\min}_{\leq \eta}({\min}_{\leq \eta}(\bm x) \cup {\min}_{\leq \eta}(\bm y))= & {\min}_{\leq \eta}(\bm x \cup \bm y)                         &
    {\min}_{\leq \eta}({\min}_{\leq \eta}(\bm x) + {\min}_{\leq \eta}(\bm y))=    & {\min}_{\leq \eta}(\bm x + \bm y) \label{eq:mineta:identity}
  \end{align}
  It follows that expressions in $\trop^+_{\leq\eta}$ can be computed
  as follows: first convert $\oplus, \otimes$ to $\cup, +$
  respectively, compute the resulting set, and apply the
  $\min_{\leq\eta}$ operator only once, on the final result.
  $\trop^+_{\leq \eta}$ is naturally ordered (see
  Prop.~\ref{prop:trop:eta:stable}) and therefore its core semiring is
  itself,
  $\trop^+_{\leq \eta}\oplus_{\leq \eta} \bm 0_{\leq \eta} =
    \trop^+_{\leq\eta}$.
  Notice that, when $\eta=0$, then we recover again
  $\trop^+_{\leq \eta} = \trop^+$.
\end{ex}

The reader may wonder why $\trop^+_p$ is defined to consist of bags of
$p+1$ numbers, while $\trop^+_{\leq \eta}$ is defined on sets.  The main
reason is for consistency with~\cite{semiring_book}.  We could have
defined either semirings on either sets or bags, and both
identities~\eqref{eq:minp:identity} and ~\eqref{eq:mineta:identity}
continue to hold, which is sufficient to prove the semiring
identities.  However, the {\em stability} property which we define and
prove later (Proposition~\ref{prop:trop:eta:stable}) holds for
$\trop^+_{\leq \eta}$ only if it is defined over sets; in contrast,
$\trop^+_p$ is stable for either sets or bags
(Proposition~\ref{prop:trop:p:stable}).

\subsubsection{Nontrivial Core Semiring}

In all our examples so far the core semiring $\bm P \oplus \bot$ is
either $\set{\bot}$ or $\bm P$.  We show next that the core semiring
may be non-trivial.  If $\bm P_1, \bm P_2$ are two POPS, then their
Cartesian product $\bm P_1 \times \bm P_2$ is also a POPS: operations
are defined element-wise, e.g.
$(x_1,x_2) \oplus (y_1,y_2) \defeq (x_1\oplus_1 y_1, x_2 \oplus_2
  y_2)$, etc, the order is defined component-wise, and the smallest
element is $(\bot_1, \bot_2)$.

\begin{ex} Consider the following two POPS.
  \begin{itemize}
    \item A naturally ordered semiring
          $\bm S = (S, \oplus_S, \otimes_S, 0_S, 1_S, \sqsubseteq_S)$.  Its
          core semiring is itself $\bm S \oplus_S 0_S=\bm S$.
    \item Any POPS $\bm P$ where addition is strict:
          $x \oplus_P \bot = \bot$.  (For example, any lifted semiring.)
          Its core semiring is $\bm P \oplus_P \bot_P = \set{\bot_P}$.
  \end{itemize}
  Consider the Cartesian product $\bm S \times \bm P$.  The smallest
  element is $(0_S,\bot_P)$, and the core semiring is
  $(\bm S \times \bm P) \oplus (0_S,\bot_P) = \bm S \times
    \set{\bot_P}$, which is a non-trivial subset of
  $\bm S \times \bm P$.
\end{ex}

%
% \subsubsection{Strictness of \texorpdfstring{$\oplus$}{oplus} and \texorpdfstring{$\otimes$}{otimes} are independent}
% We give two examples of POPS where only one of the two operators
% $\oplus, \otimes$ is strict.
%
% \begin{description}
% \item[Strict $\otimes$] In any non-trivial naturally ordered semiring
%   $\otimes$ is strict, while $\oplus$ is not.  For example, consider
%   $(\N, +, *, 0, 1, \leq)$.  Then $*$ is strict because $x * 0 = 0$,
%   while $+$ is not strict because $x + 0 \neq 0$ for $x \neq 0$.
% \item[Strict $\otimes$] Next, consider the semiring
%   $(\N \cup \set{\top}, \oplus, \otimes, \set{0}, \set{1},
%   \supseteq)$, where $\top$ the infinite set $\set{0,1,2,\ldots}$.  We
%   view each element $x \in \N$ as the singleton set $\set{x}$, and
%   define the operations $\oplus, \otimes$ set-wise:
%   $A \oplus B \defeq \setof{x+y}{x \in A, y \in B}$,
%   $A \otimes B \defeq \setof{x*y}{x \in A, y \in B}$, where we replace
%   the result with $\top$ if the resulting set is not a singleton.
%   Concretely, we have:
%   \begin{align*}
%     \set{x} \oplus \set{y} = & \set{x+y} & \set{x} \oplus \top = & \top\\
%     \set{x} \otimes \set{y} = & \set{x*y} & \set{0} \oplus \top = & \set{0} & x\neq 0:\ \set{x} \oplus \top = \top
%   \end{align*}
% %
%   Since $\top$ is the smallest element, $\oplus$ is strict, while
%   $\otimes$ is not.
% \end{description}
%
%

\section{Least Fixpoint}
\label{sec:lfp}

We review here the definition of a least fixpoint and prove some
results needed to characterize the convergence of $\name$ programs.
Fix a {\em partially ordered set} (poset), $\bm L = (L, \sqsubseteq)$.
As mentioned in Sec.~\ref{sec:pops}, in this paper we will assume
that each poset has a minimum element $\bot$, unless explicitly
specified otherwise.  We denote by $\bigvee A$, or $\bigwedge A$
respectively, the least upper bound, or greatest lower bound of a set
$A \subseteq L$, when it exists.  We assume the usual definition of a
monotone function $f$ between two posets, namely
$f(x) \sqsubseteq f(y)$ whenever $x \sqsubseteq y$.  The Cartesian
product of two posets $\bm L_1 = (L_1, \sqsubseteq_1)$ and
$\bm L_2 = (L_2,\sqsubseteq_2)$, is also the standard component-wise
ordering:
$\bm L_1 \times \bm L_2 \defeq (L_1 \times L_2, \sqsubseteq)$, where
$(x_1,x_2) \sqsubseteq (y_1,y_2)$ if $x_1 \sqsubseteq_1 y_1$ and
$x_2 \sqsubseteq_2 y_2$.  An {\em $\omega$-chain} in a poset $\bm L$
is a sequence
$x_0 \sqsubseteq x_1 \sqsubseteq x_2 \sqsubseteq \cdots$, or,
equivalently, it is a monotone function $\N \rightarrow \bm L$.  We
say that the chain is finite if there exists $n_0$ such that
$x_{n_0}=x_{n_0+1}=x_{n_0+2}=\cdots$, or, equivalently, if
$x_{n_0} = \bigvee x_n$.

Given a monotone function $f \colon \bm L \rightarrow \bm L$,  a {\em
    fixpoint} is an element $x$ such that $f(x) = x$.  We denote by
$\lfp_{\bm L}(f)$ the {\em least fixpoint} of $f$, when it exists, and
drop the subscript $\bm L$ when it is clear from the context.
Consider the following $\omega$-sequence:
\begin{align}
  f^{(0)}(\bot) \defeq & \bot & f^{(n+1)}(\bot) \defeq & f(f^{(n)}(\bot)) \label{eq:fn}
\end{align}
If $x$ is any fixpoint of $f$, then $f^{(n)}(\bot) \sqsubseteq x$;
this follows immediately by induction on $n$.  In order to ensure that
the least fixpoint exists, several authors require the semiring to be
$\omega$-complete and $f$ to be $\omega$-continuous.  In that case the
least upper bound $\bigvee_n f^{(n)}(\bot)$ always exists, and is
equal to $\lfp(f)$, due to Kleene's theorem;
see~\cite{davey1990introduction}.  This condition was used extensively
in the formal language
literature~\cite{MR1470001,DBLP:conf/popl/CousotC77}, and also by
Green et al.~\cite{DBLP:conf/pods/GreenKT07}.  We do not use this
condition in this paper, and will not define it formally.

Instead, we are interested in conditions that ensure that the
sequence~\eqref{eq:fn} reaches a fixpoint after a {\em finite} number
of steps, which justifies the following definition:

% \begin{defn}[ACC]
%   A poset $\bm L$ satisfies the {\em Ascending Chain Condition}, or ACC~\cite{MR1728440},
%   if it has {\em no infinite $\omega$-chains}.
%   The {\em rank} of a
%   strictly increasing chain $x_0 < x_1 < \cdots < x_k$ is $k$.  We say
%   that $\bm L$ has rank $k$ if every strictly increasing chain has
%   rank $\leq k$~\cite{MR2868112}.
%     \label{def:order:properties}
% \end{defn}

\begin{defn} \label{def:f:stable} A monotone function $f$ on $\bm L$
  (i.e. $f \colon \bm L \rightarrow \bm L$) is called $p$-stable if
  $f^{(p+1)}(\bot) = f^{(p)}(\bot)$.  The {\em stability index} of $f$
  is the minimum $p$ for which $f$ is $p$-stable.  The function $f$ is
  said to be {\em stable} if it is $p$-stable for some $p \geq 0$.
\end{defn}

% If $\bm L$ has rank $k$ then every monotone function $f$ is
% $k$-stable.  If ACC holds, then every monotone $f$ is stable.
% If a function $f$ is $p$-stable, then it has a least fixpoint and
% $\lfp(f) = f^{(p)}(\bot)$.

% In this paper we need to compute the least fixpoint of functions over
% product spaces, $\bm L_1 \times \bm L_2$, or $\bm L^D$ for some set
% $D$, ordered pointwise. The ACC immediately extends to
% products~\cite{davey1990introduction}:
%
% \begin{prop} \label{prop:l1:times:l2} Suppose posets
%   $\bm L_1, \bm L_2, \bm L$ satisfy ACC.  Then
%   $\bm L_1 \times \bm L_2$ satisfies ACC. If
%   $D$ is finite, $\bm L^D$ also satisfies ACC.
%   If $\bm L_1, \bm L_2, \bm L$ have ranks $k_1, k_2, k$ then
%   $\bm L_1 \times \bm L_2$ has rank $k_1+k_2$ and $\bm L^D$ has rank
%   $k\cdot |D|$.
% \end{prop}

If $f$ is $p$-stable, then $\lfp(f)$ exists and is equal to
$f^{(p)}(\bot)$.  Indeed, $f^{(p)}(\bot)$ is a fixpoint of $f$ by
definition, and, as mentioned earlier, it is below any fixpoint $x$ of
$f$.  In this case we will also say that the sequence~\eqref{eq:fn}
{\em converges}.

A sufficient condition for convergence often found in the literature
is the {\em Ascending Chain Condition}, or ACC, see
e.g.~\cite{MR1728440,DBLP:journals/jacm/EsparzaKL10}: the poset
$\bm L$ satisfies ACC if it has {\em no infinite $\omega$-chains},
meaning that every strictly increasing chain
$x_0 \sqsubset x_1 \sqsubset x_2 \sqsubset \cdots$ must be finite.  If
$\bm L$ satisfies ACC, then every function $\bm f$ on $\bm L$ is
stable, and thus has a least fixpoint.
% An even stricter condition
% than ACC is the following.  Define the {\em rank} of a strictly
% increasing chain $x_0 < x_1 < \cdots < x_k$ to be $k$.  We say that
% $\bm L$ has rank $k$ if every strictly increasing chain has rank
% $\leq k$~\cite{MR2868112}, and we say that it has finite rank if there
% exists $k$ s.t. $\bm L$ has rank $k$.  Every finite poset has finite
% rank, and every poset with finite rank satisfies ACC.  Moreover, these
% notions extend naturally to cartesian products:
One can also check that, if $\bm L_1, \ldots, \bm L_n$ satisfy ACC,
then so does $\bm L \defeq \bm L_1 \times \cdots \times \bm L_n$.
% one can each that $\bm L$ satisfies ACC if each $\bm L_i$ satisfies
% ACC, $i=1,n$, and if each $\bm L_i$ has a finite rank $k_i$, then
% $\bm L$ has finite rank $k_1 + \cdots +k_n$.
However, as we will see shortly in Sec.~\ref{sec:complexity}, when $f$
is a polynomial, then the ACC is only a sufficient, but not necessary
condition for stability.

In this paper we study the stability of functions on a product of
posets. To this end, we start by considering two posets, $\bm L_1$ and
$\bm L_2$ with minimum elements $\bot_1$ and $\bot_2$ respectively,
and two monotone functions:
\begin{align*}
  f : & \bm L_1 \times \bm L_2 \rightarrow \bm L_1 &
  g : & \bm L_1 \times \bm L_2 \rightarrow \bm L_2
\end{align*}
Let $h$ be the vector-valued function with components $f$ and $g$, i.e.
$h \defeq (f,g) : \bm L_1 \times \bm L_2 \rightarrow \bm L_1 \times
  \bm L_2$.  Our goal is to compute the fixpoint of $h$ by using the
fixpoints of $f$ and $g$.  We start with a simple case:
%The following lemma essentially says that, if $g$ is only a function of $L_2$,
%then we can get to a fixpoint $(\bar x, \bar y)$ of the function $h(x,y)$ by iterating
%on $y$ to get to a fixpoint $\bar y$ of $g$, and, fixing $y = \bar y$, iterating
%on $x$ to get to $(\bar x, \bar y)$.

\begin{lmm} \label{lemma:fixpoints:particular} Assume that $g$ does
  not depend on the first argument, i.e.
  $g: \bm L_2 \rightarrow \bm L_2$.  If $p,q$ are two numbers such
  that $g$ is $q$-stable and, denoting
  $\bar y \defeq g^{(q)}(\bot_2)$, the function
  $F(x) \defeq f(x,\bar y)$ is $p$-stable, then $\lfp(h)$ exists and
  is equal to $\lfp(h) = (\bar x, \bar y)$ where
  $\bar x \defeq F^{(p)}(\bot_1)$.  Moreover, $h$ is $p+q$-stable.
\end{lmm}

\begin{proof}
  We verify that $(\bar x, \bar y)$ is a fixpoint of $h$, by direct
  calculation:
  $h(\bar x, \bar y) = (f(\bar x, \bar y), g(\bar y)) = (F(\bar x),
    g(\bar y)) = (\bar x, \bar y)$.

  Next, we show that the stability index of $h$ is at most $p+q$. For convenience, define:
  \begin{align*}
    y_0 \defeq       & \bot_2,           & \forall \ell\geq 0:\ \ y_{\ell+1} \defeq     & g(y_\ell),  \\
    x_0 \defeq       & \bot_1,           & \forall k\geq 0:\ \ x_{k+1} \defeq           & f(x_k,y_q), \\
    (a_0,b_0) \defeq & (\bot_1, \bot_2), & \forall n\geq 0:\ \ (a_{n+1},b_{n+1}) \defeq & h(a_n,b_n).
  \end{align*}
  By definition, $\bar x = x_p$ and $\bar y = y_q$. We claim the
  following three statements:
  \begin{align}
    \forall n:\ \ (a_n,b_n) \sqsubseteq                             & (\bar x, \bar y) \label{eq:fg:claim:1}  \\
    \forall \ell \in \{0,\dots,q\}:\ \ (\bot_1, y_\ell) \sqsubseteq & (a_\ell,b_\ell) \label{eq:fg:claim:2}   \\
    \forall k \in \{0,\dots,p\}:\ \ (x_k, y_q) \sqsubseteq          & (a_{q+k},b_{q+k}) \label{eq:fg:claim:3}
  \end{align}
  Assuming the claims hold, by setting $k=p$ in Eq.~\eqref{eq:fg:claim:3} we get
  $(\bar x, \bar y) = (x_p, y_q) \sqsubseteq (a_{p+q},b_{p+q})$,
  which, together with inequality~\eqref{eq:fg:claim:1}, proves that
  $(a_{p+q}, b_{p+q})=(\bar x, \bar y)$ and that $h$ is $p+q$-stable.

  Now we prove the claims.
  \begin{itemize}
    \item Eq.~\eqref{eq:fg:claim:1} is immediate, because $(\bar x, \bar y)$
          is a fixpoint of $h$, and $(a_n,b_n) = h^{(n)}(\bot_1,\bot_2)$ is below
          any fixpoint of $h$ (due to monotonicity of $h$).
    \item To prove Eq~\eqref{eq:fg:claim:2}, we claim that
          $y_\ell = b_\ell$ for all $\ell$.  Indeed,
          $(a_{\ell+1},b_{\ell+1}) = h(a_\ell,b_\ell) =
            (f(a_\ell,b_\ell),g(b_\ell))$, which implies $b_{\ell+1}=g(b_\ell)$,
          which means that $y_\ell$ and $b_\ell$ are the same sequence.
    \item Finally, we show Eq.~\eqref{eq:fg:claim:3} by induction on $k$.  The
          base case, $k=0$, follows from
          $(x_0,y_q) = (\bot_1, y_q) \sqsubseteq (a_q, b_q)$ by
          Eq~\eqref{eq:fg:claim:2}.  Assuming the claim holds for $k$, we have
          $(x_{k+1},y_q) = (f(x_k,y_q),y_q) \sqsubseteq
            (f(a_{q+k},b_{q+k}),b_{q+k}) \sqsubseteq
            (f(a_{q+k},b_{q+k}),g(b_{q+k})) = h(a_{q+k},b_{q+k}) =
            (a_{q+k+1},b_{q+k+1})$, proving that the claim holds for $k+1$.
  \end{itemize}
\end{proof}

Next, we generalize the result to the case when both functions depend
on both inputs: $f : \bm L_1 \times \bm L_2 \rightarrow \bm L_1$ and
$g : \bm L_1 \times \bm L_2 \rightarrow \bm L_2$. We prove:

\begin{lmm} \label{lemma:fixpoints:general} Assume that $p,q$ are two
  numbers such that, for each $u \in \bm L_1$, the function
  $g_u(y)\defeq g(u,y)$ is $q$-stable, and the function
  $F(x) \defeq f(x,g_x^{(q)}(\bot_2))$ is $p$-stable.  Then the
  following hold:
  \begin{enumerate}
    \item \label{item:lemma:fixpoints:general:1} The function $h$ is
          $(pq+p+q)$-stable, and has the least fixpoint $(\bar x, \bar y)$,
          where
          \begin{align}
            \bar x  \defeq & F^{(p)}(\bot_1) & \bar y \defeq & g_{\bar x}^{(q)}(\bot_2) \label{eq:xstar:ystar}
          \end{align}
    \item \label{item:lemma:fixpoints:general:2} Further assume that
          $f,g$ also satisfy the symmetric conditions:
          $f_v(x) \defeq f(x,v)$ is $p$-stable for all $v \in \bm L_2$, and
          $G(y) \defeq g(f_y^{(p)}(\bot_1),y)$ is $q$-stable.  Then, denoting
          $(a_n,b_n) \defeq h^{(n)}(\bot_1, \bot_2)$, the following equalities
          hold:
          \begin{align}
            a_{pq+p} = & \bar x & b_{pq+q} = & \bar y  \label{eq:xstar:ystar:2}
          \end{align}
          In particular, $h$ is  $pq+\max(p,q)$-stable.
  \end{enumerate}
\end{lmm}

\begin{proof} We start by proving
  Item~(\ref{item:lemma:fixpoints:general:1}), generalizing the proof of
  Lemma~\ref{lemma:fixpoints:particular}.  To verify that
  $(\bar x, \bar y)$ is a fixpoint of $h$,
  we need to show that $f(\bar x, \bar y) = \bar x$ and $g(\bar x, \bar y) = \bar y$.
  Indeed, these follow from $q$-stability of $g_{\bar x}$ and $p$-stability of $F$:
  \begin{align*}
    g(\bar x, \bar y) & = g_{\bar x}(\bar y) =g_{\bar x}(g_{\bar x}^{(q)}(\bot_2))=g_{\bar
    x}^{(q+1)}(\bot_2)=g_{\bar x}^{(q)}(\bot_2)=\bar y                                           \\
    f(\bar x, \bar y) & = f(\bar x, g_{\bar x}^{(q)}(\bot_2)) = F(\bar x) = F(F^{(p)}(\bot_1)) =
    F^{(p)}(\bot_1) = \bar x
  \end{align*}

  To complete the proof of Item~(\ref{item:lemma:fixpoints:general:1}), we next prove that
  $h$ is $(pq+p+q)$-stable. To this end, define the following sequences:
  \begin{align*}
    x_0 \defeq        & \bot_1           & \forall k \in \{0,\dots,p-1\}:\ \ x_{k+1} \defeq        & f(x_k,y_{k,q})    \\
    y_{k,0} \defeq    & \bot_2           & \forall \ell \in \{0,\dots,q-1\}:\ \ y_{k,\ell+1}\defeq & g(x_k,y_{k,\ell}) \\
    (a_0, b_0) \defeq & (\bot_1, \bot_2) & \forall n \geq 0:\ \ (a_{n+1},b_{n+1}) \defeq           & h(a_n,b_n)
  \end{align*}
  The sequences $x_k$ and $y_{k,\ell}$ are illustrated in
  Fig.~\ref{fig:lfp:fg}.
  We claim that the sequences satisfy the following two properties:
  \begin{align}
    \forall n: \ \ (a_n,b_n)                                                        & \sqsubseteq (\bar x, \bar y) \label{eq:fg:general:claim:1}                                                                  \\
    \forall k\in \{0,\dots,p\}, \forall \ell \in \{0,\dots,q\}:\ \ (x_k,y_{k,\ell}) & \sqsubseteq (a_n,b_n)                                      &  & \mbox{where } n = k(q+1)+\ell \label{eq:fg:general:claim:2}
  \end{align}
  Before proving them, we show how they help complete the proof of both
  (\ref{item:lemma:fixpoints:general:1}) and
  (\ref{item:lemma:fixpoints:general:2}).
  \begin{itemize}
    \item By setting $(k,\ell) \defeq (p,q)$ and $n = p(q+1)+q = pq+p+q$ in
          Eq.~\eqref{eq:fg:general:claim:2} we obtain
          $(\bar x, \bar y) = (x_p,y_{p,q}) \sqsubseteq (a_n,b_n)$, which,
          together with Eq.~\eqref{eq:fg:general:claim:1} proves that
          $(\bar x, \bar y) = (a_n,b_n)$ and, therefore, $h$ is $pq+p+q$-stable.
          This completes the proof of Item~(\ref{item:lemma:fixpoints:general:1}) of the lemma.

    \item For Item~(\ref{item:lemma:fixpoints:general:2}), we notice that if we
          set $(k,\ell) \defeq (p,0)$ and $n=p(q+1)=pq+p$ in
          Eq.~\eqref{eq:fg:general:claim:2}, then we obtain
          $(\bar x, y_{p,0}) = (x_p,y_{p,0}) \sqsubseteq (a_n,b_n)$, which
          implies $\bar x = a_n = a_{pq+p}$.  By switching the roles of $f, g$
          we also obtain $\bar y = b_{pq+q}$, which proves the lemma.  In
          particular, $h$ is $pq+\max(p,q)$-stable.
  \end{itemize}

  We now prove the two claims~\eqref{eq:fg:general:claim:1} and~\eqref{eq:fg:general:claim:2}.
  The first claim~\eqref{eq:fg:general:claim:1} is immediate,
  because $(\bar x, \bar y)$ is a fixpoint of $h$ and
  $(a_n,b_n) = h^{(n)}(\bot_1, \bot_2)$ is below any fixpoint.

  For the second claim in Equation~\eqref{eq:fg:general:claim:2},
  refer to Fig.~\ref{fig:lfp:fg} for some intuition; in particular, note that the
  mapping $(k,\ell) \mapsto n \defeq k(q+1)+\ell$ is injective for
  $k \in \{0,\dots,p\}$, $\ell \in \{0,\dots,q\}$, and $n$ represents the position of
  $(x_k,y_{k,\ell})$ in the sequence defined in the figure.
  Note also that, $y_{k,q} = \bar y$ for all $k\in \{0,\dots,p\}$ and $x_p = \bar x$;
  and, $a_n \sqsubseteq f(a_n,b_n)$ and $b_n \sqsubseteq g(a_n,b_n)$: this
  follows from $h$ being monotone: $h^{(n)}(\bot_1,\bot_2) \sqsubseteq h^{(n+1)}(\bot_1,\bot_2)$.
  We prove Eq.~\eqref{eq:fg:general:claim:2} by induction on $n$.

  \begin{figure*}
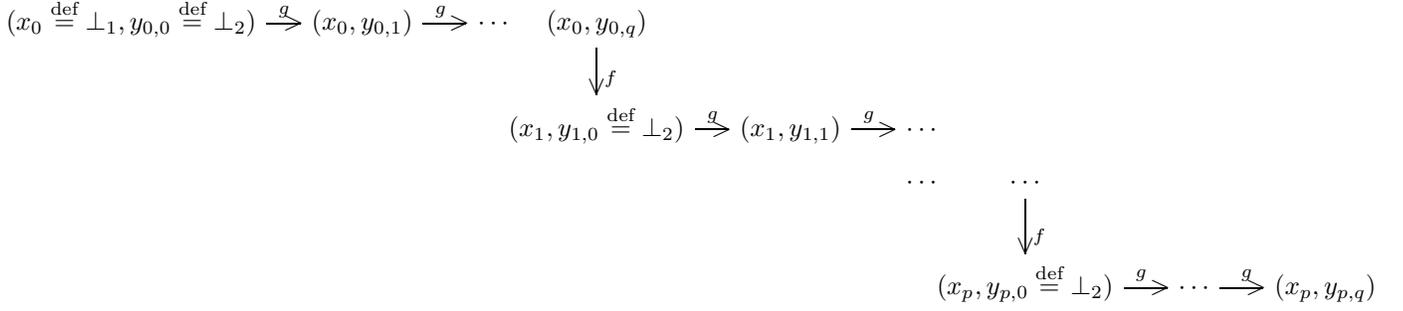
\centering
    \begin{diagram}
      (x_0\defeq \bot_1,y_{0,0}\defeq \bot_2) & \rTo^g & (x_0, y_{0,1}) & \rTo^g & \cdots & (x_0,y_{0,q}) \\
      & & & & & \dTo_f \\
      & & & & & (x_1,y_{1,0}\defeq \bot_2) & \rTo^g & (x_1,y_{1,1}) & \rTo^g & \cdots \\
      & & & & &  &  & & & \cdots &\cdots\\
      & & & & & & & & & & \dTo_f \\
      & & & & & & & & & & (x_p,y_{p,0}\defeq \bot_2) & \rTo^g & \cdots & \rTo^g & (x_p, y_{p,q})
    \end{diagram}
    \caption{Computing the fixpoint of $(f,g)$.}
    \Description{Computing the fixpoint of $(f,g)$.}
    \label{fig:lfp:fg}
  \end{figure*}

  The base case when $n=0$ then $k=\ell=0$ is trivial, as both sides of
  Eq.~\eqref{eq:fg:general:claim:2} are $(\bot_1,\bot_2)$.
  For the inductive case, assume $n > 0$, and let $k,\ell$ be the unique values s.t.
  $n = k(q+1)+\ell$. Consider two cases, corresponding to whether we are taking a horizontal
  step or a vertical step in Fig.~\ref{fig:lfp:fg}:

  {\bf Case 1:} $\ell > 0$.
  Then the pre-image of $n-1$ is $k,\ell-1$, in other words
  $(n-1) = k(q+1) + (\ell-1)$, and we have
  $(x_k,y_{k,\ell-1})\sqsubseteq (a_{n-1},b_{n-1})$ by induction
  hypothesis for $n-1$.  It follows that
  \begin{multline*}
    (x_k,y_{k,\ell}) = (x_k,g(x_k,y_{k,\ell-1})) \sqsubseteq (a_{n-1},g(a_{n-1},b_{n-1}))
    \sqsubseteq (f(a_{n-1},b_{n-1}),g(a_{n-1},b_{n-1})) = (a_n,b_n)
  \end{multline*}
  where we used the induction hypothesis and the fact that
  $a_{n-1} \sqsubseteq f(a_{n-1},b_{n-1})$.

    {\bf Case 2:} $\ell=0$.  Then the pre-image of $n-1$ is $(k-1,q)$, in
  other words $(n-1) = (k-1)(q+1) + q$, and we have
  $(x_{k-1},y_{k-1,q}) \sqsubseteq (a_{n-1},b_{n-1})$ by induction
  hypothesis.  It follows:
  \begin{multline*}
    (x_k,y_{k,0}) = (x_k,\bot_2) = (f(x_{k-1},y_{k-1,q}),\bot_2)
    \sqsubseteq (f(x_{k-1},y_{k-1,q}),y_{k-1,q}) \\
    \sqsubseteq (f(a_{n-1},b_{n-1}),b_{n-1}) \sqsubseteq (f(a_{n-1},b_{n-1}),g(a_{n-1},b_{n-1})) =(a_n,b_n)
  \end{multline*}

  This completes the proof of Eq.~\eqref{eq:fg:general:claim:2}.

\end{proof}

Next, we present Theorem~\ref{th:fixpoints:general}, which generalizes
Lemma~\ref{lemma:fixpoints:general} from 2 functions to $n$ functions.
Recall that in the lemma we had to consider two derived functions
$g_u$ and $F$ from $f$ and $g$.  When generalizing to $n$ functions,
the number of derived functions becomes unwieldy, and it is more
convenient to state the theorem for a clone of functions.  A {\em
    clone}~\cite{DBLP:conf/dagstuhl/2008coc} over $n$ posets
$\bm L_1, \ldots, \bm L_n$ is a set of functions $\calC$ where (1)
each element $f \in \calC$ of the form
$f : \bm L_{j_1} \times \cdots \times \bm L_{j_k} \rightarrow \bm
  L_i$, where $1 \leq j_1 < j_2 < \ldots < j_k \leq n$ and
$1 \leq i \leq n$, is monotone, (2) $\calC$ contains all projections
$\bm L_{j_1} \times \cdots \times \bm L_{j_k} \rightarrow \bm
  L_{j_i}$, and (3) $\calC$ is closed under composition, i.e. it
contains the function $g\circ (f_1, \ldots, f_k)$ whenever
$f_1, \ldots, f_k, g \in \calC$ and their types make the composition
correct.  We call $\calC$ a {\em c-clone} if it also contains all
constant functions: $g_u \colon () \rightarrow \bm L_i$,
$g_u() \defeq u$, for every fixed $u \in L_i$.

For a simple illustration of a c-clone, consider a POPS $\bm P$, and
define $\calC$ to consist of all multivariate polynomials over
variables $x_1, \ldots, x_n$.  More precisely, set
$\bm L_1= \cdots= \bm L_n \defeq \bm P$, and, for all choices of
indices $1 \leq j_1 < j_2 < \cdots < j_k \leq n$ and
$1 \leq i \leq n$, let $\calC$ contain all polynomials
$f(x_{j_1}, \ldots, x_{j_k})$, viewed as functions
$\bm L_{j_1}\times \cdots \times \bm L_{j_k} \rightarrow \bm L_i$.
Then $\calC$ is a c-clone.

\begin{thm} \label{th:fixpoints:general} Let $\calC$ be a c-clone of
  functions over $n$ posets $\bm L_1, \ldots, \bm L_n$, and assume
  that, for every $i \in [n]$ where we denote by $[n]$ the integers from 1 to $n$, every function
  $f \colon \bm L_i \rightarrow \bm L_i$ in $\calC$ is $p_i$-stable.
  Assume w.l.o.g. that $p_1 \geq p_2 \geq \cdots \geq p_n$.  Let
  $f_1, \ldots, f_n$ be functions in $\calC$, where
  $f_i : \bm L_1 \times \cdots \times \bm L_n \rightarrow \bm L_i$,
  and define the function $h \defeq(f_1, \ldots, f_n)$.  Then $h$ is
  $p$-stable, where $p \defeq \sum_{k=1,n}\prod_{i=1,k}p_i$.
  Moreover, this upper bound is tight: there exist posets
  $\bm L_1, \ldots, \bm L_n$, a c-clone $\calC$, and functions
  $f_1, \ldots, f_n$, such that $p$ is the stability index of $h$.
\end{thm}
\begin{proof} We defer the lower bound to Appendix~\ref{app:fixpoint},
  and prove here the upper bound.  It will be
  convenient to define the following expression, for every $m \geq 0$
  and numbers $a_1, \ldots, a_m$:
  \begin{align*}
    E_m(a_1,\ldots,a_m) \defeq & a_1 + a_1a_2 + a_1a_2a_3 \cdots + a_1a_2 \cdots a_m = \sum_{i=1,m}\prod_{j=1,i}a_j
  \end{align*}
  Note that if we permute the sequence $a_1, \ldots, a_m$, the
  expression $E_m(a_1, \ldots, a_m)$ is maximized when the sequence is
  decreasing, $a_1 \geq a_2 \geq \cdots \geq a_m$.

  We prove by induction on $n$ that $h$ is $E_n(p_1, \ldots, p_n)$-stable.
  When $n=1$ then the statement holds
  vacuously, because a function $f \colon \bm L_1 \rightarrow \bm L_1$
  is $p_1$-stable by assumption.  Assume $n > 1$ and let
  $f_1, \ldots, f_n$ be as in the statement of the theorem.

  Fix an arbitrary dimension $i \in [n]$. Let
  $\bm L_{-i} \defeq \bm L_1 \times \cdots \times \bm L_{i-1} \times
    \bm L_{i+1} \cdots \times \bm L_n$ be the product of all posets
  other than $\bm L_i$.  Given
  $\bm x = (x_1, \ldots, x_n) \in \bm L_1 \times \ldots \times \bm
    L_n$, denote by
  \begin{align}
    x_{-i} & \defeq (x_1, \ldots, x_{i-1}, x_{i+1}, \ldots, x_n) \in \bm L_{-i}
  \end{align}
  the vector consisting of all coordinates other than $i$.  Define the
  functions $f: \bm L_i \times \bm L_{-i} \rightarrow \bm L_i$ and
  $g: \bm L_i \times \bm L_{-i} \rightarrow \bm L_{-i}$:
  \begin{align*}
    f(x_i,x_{-i}) & \defeq f_i(x)                                                   &
    g(x_i,x_{-i}) & \defeq (f_1(x), \ldots, f_{i-1}(x), f_{i+1}(x), \ldots, f_n(x))
  \end{align*}
  Then $h$ can be written as
  $h(\bm x) = (f(x_i,x_{-i}),g(x_i,x_{-i}))$ (with some abuse,
  assuming $f(x_i,x_{-i})$ is moved from the 1-st to the $i$-th
  position).  The assumptions of Lemma~\ref{lemma:fixpoints:general},
  including those of Item~(\ref{item:lemma:fixpoints:general:2}) are
  satisfied.  For example, given $x_i \in \bm L_i$, the function
  $g_{x_i}(x_{-i}) \defeq g(x_i, x_{-i})$ is in the c-clone $\calC$
  and has type $\bm L_{-i} \to \bm L_{-i}$; hence, it is
  $q \defeq E_{n-1}(p_1,\ldots, p_{i-1}, p_{i+1}, \ldots,
    p_n)$-stable, by induction hypothesis.  Similarly, the function
  $F(x_i) \defeq f(x_i,g_{x_i}^{(q)}(\bot))$ is in the c-clone $\calC$
  and has type $\bm L_i \rightarrow \bm L_i$, hence it is $p_i$-stable
  by the assumption of the theorem.  The conditions for
  Item~(\ref{item:lemma:fixpoints:general:2}) of
  Lemma~\ref{lemma:fixpoints:general} are verified similarly.

  Lemma~\ref{lemma:fixpoints:general}  implies two things.  First, $h$ has a least fixpoint,
  denoted by $\lfp(h) = (\bar x_1, \ldots, \bar x_n)$.  Second, from
  Item~(\ref{item:lemma:fixpoints:general:2}) of the lemma and from induction hypothesis, the
  $i$-component of $h^{(r)}(\bot)$ reaches the fixpoint $(h^{(r)}(\bot))_i = \bar x_i$
  when
  \[
    r = p_iq+p_i=p_i \cdot E_{n-1}(p_1,\ldots, p_{i-1}, p_{i+1},\ldots, p_n) + p_i \leq E_n(p_1,\ldots,p_n).
  \]
  Note that once the fixpoint in a dimension is reached, it stays fixed.
  Since $i$ was arbitrary, $h$ reaches the fixpoint in all dimensions after
  $E_n(p_1,\dots,p_n)$ iterations.
\end{proof}

\section{\texorpdfstring{$\textsf{DATALOG}^{\circ}$}{DATALOGO}}
\label{sec:datalogo}

We define here the language $\name$, which generalizes datalog from
traditional relations to $\bm P$-relations, for some POPS $\bm P$.  As
in datalog, the input relations to the program will be called
Extensional Database Predicates, EDB, and the computed relations will
be called Intensional Database Predicates, IDB.  Each EDB can be
either a $\bm P$-relation, or standard relation, i.e. a $\B$-relation,
and we denote by $\sigma \defeq \set{R_1, \ldots, R_m}$ and
$\sigma_\B \defeq \set{B_1, \ldots, B_k}$ the two vocabularies.  All
IDBs are $\bm P$-relations, and their vocabulary is denoted by
$\tau = \set{T_1, \ldots, T_n}$.

% We consider next a slight generalization of sum-product rules
% from Definition~\ref{def:sum:product}.  A {\em sum-product rule} is a
% rule like \eqref{eq:t:monomial} where the atoms are over the
% vocabulary $\sigma \cup \tau$, and the predicate
% $\Phi$ is any domain-independent\footnote{Domain-independence is
%   undecidable in general, but is usually enforced syntactically, by
%   requiring the formula to be range-restricted,
%   see~\cite{DBLP:books/aw/AbiteboulHV95}.} FO formula over the
% vocabulary $\sigma_\B$.  The semantics of a rule $\rho$ is still given
% by Equation~\eqref{eq:jt}, where the sentence $\Phi[\theta]$ is now an
% FO sentence over the Boolean EDBs.  A {\em sum-sum-product rule} is a
% sum of sum-product expressions, as before.

A $\name$ program $\Pi$ consists of $n$ sum-sum-product rules
$r_1, \ldots, r_n$ (as in Definition~\ref{def:sum:sum:product}), where
each rule $r_i$ has the IDB $T_i$ in the head:
\begin{align}
  r_1: &  & T_1(\cdots) \cd & E_{11}\oplus E_{12} \oplus \cdots \nonumber   \\
       &  &                 & \ldots \label{eq:basic:datalogo}              \\
  r_n: &  & T_n(\cdots) \cd & E_{n1} \oplus E_{n2} \oplus \cdots, \nonumber
\end{align}
and each $E_{ij}$ is a sum-product expression as
in~\eqref{eq:t:monomial}.  The program $\Pi$ is said to be {\em
    linear} if each sum-product expression $E_{ij}$ contains at most one
IDB predicate.
% Recall that Definition~\ref{def:sum:product} requires every rule to
% be {\em safe}, meaning that every variable in the head of a rule
% $r_i$ must occur in some atom of every sum-product expression
% $E_{i1}, E_{i2}, \ldots$

\subsection{Least Fixpoint of the Immediate Consequence Operator}

\label{subsec:ico}

The {\em Immediate Consequence Operator} (ICO) associated to a program
$\Pi$ is the function
$F : \inst(\sigma,D,\bm P) \times \inst(\sigma_\B,D,\B) \times
  \inst(\tau,D,\bm P) \rightarrow \inst(\tau,D,\bm P)$, that takes as
input an instance $(I, I_\B)$ of the EDBs and an instance $J$ of the
IDBs, and computes a new instance $F(I,I_\B,J)$ of the IDBs by
evaluating each sum-sum-product rule.  By fixing the EDBs, we will
view the ICO as function from IDBs to IDBs, written as $F(J)$. We
define the {\em semantics} of the $\name$
program~\eqref{eq:basic:datalogo} as the least fixpoint of the ICO
$F$, when it exists.

\begin{algorithm}[th]
  $J^{(0)} \leftarrow \bot\mbox{;}$ \mbox{\hspace{2cm} // In a naturally ordered  semiring this becomes $J^{(0)} \leftarrow 0$}\\
  \For {  $t \leftarrow 0$   {\bf to}   $\infty$ }
  {
    {$J^{(t+1)} \leftarrow F(J^{(t)})$}\;
    \If { $J^{(t+1)} = J^{(t)}$ }{
      Break
    }
  }
  \Return $J^{(t)}$
  \caption{Na\"ive evaluation for $\name$
    %       \\ In Sec.~\ref{sec:semi:naive} the POPS $\bm P$
    %       is a naturally ordered semiring, and the first line becomes
    %       $J^{(0)} \leftarrow 0$.
  }
  \label{algo:naive}
\end{algorithm}

The Na\"ive Algorithm for evaluating $\name$ is shown in
Algorithm~\ref{algo:naive}, and it is quite similar to that for
standard, positive datalog with set semantics.  We start with all IDBs
at $\bot$, then repeatedly apply the ICO $F$, until we reach a
fixpoint.
The algorithm computes the increasing sequence
$\bot \sqsubseteq F(\bot) \sqsubseteq F^{(2)}(\bot)
  \sqsubseteq \cdots$ When the algorithm terminates we say that it {\em
    converges}; in that case it returns the least fixpoint of $F$.
Otherwise we say that the algorithm {\em diverges}.

% We will show in Sec.~\ref{sec:semi:naive} that, when evaluated over a
% naturally ordered semiring with some additional properties, the naive
% algorithm can be improved to a semi-na\"ive algorithm; in that case,
% $\bot$ coincides with $\bm 0$, hence the first line of the na\"ive
% algorithm becomes $I^{(0)} \leftarrow \bm 0$.

\subsection{Convergence of $\name$ Programs}

\label{subsec:five:cases}

While every pure datalog program is guaranteed to have a least
fixpoint, this no longer holds for $\name$ programs.  As we mentioned
in the introduction, there are five possibilities, depending on the
POPS $\bm P$:

\begin{enumerate}[label=(\roman*)]
  \item $\bigvee_t J^{(t)}$ is not a fixpoint of the ICO; in this case
        we say that the program {\em diverges}.
        % ADDED DURING THE REVISION:
        For example, suppose the POPS consists of $\N \times \N$, with
        pairwise addition and multiplication, i.e.
        $(x,y) \oplus (u,v)=(x+u,y+v)$ and similarly for $\otimes$,
        and with the lexicographic order $(x,y)\sqsubseteq (u,v)$
        defined as $x<u$ or $x=u$ and $y \leq v$.  If the ICO of a
        program is the function $F(x,y) = (x,y+1)$, then
        $\bigvee_{t\geq 0} F^{(t)}(0,0)=\bigvee_{t \geq 0} (0,t) =
        (1,0)$ is not a fixpoint of $F$, because $F(1,0)=(1,1)$; in
        fact, $F$ has no fixpoint.
        % For example, if we
        %   interpret the program~\eqref{eq:datalogo:linear} over $\N$ and set
        %   $c > 1$, then the sequence of IDBs computed by the na\"ive algorithm
        %   is $0 < 1+c < 1+c+c^2 < \cdots$, which has no least upper bound in
        %   $\N$.  One can construct POPS where $\bigvee_n J^{(n)}$ always
        %   exists, but is not the least fixpoint.
  \item $\bigvee_n J^{(n)}$ is always the least fixpoint, but the
        na\"ive algorithm does not always terminate.  With some abuse we say
        also in this case that the program {\em diverges}.
        %% ADDED DURING THE REVISION:
        For a simple example, consider the semiring
        $\N \cup \set{\infty}$ and the function $F(x) = x+1$.  Its
        fixpoint is $\infty$, but it is not computable in a finite
        number of steps.
        % lub is not reached in a finite number of steps; equivalently, the
        % sequence $J^{(n)}$ is strictly increasing.  With some abuse we
        % also say that the program {\em diverges}.  For example, if we
        % interpret the program~\eqref{eq:datalogo:linear} over
        % $\N \cup \set{\infty}$, then any increasing sequence has a lub,
        % but if the lub is $\infty$ then the na\"ive algorithm will not
        % terminate in a finite number of steps.
  \item The na\"ive algorithm always terminates, in which case we say
        that it {\em converges}.  The number of steps depends on the input
        EDB database, meaning {\em both} the number of ground atoms in the
        EDB, and their values in $\bm P$.
  \item The na\"ive algorithm always terminates in a number of steps
        that depends only on the number of ground atoms in the EDB, but not
        on their values.
  \item The na\"ive algorithm always terminates in a number of steps
        that is polynomial in the number of ground atoms in the EDB.
\end{enumerate}

In this paper we are interested in characterizing the POPS that ensure
that every $\name$ program converges.  At a finer level, our goal is
to characterize precisely
cases~\ref{item:converge:3}-\ref{item:converge:5}.  We will do this in
Sec.~\ref{sec:complexity}, and, for that purpose, we will use an
equivalent definition of the semantics of a $\name$ program, namely as
the least fixpoint of a tuple of polynomials, obtained by grounding
the program.

\subsection{Least Fixpoint of the Grounded Program}

\label{subsec:grounded:program}

In this paper we consider an equivalent semantics of $\name$, which
consists of first grounding the program, then computing its least
fixpoint.

Fix an EDB instance $I, I_\B$, and let $D_0 \subseteq D$ be the finite
set consisting of its active domain plus all constants occurring in
the program $\Pi$.  Let $M = |\ground(\sigma,D_0)|$ and
$N = |\ground(\tau,D_0)|$ be the number of ground tuples of the EDBs
and the IDBs respectively.  We associate them in 1-to-1 correspondence
with $M+N$ POPS variables $z_1, \ldots, z_M$ and $x_1, \ldots, x_N$,
and use the same notation as in Sec.~\ref{subsec:sum-products} by
writing $x_{T_i(\bm a)}$ for the variable associated to the ground
tuple $T_i(\bm a)$.

Consider a rule $T_i(\cdots) \cd E_{i1} \oplus E_{i2} \oplus \cdots$
of the $\name$ program, with head relation $T_i$.  A {\em grounding}
of this rule is a rule of the form:
\begin{align*}
  x_{T_i(\bm a)} \cd & f_{T_i(\bm a)}(z_1, \ldots, z_M, x_1, \ldots, x_N)
\end{align*}
where $T_i(\bm a) \in \ground(T_i,D_0)$ is a ground tuple, and
$f_{T_i(\bm a)}$ is the {\em provenance polynomial} (defined in
Sec.~\ref{subsec:sum-products}) of the rule's body
$E_{i1} \oplus E_{i2} \oplus \cdots$ Since the value of each EDB
variable $z_{R_i(\bm u)}$ is known, we can substitute it with its
value, and the provenance polynomial simplifies to one that uses only
IDB variables $x_j$; we will no longer refer to the EDB variables
$z_i$.  The {\em grounded program} consists of all $N$ groundings, of
all rules.  Using more friendly indexes, we write the grounded program
as:
\begin{align}
  x_1 \cd & f_1(x_1, \ldots, x_N) \nonumber \\
          & \ldots \label{eq:grounded:pi}   \\
  x_N \cd & f_N(x_1, \ldots, x_N) \nonumber
\end{align}
where each $f_i$ is a multivariate polynomial in the variables
$x_1, \ldots, x_N$.  We write $\bm f = (f_1, \ldots, f_N)$ for the
vector-valued function whose components are the $N$ provenance
polynomials, and define the semantics of the $\name$ program as its
least fixpoint, $\lfp(\bm f)$, when it exists,
where, as usual, we identify the tuple $\lfp(\bm f) \in \bm P^N$ with
an IDB instance $\lfp(\bm f) \in \inst(\tau,D_0,\bm P)$.  By
definition, $\lfp(\bm f)$ is equal to the least fixpoint of the ICO,
as defined in Sec.~\ref{subsec:ico}.

\subsection{Examples}

\label{subsec:datalogo:examples}

We illustrate $\name$ with two examples.  When the POPS $\bm P$ is a
naturally ordered semiring, then we will use the following {\em
    indicator function} $[C]_0^{1}$, which maps a Boolean condition $C$
to either $0 \in \bm P$ or $1 \in \bm P$, depending on whether $C$ is
false or true.  We write the indicator function simply as $[C]$, when
the values $0,1$ are clear from the context.  An indicator function
can be desugared by replacing
$\setof{[C] \otimes P_1 \otimes \cdots \otimes P_k}{\Phi}$ with
$\setof{P_1 \otimes \cdots \otimes P_k}{\Phi \wedge C}$.  When $\bm P$
is not naturally ordered, then we will not use indicator functions,
see Example~\ref{ex:conditional:sum:product}.

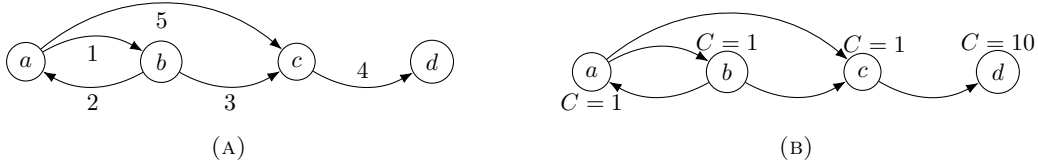
\begin{figure}
  \begin{subfigure}[b]{0.45\textwidth}
    \centering
    \begin{tikzpicture}[scale = 0.9, every node/.style={transform shape}]
      \node [draw, circle] at (0,0) (a) {$a$};
      \node [draw, circle] at (2,0) (b) {$b$};
      \node [draw, circle] at (4,0) (c) {$c$};
      \node [draw, circle] at (6,0) (d) {$d$};
      \draw [bend left,-Latex] (a) to node [auto, swap] {1} (b);
      \draw [bend left,-Latex] (b) to node [auto] {2} (a);
      \draw [bend right,-Latex] (b) to node [auto, swap] {3} (c);
      \draw [bend right,-Latex] (c) to node [auto] {4} (d);
      \draw [bend left=40,-Latex] (a) to node [auto, swap] {5} (c);
    \end{tikzpicture}
    \caption{}
  \end{subfigure}
  \begin{subfigure}[b]{0.45\textwidth}
    \centering
    \begin{tikzpicture}[scale = 0.9, every node/.style={transform shape}]
      \node [draw, circle] at (0,0) (a) {$a$};
      \node at (a) [below=.2, anchor = north] {$C=1$};
      \node [draw, circle] at (2,0) (b) {$b$};
      \node at (b) [above=.2, anchor = south] {$C=1$};
      \node [draw, circle] at (4,0) (c) {$c$};
      \node at (c) [above right=.25, anchor = south] {$C=1$};
      \node [draw, circle] at (6,0) (d) {$d$};
      \node at (d) [above=.2, anchor = south] {$C=10$};
      \draw [bend left,-Latex] (a) to (b);
      \draw [bend left,-Latex] (b) to (a);
      \draw [bend right,-Latex] (b) to (c);
      \draw [bend right,-Latex] (c) to (d);
      \draw [bend left=40,-Latex] (a) to (c);
    \end{tikzpicture}
    \caption{}
  \end{subfigure}
  \caption{A graph illustrating Example~\ref{ex:reachability:sssp} (a)
    and Example~\ref{ex:sum1:sum2} (b)}
  \Description{A graph illustrating Example~\ref{ex:reachability:sssp} (a)
    and Example~\ref{ex:sum1:sum2} (b)}
  \label{fig:simple:graph}
\end{figure}

\begin{ex} \label{ex:reachability:sssp} Let the EDB and IDB
  vocabularies be $\sigma = \set{E}$ and $\tau = \set{L}$, where $E$
  is binary and $L$ is unary.  Consider the following $\name$ program:
  \begin{align}
    L(X) \cd & [X=a] \oplus \bigoplus_Z \left(L(Z) \otimes E(Z,X)\right) \label{eq:reachability}
  \end{align}
  where $a \in D$ is some constant in the domain.  We show three
  different interpretations of the program, over three different
  naturally ordered semirings.  First, we interpret it over the
  semiring of Booleans.  In this case, the program can be written in a
  more familiar notation:
  \begin{align*}
    L(X) \cd & [X=a]_0^1 \vee \exists_Z \left(L(Z) \wedge E(Z,X)\right)
  \end{align*}
  This is the reachability program, which computes the set of nodes
  $X$ reachable from the node $a$.  The indicator function $[X=a]_0^1$
  returns $0$ or $1$, depending on whether $X\neq a$ or $X=a$.

  Next, let's interpret it over $\trop^+$.  In that case, the
  indicator function $[X=a]_\infty^0$ returns $\infty$ when $X\neq a$,
  and returns $0$ when $X=a$.  The program becomes:
  \begin{align*}
    L(X) \cd & \min\left((\texttt{if } X=a\texttt{ then } 0 \texttt{ else } \infty),  \min_Z(L(Z) + E(Z,X))\right)
  \end{align*}
  This program solves the Single-Source-Shortest-Path (SSSP) problem with source
  vertex $a$.
  Consider the graph in Fig.~\ref{fig:simple:graph}(a).  The active
  domain consists of the constants $a,b,c,d$, and the na\"ive evaluation
  algorithm converges after 5 steps, as shown here:
  \begin{align*}
     &
    \begin{array}{l|c|c|c|c|} \cline{2-5}
              & L(a)   & L(b)   & L(c)   & L(d)   \\ \cline{2-5}
      L^{(0)} & \infty & \infty & \infty & \infty \\ \cline{2-5}
      L^{(1)} & 0      & \infty & \infty & \infty \\ \cline{2-5}
      L^{(2)} & 0      & 1      & 5      & \infty \\ \cline{2-5}
      L^{(3)} & 0      & 1      & 4      & 9      \\ \cline{2-5}
      L^{(4)} & 0      & 1      & 4      & 8      \\ \cline{2-5}
      L^{(5)} & 0      & 1      & 4      & 8      \\ \cline{2-5}
    \end{array}
  \end{align*}

  Third, let's interpret it over $\trop^+_p$, defined in
  Example~\ref{ex:trop:p}.  Assume for simplicity that $p=1$.  In that
  case the program computes, for each node $X$, the bag
  $\bag{l_1,l_2}$ of the lengths of the two shortest paths from $a$ to
  $X$.  The indicator function $[X=a]$ is equal to $\bag{0,\infty}$
  when $X=a$, and equal to $\bag{\infty,\infty}$ otherwise.  The
  reader may check that the program converges to:
  \begin{align*}
    L(a) = & \bag{0,3} & L(b) = & \bag{1,4} &
    L(c) = & \bag{4,5} & L(d) = & \bag{8,9}
  \end{align*}

  Finally, we can interpret it over $\trop^+_{\leq \eta}$, the
  semiring in Example~\ref{ex:trop:eta}.  In that case the program
  computes, for each $X$, the set of all possible lengths of paths
  from $a$ to $X$ that are no longer than the shortest path plus
  $\eta$.
\end{ex}

\begin{ex} \label{ex:sum1:sum2} A classic problem that requires the
  interleaving of recursion and aggregation is the bill-of-material
  (see, e.g., \cite{DBLP:conf/amw/ZanioloYIDSC18}), where we are asked
  to compute, for each part $X$, the total cost of $X$, of all
  sub-parts of $X$, all sub-sub-parts of $X$, etc.  The EDB and IDB
  schemas are $\sigma_\B = \set{E}$, $\sigma = \set{C}$,
  $\tau=\set{T}$.  The relation $E(X,Y)$ is a standard, Boolean
  relation, representing the fact that ``$X$ has a subpart $Y$''; $C(X)$
  is an $\N$-relation or an $\R_\bot$-relation (to be discussed
  shortly) representing the cost of $X$; and $T(X)$ is the {\em total
      cost} of $X$, that includes the cost of all its (sub-)subparts.
  The $\name$ program is:
  \begin{align*}
    T(X) \cd & C(X) + \sum_Y \setof{T(Y)}{E(X,Y)}
  \end{align*}
  When the graph defined by $E$ is a tree then the program computes
  correctly the bill-of-material.  We are interested, however, in what
  happens when the graph encoded by $E$ has cycles, as illustrate with
  Fig.~\ref{fig:simple:graph}(b).  The grounded program
  is:\footnote{Strictly speaking, we should have introduced POPS
    variables, $x_{T(a)}, x_{T(b)}, \ldots$, but, to reduce clutter,
    we show here directly the grounded atoms instead of their
    corresponding POPS variable.}
  \begin{align*}
    T(a) \cd & C(a) + T(b) + T(c) \\
    T(b) \cd & C(b) + T(a) + T(c) \\
    T(c) \cd & C(c) + T(d)        \\
    T(d) \cd & C(d)
  \end{align*}
  We consider two choices for the POPS.  First, the naturally ordered
  semiring $(\N, +, *, 0, 1)$. Here the program diverges, since the
  na\"ive algorithm will compute ever increasing values for $T(a)$ and
  $T(b)$, which are on a cycle.  Second, consider the lifted reals
  $\R_\bot = (\R \cup \set{\bot},+,*,0,1,\sqsubseteq)$.  Now the
  program converges in 3 steps, as can be seen below:
  \begin{align*}
     &
    \begin{array}{l|c|c|c|c|} \cline{2-5}
          & T(a) & T(b) & T(c) & T(d) \\ \cline{2-5}
      T_0 & \bot & \bot & \bot & \bot \\ \cline{2-5}
      T_1 & \bot & \bot & \bot & 10   \\ \cline{2-5}
      T_2 & \bot & \bot & 11   & 10   \\ \cline{2-5}
      T_3 & \bot & \bot & 11   & 10   \\ \cline{2-5}
    \end{array}
  \end{align*}
\end{ex}

\subsection{Extensions}

\label{subsec:extensions:datalogo}

We discuss here several extensions of $\name$ that we believe are
needed in a practical implementation.

  {\bf Case Statements} Sum-products can be extended w.l.o.g. to include
case statements of the form:
\begin{align*}
  T(x_1, \ldots, x_k) & \cd \texttt{case } C_1: E_1; \ \ \ C_2: E_2;\cdots; [\texttt{ else } E_n]
\end{align*}
where $C_1, C_2, \ldots$ are conditions and $E_1, E_2, \ldots$ are
sum-product expressions.  This can be desugared to a sum-sum-product:
\begin{align*}
  T(x_1, \ldots, x_k) & \cd \setof{E_1}{C_1} \oplus \setof{E_2}{\neg C_1 \wedge C_2} \oplus\cdots \oplus \setof{E_n}{\neg C_1 \wedge \neg C_2 \cdots}
\end{align*}
and therefore the least fixpoint semantics and our convergence results
in Sec.~\ref{sec:complexity} continue to hold.
For example, we may compute the prefix-sum of a vector $V$ of length
100 as follows:
\begin{align*}
  W(i) & \cd \texttt{case } i=0: V(0);\ \ \ \ i<100: W(i-1) + V(i);
\end{align*}

{\bf Multiple Value Spaces} Our discussion so far assumed that all
rules in a $\name$ program are over a single POPS.  In practice one
often wants to perform computations over multiple POPS.  In that case
we need to have some predefined functions mapping between various
POPS; if these are monotone, then the least fixpoint semantics still
applies, otherwise the program needs to be {\em stratified}.  We
illustrate with an example, which uses two POPS: $\R_+$ and $\B$.

\begin{ex} We illustrate the {\em company control} example
  from~\cite[Example 3.2]{DBLP:conf/pods/RossS92}.
  $S(X,Y)=n \in \R_+$ represents the fact that company $X$ owns a
  proportion of $n$ shares in company $Y$.  We want to compute the
  predicate $C(X,Y)$, representing the fact that the company $X$ {\em
      controls} company $Y$, where control is defined as follows: $X$
  controls $Y$ if the sum of shares it owns in $Y$ plus the sum of
  shares in $Y$ owned by companies controlled by $X$ is $>0.5$.  The
  program is adapted directly from~\cite{DBLP:conf/pods/RossS92}:
  \begin{align*}
    CV(X,Z,Y) \cd & [X=Z]*S(X,Y) + [C(X,Z)]*S(Z,Y)              \\
    T(X,Y) \cd    & \sum_Z \setof{CV(X,Z,Y)}{\text{Company}(Z)} \\
    C(X,Y) \cd    & [T(X,Y) > 0.5]
  \end{align*}
  The value of $CV(X,Z,Y)$ is the fraction of shares that $X$ owns in
  $Y$, through its control of company $Z$; when $X=Z$ then this
  fraction includes $S(X,Y)$.  The value of $T(X,Y)$ is the total
  amount of shares that $X$ owns in $Y$.  The last rule checks whether
  this total is $>0.5$: in that case, $X$ controls $Y$.

  The EDB and IDB vocabularies are $\sigma = \set{S}$,
  $\sigma_\B = \set{\text{Company}}$,
  $\tau = \set{CV, T}, \tau_\B = \set{C}$.  The IDBs $CV, T$ are
  $\R_+$-relations, $C$ is a standard $\B$-relation.  The mapping
  between the two POPS is achieved by the indicator function
  $[\Phi] \in \R_+$, which returns 0 when the predicate $\Phi$ is
  false, and 1 otherwise.  All rules are monotone, w.r.t. to the
  natural orders on $\R_+$ and $\B$, and, thus, the least-fixpoint
  semantics continues to apply to this program. But the results in
  Sec.~\ref{sec:complexity} apply only to fixpoints of polynomials,
  while the grounding of our program is no longer a polynomial due to the use of the indicator functions.
\end{ex}

% {\bf Stratification} When using non-monotone functions between
% different POPS, one needs to stratify the $\name$, as is custom for
% datalog with negation.

{\bf Interpreted functions over the key-space} A practical language
needs to allow interpreted functions over the key space, i.e. the
domain $D$, as illustrated by this simple example:
\begin{align*}
  \texttt{Shipping}(cid,\texttt{date}+1) \cd & \texttt{Order}(cid,\texttt{date})
\end{align*}
Here $\texttt{date}+1$ is an interpreted function applied to
$\texttt{date}$.  Interpreted functions over $D$ may cause the active
domain to grow indefinitely, leading to divergence; our results in
Sec.~\ref{sec:complexity} apply only when the active domain is fixed,
enabling us to define the grounding~\eqref{eq:grounded:pi} of the
$\name$ program.

  {\bf Keys to Values} Finally, a useful extension is to allow key
values to be used as POPS values, when the types are right.  For
example, if $\texttt{Length}(X,Y,C)$ is Boolean relation, where a
tuple $(X,Y,C)$ represents the fact that there exists a path of length
$C$ from $X$ to $Y$, then we can compute the length of the
shortest path as the following rule of the tropical semiring
$\trop^+$:
\begin{align*}
  \texttt{ShortestLength}(X,Y) & \cd \min_C \left([\texttt{Length}(X,Y,C)]_\infty^0 + C\right)
\end{align*}
The key variable $C$ became an atom over the tropical semiring.

\section{Characterizing the Convergence of \texorpdfstring{$\name$}{Datalogo}}
\label{sec:complexity}

In this section, we prove our main result, Theorem~\ref{th:main:intro}.
As we saw in Sec.~\ref{sec:intro} (and again in
Sec.~\ref{subsec:five:cases}), there are five different possibilities
for the divergence/convergence of $\name$ programs.  Our results in
this section concern the last three cases, when every $\name$ program
converges.  Recall that by ``converge'', we mean that the na\"ive algorithm
terminates in a finite number of steps; we are not interested in
``convergence in the limit''.

Throughout this section, we will assume that the $\name$ program has
been grounded, as in Sec.~\ref{subsec:grounded:program}, and therefore
our task is to study the convergence of a tuple of polynomials,
see~\eqref{eq:grounded:pi}.  To reduce clutter when writing
polynomials, we will follow the convention in
Sec.~\ref{subsec:polynomial} and use the symbols $+, \cdot$ instead of
$\oplus, \otimes$, while keeping in mind that they represent abstract
operations in a POPS.

Consider the following grounded $\name$ program, with a single
variable $x$ and a single rule:
%
%
% As discussed , there are four different
% scenarios for the convergence property of $\name$.  We start by
% illustrating them on a simple example.  Consider the single-rule
% $\name$ program in Example~\ref{ex:reachability:sssp}
% (Eq.~\eqref{eq:reachability}).  Suppose the graph has a single node
% with a loop, $a \rightarrow a$.  We associate the POPS variable $x$ to
% the ground atom $L(a)$ and denote $c \defeq E(a,a) \in \bm P$, then
% the grounded program becomes:
%
\begin{align}
  x & \cd 1 + cx  \label{eq:datalogo:linear}
\end{align}
We need to compute the least fixpoint of $f(x) := 1+cx$.  When the
POPS $\bm P$ is a naturally ordered semiring, then its smallest
element is $\bot = 0$ and the na\"ive algorithm computes the sequence:
\begin{align*}
  f^{(0)}(0) := & 0 & f^{(1)}(0) := & 1 & f^{(2)}(0) := & 1 + c & f^{(3)}(0) := & 1 + c + c^2 &  & \ldots & f^{(q)}(0) := & 1 + c + \cdots + c^{q-1}
\end{align*}
If we want {\em every} $\name$ program to converge on $\bm P$, then
surely so must our program~\eqref{eq:datalogo:linear}.  In essence,
the main result in this section is that the converse holds too:
if~\eqref{eq:datalogo:linear} converges on $\bm P$ for any choice of
$c$, then every $\name$ program converges on $\bm P$.  For example,
~\eqref{eq:datalogo:linear} diverges on the semiring $\N$, because, if
$c = 2$, then $f^{(q)}(0) = 1+2+2^2+\cdots+2^{q-1}\rightarrow \infty$.
On the other hand this program converges on the semiring $\trop^+$
and, therefore, {\em every} $\name$ program converges on $\trop^+$.

In the rest of this section, we prove Theorem~\ref{th:main:intro}.  We
start by assuming that the POPS $\bm P$ is a naturally ordered
semiring $\bm S$, then extend the result to arbitrary POPS.

\subsection{Definition of Stable Semirings}

The following notations and simple facts can be found
in~\cite{semiring_book}. Fix a semiring $\bm S$.  For every
$c \in \bm S$ and $p \geq 0$ define:
\begin{align}
  c^{(p)} \defeq & 1+c+c^2 + \cdots + c^p \label{eq:a:p}
\end{align}

\begin{defn} \label{def:a:p:stable} An element $c \in \bm S$ is {\em
      $p$-stable} if $c^{(p)}=c^{(p+1)}$.  We say that $c$ is {\em
      stable} if there exists $p$ such that $c$ is $p$-stable.

  We call the semiring $\bm S$ {\em stable} if every element is
  stable.  We call it {\em uniformly stable} if there exists
  $p \geq 0$ such that every $c \in \bm S$ is $p$-stable; in that case
  we also call $\bm S$ a {\em $p$-stable semiring}.
\end{defn}
Note that, an equivalent definition of $c$ being $p$-stable is that
\begin{align}
  c^{(p)} = c^{(q)} &  & \text{for all } q > p, \label{eqn:c:p:q}
\end{align}
this can be seen by induction and from the fact that $c^{(q)} = 1 + c \mult c^{(q-1)}$.

We recall here a very nice and useful result from~\cite{semiring_book}.
A short proof is supplied for completeness.

\begin{prop} \label{prop:stable:ordered}
  If $1$ is $p$-stable then $\bm S$ is naturally ordered.  In
  particular, every stable semiring is naturally ordered.
\end{prop}
\begin{proof}
  If $1$ is $p$-stable, then $1^{(p+1)}=1^{(p)}$, which means that
  $1+1+\cdots+1 (\mbox{$p+1$ times})=1+\cdots+1 (\mbox{$p$ times})$,
  or, in short, $p+1=p$.  We prove that, if $a \preceq_S b$ and
  $b \preceq_S a$ hold, then $a=b$.  By definition of $\preceq_S$,
  there exist $x,y$ such that $a+x=b$ and $b+y=a$.  On one hand we have
  $a=b+y=a+x+y$, which implies $a = a+k(x+y)$ for every $k\geq 0$, in
  particular $a = a+p(x+y)$.  On the other hand we have
  $b = a+x = a + k(x+y) + x = a + (k+1)x + ky$.  We set $k=p$ and
  obtain $b = a + (p+1)x + py = a + px + py = a +p(x+y)$ and the
  latter we have seen is $=a$, proving that $a=b$.
\end{proof}

The converse does not hold in general, for example the semiring $\N$
is naturally ordered but is not stable.  However, if $\bm S$ is both
naturally ordered and satisfies the ACC condition (see
Sec.~\ref{sec:lfp}), then it is also stable.  Here, too, the converse
fails: $\trop^+$ is $0$-stable, because $\min(0,x) = 0$, yet it does
not satisfy the ACC condition, e.g. the following is an infinitely
ascending chain in $\trop^+$: $1 > 1/2 > 1/3 > 1/4 > \cdots$

The case of a $0$-stable semiring has been most extensively studied.
Such semirings are called {\em simple} by
Lehmann~\cite{DBLP:journals/tcs/Lehmann77}, are called {\em
    c-semirings} by Kohlas~\cite{DBLP:journals/ai/KohlasW08}, and {\em
    absorptive} by Dannert et al.~\cite{DBLP:conf/csl/DannertGNT21}.  In
all cases, the authors require $1+a=1$ for all $a$ (or, equivalently,
$b + ab = b$ for all $a,b$~\cite{DBLP:conf/csl/DannertGNT21}), which
is equivalent to stating that $a$ is 0-stable, and also equivalent to
stating that $(\bm S, +)$ is a join-semilattice with maximal element
$1$.  The tropical semiring is such an example; every distributive
lattice is also a 0-stable semiring where we set $+=\vee$ and
$\mult =\wedge$.

We give next two examples of stable semirings. The first one is
$p$-stable for some $p>0$ and it is not $(p-1)$-stable.  The second
one is non-uniformly stable.  These example are adapted
from~\cite{semiring_book}.

\begin{prop} \label{prop:trop:p:stable} The semiring $\trop^+_p$
  defined in Example~\ref{ex:trop:p} is $p$-stable. Moreover,
  the bound $p$ on the stability of $\trop^+_p$ is tight.
\end{prop}

\begin{proof} Let $c \in \trop^+_p$; recall that $c$ is a bag of $p+1$
  elements in $\R_+\cup\set{\infty}$.  For every $q \geq 1$, $c^q$
  consists of the $p+1$ smallest sums of the form
  $u_{i_1}+u_{i_2}+\cdots+u_{i_q}$, where each $u_{i_j} \in c$.
  Furthermore, the value $c^{(q)} = 1 + c + c^2 + \cdots + c^q$ is
  obtained as follows: (a) consider all sums of up to $q$ numbers in
  $c$, and (b) retain the smallest $p+1$ sums, including duplicates.  We
  claim that $c^{(p)}=c^{(p+1)}$.  To prove the claim, consider such a
  sum $y \defeq u_{i_0}+u_{i_1}+\cdots+u_{i_p}$ in $c^{p+1}$.  This
  term is greater than or equal to each of the following $p+1$ terms:
  $0$, $u_{i_0}$, $u_{i_0}+u_{i_1}$, $\ldots$,
  $u_{i_0}+\cdots + u_{i_{p-1}}$, which are already included in the
  bags $1, c, c^2, \ldots, c^p$.  This proves that every new term in
  $c^{p+1}$ is redundant, proving that $c^{(p)}=c^{(p+1)}$.

  To show that the bound $p$ on the stability of $\trop^+_p$ is tight,
  it suffices to show that the 1-element $\bm 1_p = \bag{0, \infty, \dots, \infty}$
  (i.e., $p$-times $\infty$) is not $(p-1)$-stable. Indeed,
  $\bm 1_p^i = \bm 1_p$ for every $i \geq 1$ and
  $\bm 1_p + \bm 1_p^1  + \bm 1_p^2 + \dots + \bm 1_p^{p-1}
    = \bag{ 0, \dots, 0, \infty }$ (i.e., $p$-times 0) whereas
  $\bm 1_p + \bm 1_p^1  + \bm 1_p^2 + \dots + \bm 1_p^{p}
    = \bag{0, \dots, 0, 0}$  (i.e., $p+1$-times 0).
\end{proof}

\begin{prop} \label{prop:trop:eta:stable} The semiring
  $\trop^+_{\leq\eta}$ defined in Example~\ref{ex:trop:eta} is stable.
  Moreover, $\trop^+_{\leq\eta}$  is not $p$-stable for any $p$.
\end{prop}

\begin{proof}
  The proof is similar to that of Prop~\ref{prop:trop:p:stable}.
  While the elements of $\trop^+_p$ are bags, those of
  $\trop^+_{\leq\eta}$ are sets.  Let $c$ be an element of
  $\trop^+_{\leq\eta}$.  Then $c^{(q)}=1_{\leq \eta}+c+c^2+\cdots+c^q$ is obtained
  as follows: (a) consider all sums of $\leq q$ numbers; this includes
  the empty sum, whose value is $0$.  (b) retain only those sums that
  are $\leq \eta$.  If $c= \set{0}$, then $c$ is $0$-stable, so assume
  w.l.o.g. that $c$ contains some element $>0$.  Let $x_0 > 0$ be the
  smallest such element, and let $p = \lceil \frac{\eta}{x_0}\rceil$.
  We claim that $c^{(p)}=c^{(p+1)}$.  To prove the claim, consider a
  sum of $p+1$ elements $y=x_{i_0}+\cdots+x_{i_p}$ which belongs to
  $c^{(p+1)}$ but does not belong to $c^{(p)}$.  In particular,
  $x_{i_j}\neq 0$, otherwise we could drop $x_{i_j}$ from the sum and we
  had $y \in c^{(p)}$.  It follows that $y \geq (p+1)x_0 > \eta$,
  which means that $y$ is not included in $c^{(p+1)}$.  It follows
  that $c^{(p)}=c^{(p+1)}$ as required.

  To show that $\trop^+_{\leq\eta}$  is not $p$-stable for any $p$, we
  assume to the contrary that there does exist $p$ such that
  $\trop^+_{\leq\eta}$ is $p$-stable and derive a contradiction.
  To this end, choose $a \in \R_+$ such that $0 < a < \frac{\eta}{p+1}$
  and let $c = \{a\}$. Then we have
  $c^i = \{i \cdot a\}$ for every $i \geq 1$.
  Recall from  Example~\ref{ex:trop:eta} that
  $\bm 1_{\leq\eta} = \{0\}$.
  Hence,
  $c^{(p)} = \bm 1_{\leq \eta} + c^1  + c^2 + \dots + c^{p}
    = \{0, a,2a,\dots, pa\}$ whereas
  $c^{(p+1)} = \{0, a,2a,\dots, pa, (p+1) a\}$.
  Indeed, $(p+1) a \in c^{(p+1)}$ because
  $\min (c^{(p+1)}) = 0$ and
  $(p+1) a  < \eta$ due to $a < \frac{\eta}{p+1}$.
  Hence, $c^{(p)} \neq c^{(p+1)}$, which contradicts the above assumption
  that $\trop^+_{\leq\eta}$ is $p$-stable.
\end{proof}

\subsection{Convergence in Non-Uniformly Stable Semirings}
\label{subsec:non-uniformly:stable}

Let $\bm S$ be a naturally ordered semiring.  Consider a vector-valued
function $\bm f$ with component functions $(f_1, \ldots, f_N)$, over variables
$\bm x = (x_1, \ldots, x_N)$. In particular, $\bm f$ is a map
$\bm f : \bm S^N \rightarrow \bm S^N$.
Recall that $\bm f$ is said to be {\em stable} if there exists $q \geq 0$ such that
$\bm f^{(q)}(\bm 0)= \bm f^{(q+1)}(\bm 0)$. If $\bm f$ was the ICO of a recursion,
then being $q$-stable means the Na\"ive algorithm terminates in $q$ steps.
In this section, we prove a result (Theorem~\ref{th:non-unformly:stable}) which
essentially says that, if the semiring $S$ is stable, and if the ICO $\bm f$
is a vector-valued multi-variate polynomial function in $S$, then the Na\"ive algorithm
converges in a finite number of steps.
We take the opportunity to develop some terminologies
and techniques to prove stronger results in the sections that follow.

The following example illustrates how a univariate quadratic function can be shown to be
stable in a stable semiring.\footnote{A similar example can be found
  in~\cite{semiring_book} where the concept of {\em quasi square root} is introduced.
  However, the formula for the Catalan number cited in the book is incorrect.}

\begin{ex} \label{ex:running:f}
  Consider a single polynomial, i.e. $N=1$:
  \begin{align}
    f(x) & = b + ax^2 \label{eq:f:x2}
  \end{align}
  Then:
  \begin{align*}
    f^{(0)}(0) & = 0                                                                        \\
    f^{(1)}(0) & = b                                                                        \\
    f^{(2)}(0) & = b + ab^2                                                                 \\
    f^{(3)}(0) & = b + a(b+ab^2)^2 = b + ab^2 + 2a^2b^3 + a^3b^4                            \\
    f^{(4)}(0) & = b + a(b + ab^2 + 2a^2b^3 + a^3b^4)^2 = b + ab^2 + 2a^2b^3+5a^3b^4+\cdots
  \end{align*}
  (Note that we overloaded notations in the above to make the expressions simpler to parse.
  We wrote $2ab^2$ to mean $a\otimes b \otimes b \oplus a\otimes b \otimes b$.
  In particular, integer coefficients are used to denote repeated summations in the
  semiring, whose $\oplus$ and $\otimes$ operators are already simplified to $+$ and
  $\mult$.)
  It can be shown by induction that, when $q \geq n$,
  the coefficient of $a^nb^{n+1}$ is ``stabilized''
  and it is the Catalan
  number $\frac{1}{n+1} \binom{2n}{n}$:
  \begin{align}
    f^{(q)}(0) & =\sum_{n=0}^q
    \frac{1}{n+1}\binom{2n}{n}a^nb^{n+1}+\sum_{n>q}\lambda_n^{(q)}a^nb^{n+1}.
    \label{eqn:fq:series}
  \end{align}
  The coefficients $\lambda^{(q)}_n\in\N$ for $n>q$ may not be in their ``final form'' yet.
  Now, suppose the element $c := ab$ is $q$-stable. Then, from
  identity~\eqref{eqn:fq:series} we can show that $f$ is also $q$-stable.
  To see this, note that if $c$ is $q$-stable then
  $c^{(m)} = c^{(q)}$ for any $m>q$ (recall the notation defined in~\eqref{eq:a:p}).
  Thus, for any $m>q$,
  a term $a^mb^{m+1}$ in the expansion of $f^{(q)}(0)$ will be ``absorbed'' by
  the earlier terms:
  \begin{align}
    \sum_{n=0}^q a^nb^{n+1} + a^{m}b^{m+1}
    = b\left( c^{(q)} + c^m \right)
    = b\left( c^{(m-1)} + c^m \right)
    = b c^{(m)}
    = b c^{(q)}
    = \sum_{n=0}^q a^nb^{n+1}.
  \end{align}

  %  To complete the example, we are back to proving~\eqref{eqn:fq:series}.
  %  To this end, suppose
  %  \begin{align}
  %      f^{(q)}(0) &=\sum_{n=0}^\infty \lambda_n^{(q)}a^nb^{n+1}.
  %  \end{align}
  %  Then, for $q \geq 1$, from the identity $f^{(q)}(0) = b+a(f^{(q-1)})^2$ we derive
  %  \begin{align}
  %      \sum_{n=0}^\infty \lambda_n^{(q)}a^nb^{n+1}
  %                 &= b + a\left(\sum_{k=0}^\infty \lambda_k^{(q-1)} a^kb^{k+1}\right)^2
  %                 = b + \sum_{n=0}^\infty \left(\sum_{k=0}^{n-1}
  %                 \lambda_k^{(q-1)}\lambda^{(q-1)}_{n-k-1} \right) a^nb^{n+1}
  %  \end{align}
  %  Equating both sides we have
  %  \begin{align}
  %      \lambda_n^{(q)} &= \begin{cases}
  %          1 & n=0\\
  %          \sum_{k=0}^{n-1}\lambda_k^{(q-1)}\lambda_{n-k-1}^{(q-1)} & n>0.
  %      \end{cases}
  %      \label{eqn:catalan}
  %  \end{align}
  %  Thus, if we hypothesize inductively that $\lambda^{(q)}_n = \frac{1}{n+1}\binom{2n}{n}$,
  %  the $n$th Catalan number, whenever $q \geq n$, then from the convergence of
  %  $\lambda^{(q-1)}_k$ and $\lambda^{(q-1)}_{n-k-1}$, we can infer the convergence of
  %  $\lambda^{(q)}_n$ when $q \geq n$, because~\eqref{eqn:catalan} is the recurrence of
  %  the Catalan sequence.

  A simple way to see how the Catalan number shows up is to let
  $q \to \infty$ in the formal power series
  sense~~\cite{MR1676282}. In this case,
  $f^{(\omega)}(0) = b+a[f^{(\omega)}(0)]^2$,
  which is then solved directly by Newton's generalized binomial
  expansion:
  \begin{align}
    f^{(\omega)}(0)
     & = \frac{1}{2a}\left(1-\sqrt{1-4ab}\right)
    = \frac{1}{2a}\left(1- \sum_{n=0}^\infty \binom{1/2}{n}(-1)^na^nb^n\right)
    = \sum_{n=0}^\infty \frac{1}{n+1}\binom{2n}{n} a^nb^{n+1}.\label{eqn:fq:series:omega}
  \end{align}
  Our proof of Theorem~\ref{th:main:intro} is based on the two
  observations in this example.  First, we prove in this section that
  there exists a finite set of monomials, like $ab$ here, whose
  stability implies the stability of $f$.  In addition, we prove in
  the next section that the coefficients, like $\lambda_n^{(q)}$
  above, reach quickly their final values.  Notice that, in general,
  no closed form like~\eqref{eqn:fq:series:omega} exists for
  $f^{(\omega)}$; for example, if $f$ has degrees $\geq 5$ then we
  cannot hope to always obtain such closed form formulas.
\end{ex}

We next introduce several notations that we also need in the next section.
Given a positive integer $k$, a tuple $\bm z = (z_1, \dots, z_k)$ of symbols or variables,
and a tuple $\bm v =(v_1,\dots,v_k) \in \N^k$ of non-negative integers, we denote
\begin{align}
  {\bm z}^{\bm v} & := \prod_{i=1}^k z_i^{v_i},
\end{align}
where the product is the product operator of the semiring under consideration.
With this notation, the $i$-th component function of a vector-valued polynomial
function $\bm f=(f_1,\dots,f_N)$ can be written
in the following form:
\begin{align}
  f_i(\bm x) & = \sum_{\bm v \in V_i} a_{i,\bm v} \mult \bm x^{\bm v},
  \label{eqn:fi}
\end{align}
where the $a_{i,\bm v}$ are constants in the semiring's domain,
and $V_i$ is a set of length-$N$ vectors of non-negative integers.
%%%%%  Sorry, i forgot to clarify this.  We don't need this
%%%%%  assumption: if there is no free term, then simply $f^{(q)}=0$,
%%%%%  and the grammar does not have any finite derivation trees.
% Without loss of generality, we will assume that $\bm 0 \in \bm V_i$,
% which means there is always a constant term $a_{i,\bm 0}$ (which could be $0$)
% in the expansion of $f_i$.

We are interested in the formal expansion of $\bm f^{(q)}(0)$, in the same way we
expressed iterative applications of $f$ in Example~\ref{ex:running:f}.
The device to express these expansions formally is context-free languages (CFL),
as was done in a long history of work in automata theory and formal
languages~\cite{DBLP:journals/jacm/EsparzaKL10,DBLP:conf/lics/HopkinsK99}.

The non-terminals of the grammar for our CFL consist of all variables
(in $\bm x$) occurring in $\bm f$.  Every constant $a_{i,\bm v}$
(shown in~\eqref{eqn:fi}) corresponds to a distinct terminal symbol in
the grammar. For example, even if $a_{i,\bm v} = a_{i',\bm v'}$ for
$(i,\bm v) \neq (i'\bm v')$, we will consider them different symbols
in the symbol set $\Gamma$ of the CFL.  Note also that every monomial
$\bm x^{\bm v}$ has a corresponding (symbolic) coefficient
$a_{i,\bm v}$, even if the coefficient is $1$. For example $1+x^2y$
becomes $a + bx^2y$, with coefficients $a=1$ and $b=1$.  The
production rules are constructed from~\eqref{eqn:fi} as follows.  For
every monomial $a_{i,\bm v} \bm x^{\bm v}$, with
$\bm v=(v_1,\ldots,v_N) \in V_i$, there is a rule:
\begin{align}
  x_i & \to a_{i,\bm v}
  \underbrace{x_1\cdots x_1}_{v_1 \text{ times}}
  \underbrace{x_2\cdots x_2}_{v_2 \text{ times}}
  \dots
  \underbrace{x_N\cdots x_N}_{v_N \text{ times}}
  \label{eq:production}
\end{align}

Given a parse tree $T$ (from the grammar above),
define the {\em yield} $Y(T)$ of $T$ to be the product of all
terminal symbols at the leaves of $T$.
For $i \in [N]$ and a positive integer $q$, let $\calT^i_q$ denote the set of all parse
trees of the above grammar, with starting symbol $x_i$, and depth $\leq q$.
The following simple but fundamental fact was observed in~\cite{DBLP:journals/jacm/EsparzaKL10}
(for completeness, we include an inductive proof of this fact in the appendix.)

\begin{lmm}[\cite{DBLP:journals/jacm/EsparzaKL10}] \label{lmm:fundamental}
  Given integers $q\geq 0$ and $i \in [N]$, the $i$-th component of the
  vector $\bm f^{(q)}(\bm 0)$, denoted by $(\bm f^{(q)}(0))_i$, can be
  expressed in terms of the yields of short parse-trees:
  \begin{align}
    (\bm f^{(q)}(0))_i & = \sum_{T \in \calT^i_q} Y(T). \label{eqn:YT}
  \end{align}
\end{lmm}

\begin{ex} \label{ex:fundamental} Consider the following map
  $\bm f = (f_1,f_2)$:
  \begin{align}
    \begin{bmatrix}
      X \\ Y
    \end{bmatrix}
    \to
    \begin{bmatrix}
      aXY + bY + c \\
      uXY + vX + w
    \end{bmatrix}
  \end{align}
  Applying the map twice, we obtain
  \begin{align*}
    \begin{bmatrix}
      X \\ Y
    \end{bmatrix}
    \to
    \begin{bmatrix}
      aXY + bY + c \\
      uXY + vX + w
    \end{bmatrix}
    \to
    \begin{bmatrix}
      a(aXY+bY+c)(uXY+vX+w) + b(uXY+vX+w) + c \\
      u(aXY+bY+c)(uXY+vX+w) + v(aXY+bY+c) + w
    \end{bmatrix}
  \end{align*}
  This means the first component of $\bm f^{(1)}(0)$ is
  $(\bm f^{(1)}(0))_1 = c$ and the first component of
  $\bm f^{(2)}(0)$ is $(\bm f^{(2)}(0))_1 = acw+bw+c$.  The same
  result can be obtained via summing up the yields of parse trees of
  depth $\leq 1$ or $\leq 2$ respectively, of the following grammar,
  as shown in Fig.~\ref{fig:parse:tree}.
  \begin{align*}
    X \to aXY \ | \ bY \ | \ c &  &
    Y \to uXY \ | \ vX \ | \ w
  \end{align*}
\end{ex}
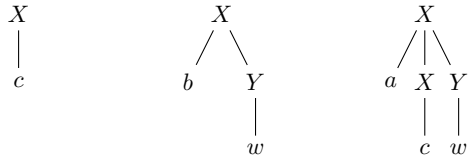
\begin{figure}[th]
  \centering
  \begin{tikzpicture}[scale = 0.9, every node/.style={transform shape}]
    \node [] at (0,1) (x1) {$X$};
    \node [] at (0,0) (c) {$c$};
    \draw [] (x1) to (c);
    \node [] at (3,1) (x2) {$X$};
    \node [] at (2.5,0) (b) {$b$};
    \node [] at (3.5,0) (y) {$Y$};
    \node [] at (3.5,-1) (w) {$w$};
    \draw [] (x2) to (b);
    \draw [] (x2) to (y);
    \draw [] (y) to (w);
    \node [] at (6,1) (x3) {$X$};
    \node [] at (5.5,0) (a) {$a$};
    \node [] at (6.5,0) (y2) {$Y$};
    \node [] at (6.5,-1) (w2) {$w$};
    \node [] at (6,0) (x4) {$X$};
    \node [] at (6,-1) (c2) {$c$};
    \draw [] (x3) to (a);
    \draw [] (x3) to (y2);
    \draw [] (y2) to (w2);
    \draw [] (x4) to (c2);
    \draw [] (x3) to (x4);
  \end{tikzpicture}
  \caption{$X$-Parse trees of depth $\leq 2$ for the grammar in
    Example~\ref{ex:fundamental}}
  \Description{$X$-Parse trees of depth $\leq 2$ for the grammar in
    Example~\ref{ex:fundamental}}
  \label{fig:parse:tree}
\end{figure}

The fact that the symbolic expansions of $\bm f^{(q)}$ can be expressed in terms of a CFL
gives us another advantage: we can make use of Parikh's theorem~\cite{MR209093} to
further characterize the terms $Y(T)$ in expression~\eqref{eqn:YT}. Note that the yield
$Y(T)$ can be thought of as a word $w$ in the CFL, where we ``concatenate" (i.e. multiply)
all terminal symbols in $T$ from left to right.
Since our multiplicative operator is commutative, only the multiplicities of the
symbols matter in differentiating two yields $Y(T)$ and $Y(T')$.
The multiplicities of the symbols are formalized by the {\em Parikh image} of the
word $w$.

More formally, consider our alphabet $\Sigma$ containing all terminal symbols in the CFL.
By renaming, we can assume $\Sigma = \{a_1,\dots,a_M\}$, and we write $\bm a =
  (a_1,\dots,a_M)$ to denote the vector of all terminal symbols.
The {\em Parikh image} of a word $w \in \Sigma^*$, denoted by $\Pi(w)$, is the vector
$\Pi(w) = (k_1,\dots,k_M) \in \N^M$ where $k_i$ is the number of occurrences of $a_i$ in
$w$.
With this notation, \eqref{eqn:YT} can be expressed as:
\begin{align}
  (\bm f^{(q)}(\bm 0))_i & = \sum_{T \in \calT^i_q} Y(T) = \sum_{T \in \calT^i_q} \bm a^{\Pi(Y(T))} \label{eqn:YT2}
\end{align}
Parikh's theorem states that the signatures $\Pi(Y(T))$ in the above expressions have a
particular format. In order to state Parikh's theorem, we need the notion of ``semi-linear
sets''.

\begin{defn}[Semi-linear sets]
  Given an integer $M>0$,
  a set $\calL \subseteq \N^M$ is said to be {\em linear} if there exist vectors
  $\bm v_0, \bm v_1, \ldots, \bm v_\ell \in \N^M$ such that:
  \begin{align*}
    \calL = & \setof{\bm v_0 + k_1 \bm v_1 + \ldots + k_\ell \bm v_\ell}{k_1, \ldots, k_\ell \in \N}
  \end{align*}
  A set $\calL \subseteq \N^M$ is called {\em semi-linear} if it is a finite
  union of linear sets.
\end{defn}

\begin{thm}[Parikh's Theorem~\cite{MR209093}]
  Let $G$ be a context-free grammar with terminal symbols $\Sigma$, and let
  $L(G) \subseteq \Sigma^*$ be the language generated by $G$.
  Then $\Pi(L(G))\subseteq \N^M$ is a semi-linear set.
\end{thm}

We now have all the tools to prove the main theorem of this section:
\begin{thm} \label{th:non-unformly:stable} If the semiring $\bm S$ is
  stable, then every polynomial function
  $\bm f: \bm S^N \rightarrow \bm S^N$ is stable.
\end{thm}
\begin{proof}
  From Parikh's theorem and identity~\eqref{eqn:YT2}, there is a {\em
      finite} collection $\calC$, where every member of $\calC$ is a
  tuple of vectors
  $(\bm v_0, \bm v_1, \dots, \bm v_\ell) \in (\N^M)^{\ell+1}$ (tuples
  in $\calC$ may have different lengths $\ell+1$), satisfying the
  following condition. For every parse-tree $T \in \calT^i_q$, there
  exists $(\bm v_0, \bm v_1, \dots, \bm v_\ell) \in \calC$, and a
  coefficient vector $(k_1,\dots,k_\ell)\in\N^\ell$ for which
  \begin{align}
    Y(T) = \bm a^{\Pi(Y(T))} = \bm a^{\bm v_0} \prod_{i=1}^\ell (\bm a^{\bm v_i})^{k_i}
  \end{align}
  Conversely, for every tuple
  $(\bm v_0, \bm v_1, \dots, \bm v_\ell) \in \calC$ and every
  coefficient vector $(k_1,\dots,k_\ell)$ there exists a parse-tree $T$ in some $\calT_q^i$
  for which the identity holds.

  Our proof strategy is as follows. We shall define below a finite set $B\subseteq\N^M$ of exponent vectors
  and express $f^{(q)}_i(0)$ by  grouping the terms $\bm a^{\bm v}$
  in~\eqref{eqn:YT2} that have the same exponent vector $\bm v\in\N^M$:
  \begin{align}
    (\bm f^{(q)}(\bm 0))_i & = \sum_{\bm v \in B} \lambda^{(q)}_{\bm v} \bm a^{\bm v} + \sum_{\bm w \notin B} \lambda^{(q)}_{\bm w} \bm a^{\bm w}, \label{eqn:B:w}
  \end{align}
  where the coefficients $\lambda_{\bm v}^{(q)}$ are integers that may change over time as
  $q$ increases:
  \begin{align}
    \lambda^{(q)}_{\bm v} = |\{
    T \in \calT^i_q : \Pi(Y(T)) = \bm v
    \}|.
    \label{eqn:lambda:q}
  \end{align}
  Here, for an element $s \in S$, and a positive integer $\lambda\in\N$, we write
  $\lambda s$ to mean $s+s+\dots +s$, $\lambda$ times. In particular, this multiplication
  of $\lambda$ and $s$ should not be confused with the multiplications of symbols in $\bm a$
  which are done over the multiplicative operator of the semiring.
  We then prove two claims, for {\em sufficiently large} $q$ the following hold:
  \begin{itemize}
    \item {\bf Claim 1} the coefficients $\lambda^{(q)}_{\bm v}\in\N$ for $\bm v \in B$ no longer
          change (they ``converge'')
    \item {\bf Claim 2} for every $\bm w \notin B$, we have
          \begin{align}
            \sum_{\bm v \in B} \lambda^{(q)}_{\bm v} \bm a^{\bm v} + \bm a^{\bm w} = \sum_{\bm v \in B} \lambda^{(q)}_{\bm v} \bm a^{\bm v} \label{eqn:claim:2:w}
          \end{align}
  \end{itemize}
  These two claims together show that $\bm f$ is stable:
  $(\bm f^{(q)}(\bm 0))_i = \sum_{\bm v \in B} \lambda^{(q)}_{\bm v} \bm a^{\bm v}$ for
  large $q$.

  We start by defining the set $B$. Fix a sufficiently large integer $p$ to be defined
  later. Define $B$ to be:~\footnote{Given a vector $\bm x =(x_1, \ldots, x_d) \in \N^d$ for some constant $d$,
  we use $\norm{\bm x}_\infty$ to denote $\max_{i \in [d]} x_i$ and
  $\norm{\bm x}_1$ to denote $\sum_{i \in [d]} x_i$.}
  \begin{align}
    B & := \{
    \bm v_0 + k_1\bm v_1 + \cdots + k_\ell \bm v_\ell \ | \
    (\bm v_0, \dots, \bm v_\ell) \in \calC \text{ and }
    \bm k = (k_1,\dots,k_\ell) \in \N^\ell \text{ where } \norm{\bm k}_\infty \leq p
    \}
    \label{eqn:B:set}
  \end{align}
  To prove Claim 1, note that, if the parse tree $T$ has depth $h$, then $\norm{\Pi(Y(T))}_1 \geq h$.
  Consequently, all parse trees whose depths are strictly greater than $\max_{\bm v \in B} \norm{\bm v}_1$
  can no longer contribute to increasing $\lambda^{(q)}_{\bm v}$ with $\bm v \in B$;
  thus, these coefficients converge after iteration number $\max_{\bm v \in B} \norm{\bm v}_1$.
  This is a finite number because~\eqref{eqn:B:set} implies:
  \begin{align}
    \max_{\bm v \in B} \norm{\bm v}_1 \leq
    \max_{(\bm v_0, \dots, \bm v_\ell) \in \calC} (1 + p \cdot \ell) \max_{i \in \{0, \ldots, \ell\}} \norm{\bm v_i}_1
  \end{align}

  Claim 2 is proved as follows. Given $\bm w \notin B$, as mentioned above, from Parikh's theorem
  we know there exists $(\bm v_0, \dots, \bm v_\ell) \in \calC$ and $\bm k \in \N^\ell$
  (but $\norm{\bm k}_\infty>p$) such that
  $\bm w = \bm v_0 + k_1\bm v_1 + \cdots + k_\ell \bm v_\ell$.
  Rewrite the monomial $\bm a^{\bm w}$ from the ``basis'' $(\bm v_0, \dots, v_\ell)$ by:
  \[
    \bm a^{\bm w} =
    \bm a^{\bm v_0} (\bm a^{\bm v_1})^{k_1} \cdots (\bm a^{\bm v_\ell})^{k_\ell}
    = m_0 m_1^{k_1} \cdots m_\ell^{k_\ell},
  \]
  where to simplify notations we define the monomials $m_i := \bm a^{\bm v_i}$.

  First, let's assume for simplicity that only one of the $k_i$ is more than $p$.
  Without loss of generality, assume $k_\ell > p$. Then, as long as $p$ is at least
  the stability index of $m_\ell$, thanks to~\eqref{eqn:c:p:q} we have
  \begin{align*}
    \sum_{i=0}^p m_0 m_1^{k_1} \cdots m_{\ell-1}^{k_{\ell-1}}m_\ell^i + m_0 m_1^{k_1} \cdots m_\ell^{k_\ell}
     & = m_0 m_1^{k_1} \cdots m_{\ell-1}^{k_{\ell-1}}m_\ell^{(p)} + m_0 m_1^{k_1} \cdots
    m_\ell^{k_\ell}                                                                               \\
     & = m_0 m_1^{k_1} \cdots m_{\ell-1}^{k_{\ell-1}}m_\ell^{(k_{\ell}-1)} + m_0 m_1^{k_1} \cdots
    m_\ell^{k_\ell}                                                                               \\
     & = m_0 m_1^{k_1} \cdots m_{\ell-1}^{k_{\ell-1}}m_\ell^{(k_{\ell})}                          \\
     & = m_0 m_1^{k_1} \cdots m_{\ell-1}^{k_{\ell-1}}m_\ell^{(p)}                                 \\
     & = \sum_{i=0}^p m_0 m_1^{k_1} \cdots m_{\ell-1}^{k_{\ell-1}}m_\ell^i.
  \end{align*}
  In particular, the term $\bm a^{\bm w}$ is ``absorbed'' by the sum of terms of the form
  $\bm a^{\bm v} = m_0 m_1^{k_1} \cdots m_{\ell-1}^{k_{\ell-1}}m_\ell^i$ for $i\leq p$.
  Since all such $\bm v$ are in $B$ (because $k_i\leq p$ for all $i<\ell$), we have just
  proved~\eqref{eqn:claim:2:w} for this simple case of $\bm w$.

  Second, when more than one of the $k_i$ is greater than $p$, we can w.l.o.g.~assume $k_\ell>p$.
  The above reasoning shows that the term
  $\bm a^{\bm w}$ is ``absorbed'' by the sum of terms $\bm a^{\bm v}$ where $\bm v$
  has one less $k_i > p$. By induction it follows that all of them will be absorbed by
  the sum $\sum_{\bm v \in B} \bm a^{\bm v}$.
\end{proof}

\subsection{Convergence in Uniformly Stable Semirings}

\label{subsec:uniformly:stable}

We consider the next case, when the semiring $\bm S$ is uniformly
stable, in other words there exists $p \geq 0$ such that every element
in $\bm S$ is $p$-stable.  In this case, we can strengthen
Theorem~\ref{th:non-unformly:stable} and prove a tighter upper bound on the
number of steps needed for convergence.  We say that the polynomial
function $\bm f:\bm S^N \rightarrow \bm S^N$ is {\em linear} if every
monomial~\eqref{eq:def:monomial} has total degree $\leq 1$.
The main theorem (Theorem~\ref{th:uniform:stable}) is proved via proving a $1$-dimensional
version of it, which is then generalized to $N$ dimensions
by applying Theorem~\ref{th:fixpoints:general}.
A univariate polynomial is of the following form:
\begin{align}
  f(x) \defeq & a_0 + a_1 x + a_2 x^2 + \cdots + a_n x^n \label{eq:f:polynomial}
\end{align}
The following lemma generalizes special cases studied by
Gondran~\cite{DBLP:journals/dm/Gondran79,semiring_book}.

\begin{lmm} \label{lmm:poly:stable} Let $\bm S$ be a $p$-stable
  semiring and let $f$ be a univariate polynomial~\eqref{eq:f:polynomial}.
  Then: (a) If $p=0$ then $f$ is $1$-stable; (b) If $f$ is linear,
  then it is $p+1$-stable; (c) In general, $f$ is $p+2$-stable.
\end{lmm}
Proving the lemma will be the bulk of work in this section. Before doing so, let us state
and prove the main theorem of the section, concerning the $N$-dimensional case.

\begin{thm} \label{th:uniform:stable}
  Assume that the semiring $\bm S$ is $p$-stable, and let
  $\bm f: \bm S^N \rightarrow \bm S^N$ be a polynomial function.
  Then:
  \begin{enumerate}
    \item \label{item:th:uniform:stable:1} The function $\bm f$ is
          $\sum_{i=1}^{N}(p+2)^i$-stable; if $\bm f$ is linear, then it is
          $\sum_{i=1}^{N}(p+1)^i$-stable.
    \item \label{item:th:uniform:stable:0} If $p=0$, then the function $\bm f$ is $N$-stable.
  \end{enumerate}
\end{thm}
\begin{proof}
  To prove item~(\ref{item:th:uniform:stable:1}), consider the
  c-clone of polynomial functions.  By Lemma~\ref{lmm:poly:stable},
  each univariate polynomial is $p+2$-stable, and therefore
  Theorem~\ref{th:fixpoints:general} implies that every polynomial
  function $S^N\rightarrow S^N$ is $\sum_{i=1,N}(p+2)^i$-stable.  If
  $\bm f$ is linear, then we consider the c-clone of linear functions,
  and derive similarly that $\bm f$ is $\sum_{i=1,N}(p+1)^i$-stable.

  For item~(\ref{item:th:uniform:stable:0}),
  when $p=0$, consider the expansion of $(\bm f^{(q)}(\bm 0))_i$ shown in~\eqref{eqn:YT2}.
  We will prove that, when $q > N$, all parse trees $T \in \calT^i_q - \calT^i_N$ are
  ``absorbed'' by those in $\calT^i_N$:
  \begin{align}
    (\bm f^{(q)}(\bm 0))_i
     & = \sum_{T \in \calT^i_q} Y(T)
    = \sum_{T \in \calT^i_N} Y(T)
    = (\bm f^{(N)}(\bm 0))_i
     &                               & \text{ when } q \geq N
    \label{eqn:YTN}
  \end{align}
  To see this, consider a parse tree $T \in \calT^i_q - \calT^i_N$. Since its depth is $>N$,
  there is a path from the root to a leaf containing a repeated variable symbol, say $x_j$
  for some $j \in [N]$.
  Let $w$ be the word that the lower copy of $x_j$ derives,
  then the higher copy of $x_j$ derives some word of the form $uwv$,
  and the tree derives the word $Y(T) = auwvb$ for some words $a,b$.
  Let $T'$ be the tree obtained from $T$ by replacing the derivation of the higher-copy of
  $x_j$ with the derivation of the lower copy of $x_j$.
  Then, $Y(T') = awb$. Now, since $p=0$, we have
  $Y(T')+Y(T) = awb + auwvb = awb(1+uv) = awb = Y(T')$.
  Repeating this process, by induction it follows that $Y(T)$ is absorbed by
  $\sum_{T \in \calT^i_N} Y(T)$.
\end{proof}

In the rest of this section, we prove the main
Lemma~\ref{lmm:poly:stable}.

\begin{proof}[Proof of Lemma~\ref{lmm:poly:stable}]
  To prove (a), we recall that in a $0$-stable semiring, $1+c = 1$ for
  every $c \in S$.  If $f$ is the polynomial in
  Eq.~\eqref{eq:f:polynomial}, then we have $f^{(1)}(0) = a_0$, and
  $f$ is $1$-stable because
  \[
    f^{(2)}(0) = a_0 + \sum_{i=1}^na_0^i a_i =
    a_0\left(1+\sum_{i=1,n}a_0^{i-1}a_i\right)=a_0 1 = a_0 = f^{(1)}(0).
  \]
  To prove (b), note that linearity means $f(x) = a_0 + a_1x$, and
  $
    f^{(n+1)}(0) = a_0 + a_1a_0 + a_1^2a_0 + \cdots + a_1^{n}a_0 =
    a_1^{(n)}a_0,
  $
  which implies $f^{(p+1)}(0)=f^{(p+2)}(0)$ since $a_1$
  is $p$-stable.

  Part (c) is the most interesting result.
  Our strategy follows that of the proof of Theorem~\ref{th:non-unformly:stable}.
  We start by writing an expansion of $f$ as was done in~\eqref{eqn:YT2} and the regrouping~\eqref{eqn:B:w}:
  \begin{align}
    f^{(q)}(0)
     & = \sum_{T \in \calT_q} Y(T)
    = \sum_{T \in \calT_q} \bm a^{\Pi(Y(T))}
    = \sum_{\bm v \in B} \lambda^{(q)}_{\bm v} \bm a^{\bm v}
    + \sum_{\bm w \notin B} \lambda^{(q)}_{\bm w} \bm a^{\bm w},
    \label{eqn:YT4}
  \end{align}
  The main difference is that in the CFL for the function~\eqref{eq:f:polynomial},
  the terminal set is $\Sigma = \{a_0,\dots,a_n\}$; and, there is only one variable, so
  we remove the index $i$ from~\eqref{eqn:YT2}: $\calT_q$ is the set of all parse trees
  of depth at most $q$, whose root is $x$.

  The exponent set $B$ is defined in exactly the same way as that in~\eqref{eqn:B:set}. However,
  taking advantage of the fact that $f$ in~\eqref{eq:f:polynomial} is univariate, we show in
  Proposition~\ref{prop:univ:B} below that the collection $\calC$ has only {\em one} specific
  tuple of vectors $(\bm v_0,\bm v_1,\dots,\bm v_n)$ (defined in~\eqref{eqn:specific:v}).
  In particular,
  \begin{align}
    B & := \{
    \bm v_0 + k_1\bm v_1 + \cdots + k_n \bm v_n \ | \
    \bm k = (k_1,\dots,k_n) \in \N^n \text{ where } \norm{\bm k}_\infty \leq p
    \}
    \label{eqn:B:set:univariate}
  \end{align}
  Finally, to show that $f$ has stability $p+2$, we prove in
  Proposition~\ref{prop:convergence:lambda} below that the sum $\sum_{\bm v \in B} \lambda^{(q)}_{\bm
      v} \bm a^{\bm v}$ is ``finalized'' (unchanged) when $q\geq p+2$.
  The fact that the remaining coefficients $\lambda^{(q)}_{\bm w}$, $\bm w \notin B$,
  are absorbed by the sum over $B$ was shown in~\eqref{eqn:claim:2:w}.
  The lemma is thus proved, modulo the proofs of the promised propositions below.
\end{proof}

The following proposition is a specialization of Parikh's theorem to the particular grammar
arising from the polynomial~\eqref{eq:f:polynomial}.
\begin{prop} \label{prop:univ:B}
  Define the following vectors in $\N^{n+1}$:
  \begin{align}
    \bm v_0 = (1, 0, \dots, 0) &  & \bm v_i = (i-1, 0, \dots, 0, 1, 0, \dots 0) &  & i \in [n].
    \label{eqn:specific:v}
  \end{align}
  Let $\calT$ be the set of {\em all} parse trees for the CFG defined for the
  polynomial~\eqref{eq:f:polynomial}, then the Parikh's images of their yields can be
  characterized precisely by:
  \begin{align}
    \{ \Pi(Y(T)) \ | \ T \in \calT \} & =
    \{ \bm v_0 + k_1\bm v_1 + \cdots + k_n \bm v_n \ | \ \bm k \in \N^n \}
  \end{align}
\end{prop}
\begin{proof}
  For the forward direction, let $T$ be any parse tree of the grammar.
  Let $k_i$ denote the number of derivation rules of the form $x \to a_i
    x\cdots x$ ($x$ occurs $i$ times on the RHS) that were used in the tree.
  Since each such rule produces one copy of $a_i$ in the yield $Y(T)$,
  it follows that $\Pi(Y(T)) = (k_0,k_1,\dots,k_n)$.
  The numbers $k_i$ are related to one another tightly.
  Consider the sub-tree of $T$ containing only the internal nodes (i.e. nodes
  corresponding to $x$, not the terminal symbols), then this sub-tree has
  exactly $k_0+\sum_{i=1}^n k_i$ nodes and $\sum_{i=1}^n i k_i$ edges.
  In a tree, the number of edges is $1$ less than the number of nodes. Hence,
  $k_0  = 1 + \sum_{i=1}^n (i-1) k_i$, which implies
  $\Pi(Y(T)) = \bm v_0 + \sum_{i=1}^n k_i \bm v_i$.

  For the backward direction, consider an arbitrary vector $\bm k = (k_1,\dots,k_n) \in
    \N^n$. We show that we can construct a parse-tree $T$ whose yield satisfies
  $\Pi(Y(T)) = \bm v_0 + \sum_{i=1}^n k_i \bm v_i$.
  We prove this by induction on $\norm{\bm k}_1$. The base case when $\norm{\bm k}_1=0$
  is trivial: the parse tree represents a single rule $x \to a_0$.
  Consider $\norm{\bm k}_1>0$, and assume $k_i > 0$.
  Let $\bm k' = (k'_1,\dots,k'_n)$ be the vector defined by $k'_j = k_j$ when $j \neq i$,
  and $k'_i = k_i-1$.
  Then, by the induction hypothesis there is a parse tree $T'$ where
  $\Pi(Y(T')) = \bm v_0 + \sum_{j=1}^n k'_j \bm v_j$.
  The parse-tree $T'$ must have used at least one rule of the form $x \to a_0$
  (otherwise the tree is infinite).
  Construct the tree $T$ from $T'$ by replacing the rule $x \to a_0$ with $x \to a_ix\cdots x$;
  and then each of the $i$ new $x$-nodes derives $a_0$.
  Then, it is trivial to verify that $\Pi(Y(T)) = \bm v_0 + \sum_{j=1}^n k_j \bm v_j$.
\end{proof}

\begin{prop} \label{prop:convergence:lambda}
  After iteration $q \geq p+2$, the sum $\sum_{\bm v \in B} \lambda^{(q)}_{\bm v} \bm a^{\bm v}$
  in~\eqref{eqn:YT4} remains unchanged.
\end{prop}
\begin{proof}
  Partition $B$ into two sets $B = B_1 \cup B_2$:
  \begin{align}
    B_1 & := \{
    \bm v_0 + k_1\bm v_1 + \cdots + k_n \bm v_n \ | \
    \bm k = (k_1,\dots,k_n) \in \N^n \text{ where } \norm{\bm k}_\infty \leq p
    \text{ and } \norm{\bm k}_1 \leq p
    \}          \\
    B_2 & := \{
    \bm v_0 + k_1\bm v_1 + \cdots + k_n \bm v_n \ | \
    \bm k = (k_1,\dots,k_n) \in \N^n \text{ where } \norm{\bm k}_\infty \leq p
    \text{ and } \norm{\bm k}_1 > p
    \}
  \end{align}
  Note that the condition $\norm{\bm k}_\infty\leq p$ in the definition of $B_1$ is
  redundant (because $\norm{\bm k}_1\leq p$ implies $\norm{\bm k}_\infty\leq p$),
  but we left it there to mirror precisely the definition of $B$
  in~\eqref{eqn:B:set:univariate}.

  Recall from~\eqref{eqn:lambda:q} that $\lambda^{(q)}_{\bm v}$ is the number of parse trees
  $T \in \calT_q$ where $\Pi(Y(T)) = \bm v$;
  furthermore, if $\bm v = \bm v_0 + \sum_{i=1}^n k_i \bm v_i$, then
  $\norm{\bm k}_1 \geq \textsf{depth}(T)$, because $\norm{\bm k}_1$ is the number of
  internal nodes  of the tree, which is at least its depth.
  Hence, when $\bm v \in B_1$, only trees of depth $\leq \norm{\bm k}_1 \leq p$ can
  contribute to increasing $\lambda^{(q)}_{\bm v}$. In other words, these coefficients are
  fixed after iteration $p$.

  It remains to show that the sum $\sum_{\bm v \in B_2} \lambda^{(q)}_{\bm v} \bm a^{\bm v}$
  is fixed after iteration $q \geq p+2$.
  To this end, note the following.
  For any element $s$ in a $p$-stable semiring, and any positive integer
  $\lambda \geq p+1$, we have
  \[
    \lambda  s := \underbrace{s+s+\cdots+s}_{\lambda  \text{ times}}
    = (1+1+\dots+1)s
    = (1+1+1^2\dots+1^{\lambda -1})s
    = 1^{(\lambda -1)}s
    = 1^{(p)}s = (p+1)s
  \]
  In particular, the sum that we want to hold fixed can be written as
  \begin{align}
    \sum_{\bm v \in B_2} \lambda^{(q)}_{\bm v} \bm a^{\bm v}
     & =
    \sum_{\bm v \in B_2} \min\{\lambda^{(q)}_{\bm v}, p+1\} \bm a^{\bm v}
  \end{align}
  It is thus sufficient to prove that, for any $\bm v \in B_2$,
  we have $\lambda_{\bm v}^{(q)} \geq p+1$ after iteration $q \geq p+2$.
  We will prove something a little stronger:

  {\bf Claim:} for any $\bm v =\bm v_0+\sum_i k_i\bm v_i \in B_2$,
  we have $\lambda_{\bm v}^{(\norm{\bm k}_\infty+2)} \geq \norm{\bm k}_1$; namely,
  there are at least $\norm{\bm k}_1$ many parse trees $T$ with depth at most
  $2+\norm{\bm k}_\infty$ and whose Parikh's image is $\Pi(Y(T)) = \bm v$.

  The claim implies what we desire, because
  for any $\bm v \in B_2$,
  $\lambda^{(p+2)}_{\bm v} \geq
    \lambda^{(\norm{\bm k}_\infty+2)}_{\bm v} \geq
    \norm{\bm k}_1 \geq p+1$.

  We prove the claim by induction on $\norm{\bm k}_\infty$.
  Consider the base case when $\norm{\bm k}_\infty=1$.
  Let $J := \{j \in [n] \ | \ k_j=1\}$ and $i = \max\{j \ | \ j \in J\}$.
  We construct parse trees by stitching together the rules (i.e. sub-trees)
  $x \to a_jx\dots x$ for $j \in J$, and fill up the leaves with $x \to a_0$ rules.
  At the root of the tree, we will apply $x \to a_ix\dots x$ for maximum flexibility,
  where there are $i$ $x$-nodes to fit the other $|J|-1$ rules in.
  Clearly, there are $\binom{i}{|J|-1}((|J|-1)!) \geq |J| \geq \norm{\bm k}_1$ distinct trees
  that can be constructed this way.

  For the inductive step, assume $\norm{\bm k}_\infty > 1$.
  Similar to the base case, define
  $J := \{j \in [n] \ | \ k_j= \norm{\bm k}_\infty \}$ and $i = \max\{j \ | \ j \in J\}$.
  Let $\bm k'$ be the vector obtained from $\bm k$ by lowering each $k_j$ by $1$, for $j \in J$;
  namely, define $k'_j = \min\{k_j, \norm{\bm k}_\infty-1\}$.
  Then, by the induction hypothesis,  there are at least $\norm{\bm k'}_1$ parse trees
  $T'$ for which $\Pi(Y(T')) = \bm v_0 + \sum_\ell k'_\ell \bm v_\ell$.
  From each of these $T'$, we construct our tree $T$ by stitching together $T'$
  and the rules (i.e. sub-trees) $x \to a_jx\dots x$, $j \in J$.
  (Implicitly, $x$-leaves are filled up with $x \to a_0$ rules.)
  For the root of $T$ we use the rule $x \to a_ix\dots x$. There are $i$ sub-tree slots to fill,
  where we can use $|J|-1$ sub-trees of the form $x \to a_jx\dots x$, $j \in J-\{i\}$,
  and $T'$, for a total of $|J|$ subtrees. (Note that $|J|\leq i$, and thus we have more slots
  than sub-trees.)

  If $i \geq 2$, then the total number of trees that can be obtained this way is
  \footnote{Note that $T'$ is distinct from any of the trees $x\to a_j x\ldots x$ for $j \in J-{i}$ because it has a different
  signature: $T'$ contains at least two internal nodes of the form $x \to a_\ell x
    \cdots x$ for $\ell>0$ due to the fact that $\norm{\bm k}_\infty < \norm{\bm
      k}_1$.}
  \begin{align}
    \norm{\bm k'}_1\binom{i}{|J|}(|J|)!
                                                 & \geq (\norm{\bm k}_1-|J|)i
    \geq (\norm{\bm k}_1-i)i \geq \norm{\bm k}_1 &                            & \text{as desired.}
  \end{align}

  If $i=1$, we have to do a bit of extra work. The total number of trees $T$ constructed
  above is only $\norm{\bm k'}_1 = \norm{\bm k}_1-1$, because the root $x \to a_1x$ has only
  one slot to fill $T'$ in.
  Let's refer to this set as $U_1$.
  We construct another set $U_2$ of trees by letting $T'$ be at the top, and find a slot to
  fit the sub-tree $x \to a_1x$ in.

  We refer to a node $x \to a_\ell x\dots x$ of a parse-tree as an $\ell$-node.
  Since $\norm{\bm k}_\infty \leq p$ and $\norm{\bm k}_1>p$, there must be some $\ell>1$
  for which $k_\ell>0$. In particular, there must be a sub-tree of $T'$ whose root node
  is an $\ell$-node with $\ell > 1$.
  In $T'$, find the left most such $\ell$-node, and the left most $0$-node ($x \to
    a_0$) below it. Replace the $x \to a_0$ rule with $x \to a_1x \to a_1a_0$ sub-tree.
  This way of construction gives us the second set $U_2$ of trees $T$ for which $\Pi(Y(T))=
    \bm v$.
  We know $|U_1|=|U_2| \geq \norm{\bm k}_1-1$.
  Furthermore, there must be at least one tree in $U_1$ that is not in $U_2$,
  because the tree $T$ in $U_1$ with the {\em longest} path containing consecutive $1$-nodes from
  the root is not in $U_2$.
  Thus $|U_1 \cup U_2| \geq \norm{\bm k}_1$, and the proof is complete.
\end{proof}

\begin{ex} \label{ex:p:plus:2:is:tight}
  Assume a semiring that is $1$-stable, and
  let $f(x) = a_0+a_2x^2 + a_3x^3 + a_4x^4$.  Then:
  \begin{align*}
    f^{(0)}(0)= & 0 & f^{(1)}(0)= & a_0 & f^{(2)}(0) = & a_0+a_0^2a_2+a_0^3a_3+a_0^4a_4
  \end{align*}
  Next, $f^{(3)}(0)$ is a longer expression that includes monomials
  like $a_0^4a_2a_3$ and $a_0^7a_2a_3a_4$.  None of these monomials
  appears in $f^{(2)}$, hence the stability index for $f$ is at least
  3.  On the other hand, $f^{(4)}(0)=f^{(3)}(0)$ because in any new
  monomial in $f^{(4)}(0)$, at least one of $a_2, a_3$, or $a_4$ has
  degree $\geq 2$, and is absorbed by other monomials already present
  in $f^{(3)}(0)$: for example, $f^{(4)}(0)$ contains the new monomial
  $a_0^5a_2^2a_3$, which is absorbed, because $f^{(4)}(0)$ also
  contains the monomials $a_0^3a_3$ and $a_0^4a_2a_3$ (they were
  already present in $f^{(3)}(0)$), and the following identity holds
  in the 1-stable semiring:
  $a_0^3a_3+a_0^4a_2a_3+a_0^5a_2^2a_3=
    a_0^3a_3\left(1+(a_0a_2)+(a_0a_2)^2\right) =
    a_0^3a_3\left(1+(a_0a_2)\right)=a_0^3a_3+a_0^4a_2a_3$.
\end{ex}

\subsection{Convergence in Stable POPS}

\label{subsec:convergence:in:pops}

We will now generalize the convergence theorems from semirings to
POPS.  Let $\bm P$ be a POPS, and recall that we assume throughout
this paper that multiplication is strict, $x \cdot \bot = \bot$.
Recall that $\bm S \defeq \bm P + \bot$ is a semiring, the {\em core
    semiring} of $\bm P$.  We call $\bm P$ {\em stable}, or {\em
    uniformly stable}, if the core $\bm S$ is stable or uniformly stable
respectively.  We generalize now Theorems~\ref{th:non-unformly:stable}
and~\ref{th:uniform:stable} from stable semirings to stable POPS.

Fix an $N$-tuple of polynomials $\bm f=(f_1, \ldots, f_N)$ over $N$
variables $\bm x = (x_1, \ldots, x_N)$. We define the following directed graph
$G_{\bm f}$.  The nodes of the graph are the variables
$x_1, \ldots, x_N$, and there exists an edge $x_i \rightarrow x_j$ if
the polynomial $f_j$ depends on $x_i$, i.e. it contains a monomial
where the factor $x_i^{\ell_i}$ has a degree $\ell_i \geq 1$.  Call a
variable $x_i$ {\em recursive} if it belongs to a cycle, or if there
exists a recursive variable $x_j$ and an edge $x_j \rightarrow x_i$.
Otherwise, we call $x_i$ {\em non-recursive}.

\begin{prop}
  If $x_i$ is recursive, then for all $q \geq 0$, $(\bm f^{(q)}(\bot))_i
    \in \bm P + \bot$.  In other words, the recursive variables cannot
  escape $\bm P + \bot$.
\end{prop}

\begin{proof}
  We prove the statement by induction on $q$.  When $q=0$ the
  statement holds because, at initialization, $(\bm f^{(0)}(\bot))_i = \bot$ for all $i$.
  Assume the statement holds for $q\geq 0$, and consider the value
  \[ (\bm f^{(q+1)}(\bot))_i = f_i((\bm f^{(q)}(\bot))_1, \ldots, (\bm
    f^{(q)}(\bot))_N),\]
  where $i$ is such that $x_i$ is recursive.  By
  assumption, $f_i$ contains some monomial
  $c \cdot x_1^{\ell_1}\cdots x_N^{\ell_N}$ that includes a recursive
  variable $x_j$, and by induction hypothesis,
  $(\bm f^{(q)}(\bot)))_j \in \bm P + \bot$.  Therefore,
  $(f^{(q+1)}(\bot))_i$ is a sum that includes the expression
  $c \cdot (\bm f^{(q)}(\bot))_1^{\ell_1} \cdots (\bm
    f^{(q)}(\bot))_N^{\ell_N}$, where the $j$-th factor is in
  $\bm P+\bot$.  Therefore $(f^{(q+1)}(\bot))_i \in \bm P + \bot$
  because both multiplication and addition with an element in
  $\bm P + \bot$ result in an element in $\bm P+\bot$, by
  $(u + \bot)v = uv + \bot v = uv+\bot$ and similarly
  $(u +\bot) + v= (u+v)+\bot$.
\end{proof}

Suppose the polynomial function $\bm f: \bm P^N \rightarrow \bm P^N$
has $N_0$ non-recursive variables and $N_1$ recursive variables, where
$N_0+N_1=N$.  We write $\bm f = (\bm g, \bm h)$, where
$\bm g : \bm P^{N_1} \times \bm P^{N_0} \rightarrow \bm P^{N_1}$ are
the polynomials associated to the recursive variables, and
$\bm h : \bm P^{N_0} \rightarrow \bm P^{N_0}$ are the rest; note that
$\bm h$ does not depend on any recursive variables.  By
Lemma~\ref{lemma:fixpoints:particular}, if $\bm g, \bm h$ are stable,
then so is $\bm f$.  We claim that the stability index of $\bm h$ is
$N_0$.  This follows easily by induction on $N_0$, using
Lemma~\ref{lemma:fixpoints:particular}.  If $N_0 = 1$ then $h(x) = c$
for some constant $c$, since it cannot depend on any variable.  Assume
$N_0 > 1$.  Since none of the variables used by $\bm h$ is recursive,
its graph $G_{\bm h}$ is acyclic, hence there is one variable without
an incoming edge.  Write $\bm h = (\bm h', h'')$, where $h''$
corresponds to the variable without incoming edge, hence it is a
constant function (doesn't depend on any variables), and $\bm h'$ is a
vector of $N_0-1$ non-recursive polynomials.  The stability index of
$\bm h'$ is $N_0-1$ by induction hypothesis, while that of $h''$ is 1,
therefore, by Lemma~\ref{lemma:fixpoints:particular}, the stability
index of $\bm h$ is $(N_0-1)+1= N_0$.

Our discussion implies:

\begin{cor} \label{cor:non-unformly:stable} (Generalization of
  Theorem~\ref{th:non-unformly:stable}) If the semiring $\bm P +\bot$
  is stable, then every polynomial function
  $\bm f : \bm P^N \rightarrow \bm P^N$ is stable.  In particular,
  every $\name$ program converges on the POPS $\bm P$.
\end{cor}

\begin{cor} \label{cor:uniform:stable} (Generalization of
  Theorem~\ref{th:uniform:stable}). Assume the POPS $\bm P$ is
  $p$-stable, and let $\bm f: \bm P^N \rightarrow \bm P^N$ be a
  polynomial function.  Then function $\bm f$ is
  $\sum_{i=0}^{N}(p+2)^i$-stable; if  $\bm f$ is linear, then it is
  $\sum_{i=0}^{N}(p+1)^i$-stable.

  In particular, every $\name$ program over the POPS $\bm P$ converges
  in a number of steps given by the expressions above, where $N$ is
  the number of grounded IDB atoms.
\end{cor}

The corollaries convey a simple intuition.  If the grounded $\name$
program is recursive, then the values of the recursive atoms cannot
escape the semiring $\bm P + \bot$; the non-recursive atoms, however,
can take values outside of this semiring.  This can be seen in
Example~\ref{ex:sum1:sum2}: when we interpret the program over the
lifted reals, $\R_\bot$, then the values of $T(a), T(b)$ remain
$\bot$, yet those of $T(c),T(d)$ take values that are $\neq \bot$.
This implies that the convergence of a recursive grounded program is
tied to the stability of $\bm R + \bot$.  The convergence of a
non-recursive grounded program, however, follows directly from the fact
that its graph is acyclic.

Finally, we prove a sufficient condition for convergence in polynomial
time.

\begin{cor} \label{cor:uniform:stable:p0} (Generalization of the case
  $p=0$ in Theorem~\ref{th:uniform:stable}) Assume the POPS $\bm P$ is
  $0$-stable, and let $\bm f: \bm P^N \rightarrow \bm P^N$ be a
  polynomial function.  Then function $\bm f$ is $N$-stable.  In
  particular, every $\name$ program over the POPS $\bm P$ converges in
  a number of steps that is polynomial in the size of the input
  database.
\end{cor}

The semirings $\B, \trop^+$ are $0$-stable; the POPS $\R_\bot$ is also
$0$-stable (because $\R_\bot + \bot$ is the trivial semiring
$\set{\bot}$, with a single element).
Corollary~\ref{cor:uniform:stable:p0} implies that $\name$ converges
in polynomial time on each of these semirings.

These three corollaries complete the proof of
Theorem~\ref{th:main:intro} in the introduction.

\subsection{Convergence in PTIME for $p$-Stable Semirings when $p>0$}
\label{sec:convergence:in:ptime}

Consider a $\name$ program over a $p$-stable POPS $\bm P$.  When $p=0$
then we know that the na\"ive algorithm converges in polynomial time.
What is its runtime when $p>0$?  Corollary~\ref{cor:uniform:stable}
gives only an exponential upper bound, and we leave open the question
whether that bound is tight.  Instead, in this section, we restrict
the $\name$ program to be a linear program, and ask whether it can be
computed in polynomial time, over a $p$-stable POPS.  We prove two
results. First, if the POPS is $\trop^+_p$ (which is $p$-stable) then
the na\"ive algorithm converges in polynomial time.  Second, for any
$p$-stable POPS, the $\name$ program can be computed in polynomial
time, by using the Floyd-Warshall-Kleene
approach~\cite{MR1059930,DBLP:journals/tcs/Lehmann77}.
We leave open the question
whether the na\"ive algorithm also converges in polynomial time.

We start with some general notations.  Fix a semiring $\bm S$, and a
linear function $F : \bm S^N \rightarrow \bm S^N$.  We can write it as
a matrix-vector product: $F(X) = AX + B$, where $A$ is an $N \times N$
matrix, and $X$, $B$ are $N$-dimensional column vectors.  After $q+1$
iterations, the na\"ive algorithm computes
$F^{(q+1)}(0) = B + AB + A^2B + \cdots + A^qB = A^{(q)}B$, where
$A^{(q)} \defeq I_N + A + A^2 + \cdots + A^q$.  The na\"ive algorithm
converges in $q+1$ steps iff $F$ is $q+1$-stable.
A matrix $A$ is called $q$-stable~\cite{semiring_book} if
$A^{(q)} = A^{(q+1)}$.  The following is easy to check: the matrix $A$
is $q$-stable iff, for every vector $B$, the linear function
$F(X)=AX+B$ is $q+1$ stable.  Our discussion implies that, in order to
determine the runtime of the na\"ive algorithm on a linear $\name$
program over a $p$-stable semiring, one has to compute the stability
index of an $N \times N$ matrix $A$ over that semiring.  Surprisingly,
no polynomial bound for the stability index of a matrix is known in
general, except for $p=0$, in which case $A$ is $N$-stable (this was
shown in~\cite{DBLP:journals/dm/Gondran79}, and also follows from our
more general Corollary~\ref{cor:uniform:stable:p0}).  We prove here a
result in the special case when the semiring is $\trop^+_p$
(introduced in Example~\ref{ex:trop:p}), which is $p$-stable.

\begin{lmm}
  \label{lemma:tropPmatrix}
  Every $N\times N$ matrix $A$ over $\trop^+_p$ semiring is
  $((p+1)N -1)$-stable.  This bound is tight, i.e., there exist
  $N\times N$-matrices over $\trop^+_p$ whose stability index is
  $(p+1)N -1$.
\end{lmm}

\begin{proof}
  We consider the $N\times N$-matrix $A$ over $\trop^+_p$ as the
  adjacency matrix of a directed graph $G$ with $N$ vertices and up to
  $p+1$ parallel edges from some vertex $i$ to $j$.  Then $A_{ij}$ is
  a bag of $p+1$ numbers representing the costs of $p + 1$ edges from
  $i$ to $j$; if fewer than $p+1$ edges exists from $i$ to $j$, then
  we complete the bag with $\infty$, intuitively saying that no
  further edge from $i$ to $j$ exists. For instance,
  $A_{ij} = \{1,2,3, \infty, \dots, \infty\}$ means that there are 3
  edges from $i$ to $j$ of length $1,2,3$, while
  $A_{ij} = \{\infty, \dots, \infty\}$ means that there is no edge
  from $i$ to $j$.

  \noindent
  {\bf Claim.} Let $k \geq 1$, let $B = A^{\ell}$ with $\ell \geq k \cdot N$, and
  let $b$ denote the minimum element in $B_{ij}$ for $1\leq i,j \leq N$.
  Moreover, let $C = I + A + A^2 + \dots + A^{\ell-1}$.
  Then  $C_{ij}$ contains at least $k$ elements which are $\leq b$.

  The upper bound of $(N+1)p-1$ follows immediately from the claim.
  To prove the claim, observe the following:
  \begin{itemize}
    \item $(A^{\ell})_{ij}$ contains the lengths of the $p+1$
          lowest-cost paths in $G$ from $i$ to $j$
          consisting of at least $kN$ edges.
    \item $C_{ij}$ contains the lengths of the $p+1$
          lowest-cost paths in $G$ from $i$ to $j$
          consisting of at most $kN-1$ edges.
  \end{itemize}
  Consider a cost-$b$ path (not necessarily simple) from $i$ to $j$ in $G$ containing at least $kN$
  edges. Every segment of length $N$ on this path contains at least a simple cycle.
  Removing the cycle yields an $ij$-path with strictly fewer edges with lower or equal
  cost. By repeating this process, we
  conclude that there are at least $k$ different $ij$-paths of cost at most $b$ with $< kN$
  edges. This observation proves the claim, and thus the upper bound.

  To prove the lower bound, consider the following matrix $A$:

  $$
    \left(
    \begin{array}{cccccc}
        \infty & A_{12} & \infty & \cdots & \cdots & \infty     \\\
        \infty & \infty & A_{23} & \infty & \cdots & \infty     \\
        \vdots &        &        & \ddots & \ddots & \vdots     \\
        \vdots &        &        &        & \ddots & \infty     \\
        \infty & \cdots & \cdots & \cdots & \infty & A_{(N-1)N} \\
        A_{N1} & \infty & \cdots & \cdots & \infty & \infty
      \end{array}
    \right)
  $$

  \noindent
  where we write $\infty$ for the bag $\{\infty,  \dots, \infty\}$, i.e.,  the 0-element
  of $\trop^+_p$. And we set $A_{12} = A_{23} = \dots = A_{(N-1)N} = A_{N1} =
    \{1, \infty, \dots$, $\infty\}$, i.e., the number 1 plus
  $p$ times $\infty$. In other words, $A$ is the adjacency matrix of the directed cycle of length $N$.
  By Claim above, we  get $A^{(N-1)}_{1N} = \{N-1,\infty, \dots,\infty\}$.
  That is,
  there exists one path from vertex 1 to vertex $N$ with $N-1$ edges and this path has
  cost (= length) $N-1$.
  Moreover, there is an edge of length $1$ back to vertex $1$.
  Hence, the $p+1$ shortest paths from vertex 1 to vertex $N$ are obtained by looping
  $0,1,2, \dots, p$ times through the vertices $1, \dots, N$ and finally going from $1$ to $N$
  along the single shortest path.
\end{proof}

The lemma immediately implies:

\begin{cor} \label{cor:convergence:trop:p} Any linear $\name$ program
  over $\trop^+_p$, converges in $(p+1)N-1$ steps, where $N=|D|^{O(1)}$ is
  the number of ground IDB tuples.  This bound is tight.
\end{cor}

Second, we consider arbitrary $p$-stable POPS $\bm P$, where $\mult$
is strict and prove that every linear $\name$ program can be computed
in polynomial time.  Gaussian elimination method, which coincides with
the Floyd-Warshall-Kleene
algorithm~\cite{MR1059930,DBLP:journals/tcs/Lehmann77}, computes the
closure $A^*$ of an $N\times N$ matrix in $O(N^3)$ time, in a closed
semiring.  This immediately extends to a $p$-stable semiring, since
every $p$-stable semiring is closed, by setting $a^* \defeq a^{(p)}$.
Our next theorem proves that essentially the same algorithm also
applies to $p$-stable POPS.

% When $\bm P$ is a {\em semiring}, then this follows from adapting
% Gaussian elimination to
% semirings~\cite{MR1059930,DBLP:journals/tcs/Lehmann77}, which
% coincides with the Floyd-Warshall-Kleene algorithm. The algorithm and
% its variants compute the closure $A^*$ of an $N\times N$ matrix in
% $O(N^3)$ time, in a closed semiring. The same principle applies to a
% $p$-stable POPS:

\begin{thm} \label{th:ptime} Let $\bm P$ be a $p$-stable POPS, strict
  ($x \cdot \bot = \bot$), and assume that both operations $+$ and
  $\cdot$ can be computed in constant time.  Let $f_1, \ldots, f_N$ be
  $N$ linear functions in $N$ variables. Then,
  $\lfp(f_1, \ldots, f_N)$ can be computed in time $O(pN + N^3)$.
\end{thm}

\begin{proof}
  Recall that a polynomial is defined as a sum of
  monomials, see Eq.~\eqref{eqn:fi}.  Thus, a {\em linear} polynomial
  is given by an expression
  $f(x_1, \ldots, x_N) = \sum_{i \in V}a_i x_i + b$, for some set
  $V \subseteq [N]$.  In other words, we cannot simply write it as
  $\sum_{i=1,N}a_ix_i + b$ because if we want to make $f$ independent
  of $x_i$, then it is not sufficient to set $a_i=0$ because
  $0 \cdot \bot = \bot$ and $x + \bot = \bot$; we need to represent
  explicitly the set of monomials $V$ in $f$.  Equivalently, $f$ is
  represented by a list of coefficients.  Let
  $g(X) = \sum_{j \in W} c_jx_j + d$ be another linear function.  We
  denote by $f[g/x_k]$ the linear function obtained by substituting
  $x_k$ with $g$ in $f$.  More precisely, if $k \not\in V$ then
  $f[g/x_k]=f$, and if $k \in V$ then we replace the term $a_k x_k$ in
  $f$ by $\sum_{j \in W} (a_kc_j) x_j$.  The representation of
  $f[g/x_k]$ can be computed in time $O(N)$ from the representations
  of $f$ and $g$.

  \begin{algorithm}[th]
    \If { $N=0$ }{ \Return $()$ }
    \If { $f_N$ is independent of $x_N$ }{
    $c(x_1,\ldots,x_{N-1})\leftarrow f_N(x_1,\ldots,x_{N-1})$}
    \If { $f_N = a_{NN}\mult x_N + b(x_1,\ldots,x_{N-1})$ }{
    $c(x_1,\ldots,x_{N-1})\leftarrow a_{NN}^{(p)}\mult
      b(x_1,\ldots,x_{N-1})+\bot$
    }
    {$(\bar x_1, \ldots, \bar x_{N-1}) \leftarrow \texttt{LinearLFP}(f_1[c/x_N],\ldots,f_{N-1}[c/x_N])$}\;
    {$\bar x_N \leftarrow c(\bar x_1, \ldots, \bar x_{N-1})$}\;
    \Return $(\bar x_1, \ldots, \bar x_N)$
    \caption{$\texttt{LinearLFP}(f_1, \ldots, f_N)$}
    \label{algo:linear:lfp}
  \end{algorithm}

  Algorithm~\ref{algo:linear:lfp} proceeds recursively on $N$.  Let
  $H \defeq (f_1, \ldots, f_N)$.  Considering the last linear function
  $f_N(x_1, \ldots, x_N)$, we write $H$ with some abuse as
  $H(x_1, \ldots, x_N) = (f(x,y),g(x,y))$, where
  $f \defeq (f_1, \ldots, f_{N-1})$ and $g \defeq f_N$, and similarly
  $x \defeq (x_1, \ldots, x_{N-1})$, $y \defeq x_N$.  Then we apply
  Lemma~\ref{lemma:fixpoints:general}.  Since $\bm S$ is $p$-stable,
  then function $g_x$ is $p+1$ stable, hence to compute the fixpoint
  using Lemma~\ref{lemma:fixpoints:general} we need to represent the
  function $c(x) \defeq g_x^{(p+1)}(\bot)$.  There are two cases. The
  first is when $f_N$ does not depend on $x_N$: then
  $g(x,y) \equiv g(x)$ does not depend on $y$, and $g_x(y)$ is a
  constant function, hence $c(x) = g_x^{(p+1)}(\bot) = g(x)$.  The
  second is when $f_N = a_{NN}\mult x_N + b(x_1, \ldots, x_{N-1})$ for
  some $a_{NN} \in \bm P$ and linear function $b$ in $N-1$ variables.
  In our new notation, $g_x(y) = a_{NN}\mult y + b(x)$. We thus get

  $g_x^{(1)}(\bot) = \bot + b(x)$,

  $g_x^{(2)}(\bot) = a_{NN} \mult (\bot + b(x)) + b(x) =
    \bot + a_{NN} \mult b(x) + b(x) = a_{NN}^{(1)} \mult b(x) + \bot$,

  $g_x^{(3)}(\bot) = a_{NN} \mult (a_{NN}^{(1)} \mult b(x) + \bot)  + b(x) =
    a_{NN}^{(2)} \mult b(x) + \bot$, etc.

  \noindent
  Hence, $c(x) = g_x^{(p+1)}(\bot) = a_{NN}^{(p)}\mult b(x) + \bot$.  Next,
  following Lemma~\ref{lemma:fixpoints:general}, we compute the
  fixpoint of $F(x) \defeq f(x,c(x))$: this is done by the algorithm
  inductively, since both $f$ and $x$ have dimension $N-1$.

  Finally, we compute the runtime.  We need $O(p)$ operations to
  compute $a_{NN}^{(p)}$, then $O(N^2)$ operations to compute a
  representation of the linear function $c$, then representations of
  the linear functions $f_1[c/x_N], \ldots, f_{N-1}[c/x_N]$.  Thus,
  the total runtime of the algorithm is
  $\sum_{n \leq N} O(p+n^2) = O(pN + N^3)$.
\end{proof}

\section{Semi-Na\"ive Optimization}
\label{sec:semi:naive}

In this section we show that the semi-na\"ive algorithm for standard
datalog can be generalized to $\name$, for certain restricted POPS.
The Na\"ive Algorithm~\ref{algo:naive} repeatedly applies the
immediate consequence operator $F$ and computes
$J^{(0)}, J^{(1)}, J^{(2)}, \ldots$, where
$J^{(t+1)} \defeq F(J^{(t)})$. This strategy is inefficient because
all facts discovered at iteration $t$ will be re-discovered at
iterations $t+1, t+2, \ldots$ The {\em semi-na\"ive} optimization
consists of a modified program that computes $J^{(t+1)}$ by first
computing only the ``novel'' facts
$\delta^{(t)} = F(J^{(t)}) - J^{(t)}$, which are then added to
$J^{(t)}$ to form $J^{(t+1)}$. Furthermore, the difference
$F(J^{(t)}) - J^{(t)}$ can be computed efficiently, {\em without}
fully evaluating $F(J^{(t)})$, by using techniques from incremental
view maintenance.
% More precisely, we will exploit the fact that $F$
% is multilinear and compute
% $\delta^{(t)} = F(J^{(t-1)}+\delta^{(t-1)}) - F(J^{(t-1)})$
% incrementally, without fully evaluating $F(J^{(t)})$.

This section generalizes the semi-na\"ive algorithm from datalog to
$\name$.  The main problem we encounter is that, while the difference
operator is well defined in the Boolean semiring, as
$x - y \defeq x \wedge \neg y$, no generic difference operator exists
in an arbitrary POPS.  In order to define a difference operator, we
will restrict the POPS to be a complete, distributive dioid.

\subsection{Complete, Distributive Dioids}
\label{subsec:dioid}

A {\em dioid} is a semiring $\bm S = (S,\oplus,\otimes,0,1)$ for which
$\oplus$ is idempotent (meaning $a \oplus a = a$).  Dioids have many applications in a wide range
of areas, see~\cite{MR1608370} for an extensive coverage.  We review
here a simple property of dioids:

%%%%%%%% We should use $\preceq$ for the natural order, as in Sec. 2
\begin{prop}
  \label{prop:dioid}
  Let $(S,\oplus,\otimes,0,1)$ be a dioid.  Then, the following hold:
  \begin{itemize}
    \item[(i)] $\bm S$ is naturally ordered, and the natural order
      coincides with the relation $\sqsubseteq$, defined by
      $a \sqsubseteq b$ if $a\oplus b=b$.
    \item[(ii)] $\oplus$ is the same as the least upper bound, $\vee$.
  \end{itemize}
  \label{prop:dioid:trivial}
\end{prop}
\begin{proof}
  Recall from Sec.~\ref{sec:pops} that the natural order is defined as
  $a\preceq b$ iff $\exists c : a\oplus c=b$.  We first show that
  $\preceq$ is the same as $\sqsubseteq$.  One direction,
  $a \sqsubseteq b$ implies $a \preceq b$ is obvious.  For the
  converse, assume $a\oplus c=b$. Then,
  $a\oplus b=a\oplus (a\oplus c)=a\oplus c=b$ due to idempotency, and
  thus $a \sqsubseteq b$.  Since $\preceq$ is a preorder,
  $\sqsubseteq$ is also a preorder.  To make it a partial order we
  only need to verify anti-symmetry, which is easily verified::
  $a\oplus b=b$ and $a\oplus b=a$ imply $a=b$.  We just proved $(i)$.

  To show $(ii)$, let $c=a\oplus b$ for some $a,b,c$.  Then
  $a\oplus c=a\oplus a\oplus b=a\oplus b=c$; thus, $a \sqsubseteq
    c$. Similarly $b \sqsubseteq c$, which means
  $a \vee b \sqsubseteq c = a\oplus b$.  Conversely, let
  $d = a \vee b$. Then, $a\oplus d=d$ and $b\oplus d=d$, which means
  $a\oplus b\oplus d=d$ and thus $a\oplus b \preceq d = a\vee b$.
\end{proof}

\begin{defn}
  A POPS $\bm S=(S,\oplus,\otimes,0,1,\sqsubseteq)$ is called a {\em
      complete, distributive dioid} if $(S,\oplus,\otimes,0,1)$ is a
  dioid, $\sqsubseteq$ is the dioid's natural order, and the ordered
  set $(S, \sqsubseteq)$ is a {\em complete, distributive lattice},
  which means that every set $A \subseteq S$ has a greatest lower
  bound $\bigwedge A$, and
  $x \vee \bigwedge A = \bigwedge \setof{x \vee a}{a \in A}$.
  In a complete, distributive dioid, the {\em difference} operator is
  defined by
  \begin{align}
    b \ominus a & \defeq \bigwedge \setof{c}{a \oplus c \sqsupseteq b} \label{eq:def:minus}
  \end{align}
\end{defn}

In order to extend the semi-na\"ive algorithm to $\name$, we require
the POPS to be a complete, distributive dioid.  There are many
examples of complete, distributive dioids:
$(2^{\bm U}, \cup, \cap, \emptyset, \bm U, \subseteq)$ is a complete,
distributive dioid, whose difference operator is exactly
set-difference
$b-a=\bigcap \setof{c}{b \subseteq a \cup c} = b \setminus a$;
$\trop^+ = (\R_+ \cup \set{\infty}, \min, +, \infty, 0, \geq)$ is also
a complete, distributive dioid, whose difference operator is defined
by Eq.~\eqref{eqn:trop:minus} in Sec.~\ref{subsec:overview}; and
$\N \cup \set{\infty}$ is also a complete, distributive dioid.  On the
other hand, $\trop^+_p$, $\trop^+_{\leq \eta}$, $\R_\bot$ are not
dioids.

The next lemma proves the only two properties of $\ominus$ that we
need for the semi-na\"ive algorithm: we need the first property in
Theorem~\ref{thm:seminaive} to prove the correctness of the algorithm,
and the second property in Theorem~\ref{thm:differential:rule} to
prove the differential rule.

\begin{lmm} \label{lemma:seminaive} Let
  $\bm S=(S,\oplus,\otimes,0,1,\sqsubseteq)$ be a complete,
  distributive dioid, and let $\ominus$ be defined
  by~\eqref{eq:def:minus}.  The following hold:
  \begin{align}
    \mbox{if } a \sqsubseteq b:\ \ \       a\oplus (b\ominus a) & = b  \label{eq:seminaive:1}                    \\
    (a\oplus b)\ominus (a\oplus c)                              & = b\ominus (a\oplus c) \label{eqn:seminaive:3}
  \end{align}
\end{lmm}
\begin{proof} We start by showing a general inequality:
  \begin{align}
    \forall a,b \in S: \ \ \  a\oplus (b\ominus a) & \sqsupseteq b   \label{eqn:properties:of:minus:a}
  \end{align}
  This follows from
  $a\oplus (b\ominus a) = a\oplus \bigwedge \setof{x}{a\oplus x
      \sqsupseteq b} = a \vee \left( \bigwedge\setof{x}{a \vee x
      \sqsupseteq b} \right)= \bigwedge \setof{a \vee x}{a \vee x
      \sqsupseteq b} \sqsupseteq b$.

  We prove now~\eqref{eq:seminaive:1}.  Denote by
  $A = \setof{x}{a \oplus x \sqsupseteq b}$, and observe that
  $b \in A$ because $a\oplus b \sqsupseteq b$.  Therefore
  $b \ominus a = \bigwedge A \sqsubseteq b$.  We also have
  $a \sqsubseteq b$ by assumption.  It follows that
  $a \oplus (b \ominus a) \sqsubseteq b \oplus b = b$.  This, together
  with~\eqref{eqn:properties:of:minus:a},
  implies~\eqref{eq:seminaive:1}.

  We prove~\eqref{eqn:seminaive:3}.  For any $x$ for which
  $a\oplus c\oplus x \sqsupseteq b$ holds, also
  $a\oplus a\oplus c\oplus x \sqsupseteq a\oplus b$ holds, which
  implies $a\oplus c\oplus x \sqsupseteq a\oplus b$ because $\oplus$
  is idempotent.  Conversely,
  $a\oplus c\oplus x \sqsupseteq a\oplus b$ implies
  $a\oplus c\oplus x\sqsupseteq b$ because $a\oplus b\sqsupseteq b$.
  Hence,
  \begin{align*}
    (a\oplus b)\ominus (a\oplus c)
     & = \bigwedge\setof{x}{a\oplus c\oplus x \sqsupseteq a\oplus b}
    = \bigwedge\setof{x}{a\oplus c\oplus x \sqsupseteq b}
    = b\ominus (a\oplus c)
  \end{align*}
\end{proof}

\subsection{The Semi-na\"ive Algorithm}

\label{subsec:differential:rule}

We describe now the semi-na\"ive algorithm for a program over a
complete, distributive dioid.  Following the notations in
Sec.~\ref{sec:datalogo}, we write $F(J)$ for the immediate consequence
operator of the program, where $J$ is an instance of the IDB
predicates.  The semi-na\"ive algorithm is shown in
Algorithm~\ref{algo:seminaive}.  It proceeds almost identically to the
Na\"ive Algorithm~\ref{algo:naive}, but splits the computation
$J \leftarrow F(J)$ into two steps: first compute the difference
$F(J)\ominus J$, then add it to $J$.  We prove:

%%% I tweaked it to align the notations with those in the Na\"Ive Algorithm
% \begin{algorithm}[t]
%      $R_0 \leftarrow \bm 0$ \ \ $\delta_0 \leftarrow \bm 0$\;
%     \For {  $t \leftarrow 1$   {\bf to}   $\infty$ }
%     {
%         {$\delta_t \leftarrow F(R_{t-1})-R_{t-1}$}\tcp*[r]{incremental computation}
%         {$R_t \leftarrow R_{t-1}+\delta_t$}\;
%         \If { $\delta_t = \bm 0$ }{
%            Break
%         }
%     }
%     \Return $ R_t$
%     \caption{Semi-na\"ive evaluation for $\name$}
%     \label{algo:seminaive}
% \end{algorithm}

\begin{algorithm}[t]
  $J^{(0)} \leftarrow \bm 0$\;
  \For {  $t \leftarrow 0$   {\bf to}   $\infty$ }
  {
    {$\delta^{(t)} \leftarrow F(J^{(t)})\ominus J^{(t)}$}\tcp*[r]{incremental computation, see Sec.~\ref{subsec:differential:rule}}
    {$J^{(t+1)} \leftarrow J^{(t)} \oplus \delta^{(t)}$}\;
    \If { $\delta^{(t)} = \bm 0$ }{
      Break
    }
  }
  \Return $J^{(t)}$
  \caption{Semi-na\"ive evaluation for $\name$}
  \label{algo:seminaive}
\end{algorithm}

\begin{thm}
  Consider a $\name$ program over a complete, distributive dioid.
  Then the Semi-na\"ive Algorithm~\ref{algo:seminaive} returns the
  same answer as the Na\"ive Algorithm~\ref{algo:naive}.
  \label{thm:seminaive}
\end{thm}

\begin{proof}
  We will use identity~\eqref{eq:seminaive:1} in
  Lemma~\ref{lemma:seminaive}.  Let $J^{(t)}$ be the sequence of IDB
  instances computed by the Semi-na\"ive
  Algorithm~\ref{algo:seminaive}, and let $\bar J^{(t)}$ be the
  sequence computed by the Na\"ive Algorithm~\ref{algo:naive}.  Recall
  that $\bar J^{(t)}$ forms an $\omega$-sequence (see
  Sec.~\ref{sec:lfp}), meaning
  $\bar J^{(0)} \sqsubseteq \bar J^{(1)} \sqsubseteq \bar J^{(2)}
    \sqsubseteq \cdots$ We prove, by induction on $t$, that
  $J^{(t)} = \bar J^{(t)}$.  The claim holds trivially for $t = 0$.
  Assuming $J^{(t)} = \bar J^{(t)}$ we prove
  $J^{(t+1)} = \bar J^{(t+1)}$.  This follows from:
  \begin{align*}
    J^{(t+1)} = & J^{(t)} \oplus (F(J^{(t)})  \ominus J^{(t)}) = \bar J^{(t)} \oplus (F(\bar J^{(t)})  \ominus \bar J^{(t)}) = F(\bar J^{(t)}) = \bar J^{(t+1)}
  \end{align*}
  The first equality is the definition of the semi-na\"ive algorithm,
  while the second equality uses the induction hypothesis
  $J^{(t)} = \bar J^{(t)}$.  The third equality is based on
  Eq.~\eqref{eq:seminaive:1},
  $\bar J^{(t)} \oplus (F(\bar J^{(t)}) \ominus \bar J^{(t)}) =
    F(\bar J^{(t)})$, which holds because
  $\bar J^{(t)} \sqsubseteq \bar J^{(t+1)} = F(\bar J^{(t)})$.
  Finally, the last equality is by the definition of the na\"ive
  algorithm.
  %     It is easy to see that $F$ is monotone, i.e. $X\sqsubseteq Y$ implies $F(X)\sqsubseteq
  %     F(Y)$. Hence, by induction we have
  %     $R_{t-1} = F^{(t-1)}(\perp) \sqsubseteq F^{(t)}(\perp)=R_t$ for all $t\geq 1$.
  %     Thus, from~\eqref{eq:seminaive:1} we have
  %     $\delta_t + R_{t-1} = (F(R_{t-1})-R_{t-1}) + R_{t-1} = F(R_{t-1})$.
\end{proof}

\subsection{The Differential Evaluation Rule}

As described, the semi-na\"ive algorithm is no more efficient than the
na\"ive algorithm.  Its advantages comes from the fact that we can
compute the difference
\begin{align}
  \delta^{(t)} \leftarrow & F(J^{(t)}) \ominus J^{(t)} \label{eq:incremental:delta:1}
\end{align}
incrementally, without computing the ICO $F$. Recall that the $\name$
program consists of $n$ rules $T_i \cd F_i(T_1, \ldots, T_n)$, one for
each IDB $T_1, \ldots, T_n$, where $F_i$ is a sum-sum-product
expression (Eq.~\eqref{eq:basic:datalogo} in Sec.~\ref{sec:datalogo}),
and the ICO is the vector of the sum-sum-product expressions,
$F = (F_1, \ldots, F_n)$.  The
difference~\eqref{eq:incremental:delta:1} consists of computing the
following differences, for $i=1,n$:
\begin{align}
  \delta_i^{(t)} \leftarrow & F_i(T_1^{(t)},\ldots,T_n^{(t)}) \ominus T_i^{(t)} \label{eq:incremental:delta:1:n}
\end{align}
For the purpose of incremental evaluation, we will assume
w.l.o.g. that each $T_j$ occurs at most once in any sum-product term
of $F_i$; otherwise, we give each occurrence of $T_j$
in~\eqref{eq:incremental:delta:1:n} a unique name, see
Example~\ref{ex:quadratic:tc} below for an illustration.  Therefore,
$F_i$ is affine\footnote{In the datalog context, ``linear'' is often used instead of ``affine''.}
in each argument $T_j$.  Notice that, when $t \geq 1$,
then $T_j^{(t)} = T_j^{(t-1)}\oplus \delta_j^{(t-1)}$ for $j=1,n$, and
$T_i^{(t)} = F_i(T_1^{(t-1)}, \ldots,T_n^{(t-1)})$.  We prove:

\begin{thm}[The Differential Rule]
  \label{thm:differential:rule} Assume that the sum-sum-product
  expression $F_i(T_1, \ldots, T_n)$ is affine in each argument $T_j$.
  Then, when $t \geq 1$, the
  difference~\eqref{eq:incremental:delta:1:n} can be computed as
  follows:
  \begin{align}
    \delta_i^{(t)} \leftarrow & \left(\bigoplus_{j=1,n}F_i(T_1^{(t)},\ldots,T_{j-1}^{(t)},\delta_j^{(t-1)},T_{j+1}^{(t-1)},\ldots,T_n^{(t-1)})\right)\ominus T_i^{(t)}\label{eq:differential:rule:f}
  \end{align}
  Furthermore, if the sum-sum-product $F_i$ can be written as
  $F_i(T_1, \ldots, T_n) = E_i \oplus G_i(T_1, \ldots, T_n)$, where
  $E_i$ is independent of $T_1, \ldots, T_n$ (i.e. it depends only on
  the EDBs), then Eq.~\eqref{eq:differential:rule:f} can be further
  simplified to:
  \begin{align}
    \delta_i^{(t)} \leftarrow & \left(\bigoplus_{j=1,n}G_i(T_1^{(t)},\ldots,T_{j-1}^{(t)},\delta_j^{(t-1)},T_{j+1}^{(t-1)},\ldots,T_n^{(t-1)})\right)\ominus T_i^{(t)}\label{eq:differential:rule}
  \end{align}
\end{thm}

The significance of the theorem is that it replaces the expensive
computation $F_i(T_1^{(t)}, \ldots, T_n^{(t)})$
in~\eqref{eq:incremental:delta:1:n} with several computations
$F_i(\cdots)$ (or $G_i(\cdots)$), where one argument is $\delta_j$
instead of $T_j$.  This is usually more efficient, because $\delta_j$
is much smaller than $T_j$.

\begin{proof}
  In order to reduce clutter, we will write $T_j$ for $T_j^{(t-1)}$
  and $\delta_j$ for $\delta_J^{(t-1)}$.  Therefore
  $T_j^{(t)} = T_j \oplus \delta_j$, and the
  difference~\eqref{eq:incremental:delta:1:n} becomes:
  \begin{align}
    \delta_i^{(t)} \leftarrow & F_i(T_1 \oplus \delta_1, \ldots, T_n\oplus\delta_n)\ominus F_i(T_1, \ldots, T_n)\label{eq:incremental:delta:3}
  \end{align}
  Fix one IDB predicate $T_j$.  Since $F_i$ is affine in $T_j$, we can
  write it as $F_j(\ldots, T_j, \ldots) = E_{ij} \oplus G_{ij}(T_j)$
  where $E_{ij}$ does not contain $T_j$, and each sum-product in
  $G_{ij}$ contains exactly one occurrence of $T_j$.  This
  representation of $F_i$ depends on the IDB $T_j$, hence we index
  $E_{ij}, G_{ij}$ by both $i$ and $j$.  Therefore,
  \begin{align*}
    F_i(\ldots, T_j \oplus \delta_j, \ldots) = & E_{ij} \oplus  G_{ij}(T_j\oplus\delta_j) = E_{ij} \oplus G_{ij}(T_j) \oplus  G_{ij}(\delta_j)                                         \\
    =                                          & (E_{ij} \oplus G_{ij}(T_j)) \oplus (E_{ij} \oplus  G_{ij}(\delta_j)) = F_i(\ldots, T_j, \ldots) \oplus F_i(\ldots,  \delta_j, \ldots)
  \end{align*}
  because $\oplus$ is idempotent, $E_{ij} \oplus E_{ij} = E_{ij}$.
  Therefore, we have:
  \begin{align*}
    F_i(T_1 \oplus \delta_1, \ldots, T_n\oplus\delta_n) = & F_i(T_1,T_2,\ldots,T_n)                                  \\
    \oplus                                                & F_i(\delta_1,T_2,\ldots,T_n)                             \\
    \oplus                                                & F_i(T_1\oplus\delta_1,\delta_2,\ldots,T_n)               \\
                                                          & \ldots                                                   \\
    \oplus                                                & F_i(T_1\oplus\delta_1,T_2\oplus\delta_2,\ldots,\delta_n)
  \end{align*}
  We substitute this in the difference~\eqref{eq:incremental:delta:3},
  then cancel the first term $F_i(T_1, \ldots, T_n)$ by using
  identity~\eqref{eqn:seminaive:3}, and obtain
  \begin{align*}
    \delta_i^{(t)} \leftarrow & \left(\bigoplus_{j=1,n}F_i(T_1\oplus\delta_1,\ldots,T_{j-1}\oplus\delta_{j-1},\delta_j,T_{j+1},\ldots,T_n)\right)\ominus F_i(T_1, \ldots, T_n)
  \end{align*}
  This proves equality~\eqref{eq:differential:rule:f}.

  To prove~\eqref{eq:differential:rule}, we observe that both
  $F_i(T_1^{(t)},\ldots,T_{j-1}^{(t)},\delta_j^{(t-1)},T_{j+1}^{(t-1)},\ldots,T_j^{(t-1)})
    = E_i \oplus G_i(\cdots)$ and $T_i^{(t)} = E_i \oplus G_i(\cdots)$
  contain the common term $E_i$, and therefore we can use
  identity~\eqref{eqn:seminaive:3} to cancel $E_i$, replacing each
  $F_i$ with $G_i$.
\end{proof}

We end this section with a short example.

\begin{ex}[Quadratic transitive closure]
  Consider the following non-linear datalog program computing the
  transitive closure:
  \begin{align*}
    T(x,y) & \cd E(x,y) \vee (\exists z : T(x,z) \wedge T(z,y))
  \end{align*}
  The semi-na\"ive algorithm with the differential
  rule~\eqref{eq:differential:rule} proceeds as follows.  It
  initializes $T^{(0)} = \emptyset$ and $\delta^{(0)} = E$, then
  repeats the following instructions for $t=1,2,\ldots$
  \begin{align*}
    \delta^{(t)}(x,y)
                   & \leftarrow \left(\exists z : \delta^{(t-1)}(x,z) \wedge T^{(t-1)}(z,y)\right) \vee
    \left(\exists z : T^{(t)}(x,z) \wedge \delta^{(t-1)}(z,y)\right)
    \setminus T^{(t)}(x,y)                                                                              \\
    T^{(t+1)}(x,y) & \leftarrow T^{(t)}(x,y) \vee \delta^{(t)}(x,y)
  \end{align*}
  \label{ex:quadratic:tc}
\end{ex}

\subsection{Discussion}

The semi-na\"ive algorithm immediately extends to stratified $\name$
(see Sec.~\ref{subsec:extensions:datalogo}), by applying the algorithm
to each stratum.

Our differential rule~\eqref{eq:differential:rule:f}
or~\eqref{eq:differential:rule} is not the only possibility for
incremental evaluation.  It has the advantage that it requires only
$n$ computations of the sum-sum-product expression $F_i$, where $n$ is
the maximum number of IDBs that occur in any sum-product of $F_i$.
But these expressions use both the previous IDBs $T_j^{(t-1)}$ and
current IDBs $T_j^{(t)}$, which, one can argue, is not ideal, because
the newer IDB instances are slightly larger than the previous ones.
An alternative is to use the $2^n-1$ discrete differentials of $F_i$,
which use only the previous IDBs $T_j^{(t-1)}$, at the cost of
increasing the number of expressions.  Yet a more sophisticated
possibility is to use higher order differentials, as pioneered by the
DBToaster project~\cite{DBLP:journals/vldb/KochAKNNLS14}.

The techniques described in this section required the POPS to be a
complete, distributive dioid.  The attentive reader may have observed
that the Semi-na\"ive Algorithm~\ref{algo:seminaive} only needs to
compute a difference $b \ominus a$ when $a \sqsubseteq b$.  This opens
the question whether the semi-na\"ive algorithm can be extended beyond
complete, distributive dioids, with a more restricted definition of
$\ominus$.  We leave this for future work.

\section{POPS and the Well-Founded Model}
\label{sec:fitting}

Our main motivation in this paper was to extend datalog to allow the
interleaving of aggregates and recursion.  We have seen that by
choosing different POPS, one can capture different kinds of
aggregations.  A large literature on datalog is concerned however with
a different extension: the interleaving of negation and recursion.  It
turns out that a certain POPS can also be used to interpret datalog
programs with negation; we give the details in this section.

Let's briefly review the most important approaches to extend datalog
with negation.  The simplest is {\em stratified datalog}, which is the
most commonly used in practice, but has limited expressive
power~\cite{DBLP:journals/iandc/Kolaitis91}.  Gelfond and
Lifschitz~\cite{DBLP:conf/iclp/GelfondL88} introduced {\em stable
  model semantics}, which has been used in constraint satisfaction
systems like DLV~\cite{DBLP:journals/tocl/LeonePFEGPS06}, but has high
data complexity, making it unsuitable for large scale data analytics.
One possibility is to simply allow the ICO to be
non-monotone~\cite{DBLP:journals/jcss/KolaitisP91}, but this leads to
an operational, non-declarative semantics.  The extension most related
to our paper is the {\em well-founded semantics}, introduced by Van
Gelder et al.~\cite{DBLP:journals/jacm/GelderRS91}, and refined
in~\cite{DBLP:conf/pods/Gelder89,DBLP:conf/pods/Przymusinski89,DBLP:journals/jlp/Fitting93,DBLP:journals/jlp/Fitting91}
(see the survey~\cite{DBLP:journals/tcs/Fitting02}).  This semantics
can be described in terms of an alternating
fixpoint~\cite{DBLP:conf/pods/Gelder89}, and therefore it is
tractable.  Przymusinski~\cite{DBLP:journals/fuin/Przymusinski90}
extended the semantics from 2-valued to 3-valued (thus, allowing atoms
to be true, or false, or undefined) and proved that every program has
a 3-valued well-founded model that is also the minimal 3-valued stable
model.  Fitting~\cite{DBLP:journals/jlp/Fitting85a} extended this
result to allow 4-valued logics (to handle contradiction) and other
bilattices.  In this section we briefly review the well-founded model,
and illustrate its connection to the least fixpoint semantics of
$\name$ on the POPS THREE, introduced in Sec.~\ref{subsec:three:pops}.

\subsection{Review of the Well-founded Model}

Instead of a formal treatment, we prefer to describe the well-founded
model on a particular example: the win-move game, which has been
extensively studied in the literature as a prototypical datalog
program with negation that cannot be
stratified~\cite{DBLP:books/aw/AbiteboulHV95}. Two players play the
following pebble game on a graph $E(X,Y)$.  The first player places
the pebble on some node. Players take turn and move the pebble, which
can only be moved along an edge.  A player who cannot move loses.  The
problem is to compute the set of winning position for the first
player.  This set is expressed concisely by the following rule:
\begin{align}
  Win(X) \cd &  \exists_Y(E(X,Y) \wedge \neg Win(Y)) \label{eq:win-move}
\end{align}
This is a datalog program {\em with} negation.  Its ICO is no longer
monotone, and the standard least fixpoint semantics no longer
applies.

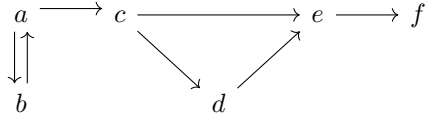
\begin{figure}
  \centering
  \[\begin{tikzcd}
	a & c && e & f \\
	b && d
	\arrow[shift right=1, from=1-1, to=2-1]
	\arrow[shift right=1, from=2-1, to=1-1]
	\arrow[from=1-2, to=1-4]
	\arrow[from=1-2, to=2-3]
	\arrow[from=2-3, to=1-4]
	\arrow[from=1-4, to=1-5]
	\arrow[shift left=1, from=1-1, to=1-2]
\end{tikzcd}\]
% https://q.uiver.app/?q=WzAsNixbMCwwLCJhIl0sWzAsMSwiYiJdLFsxLDAsImMiXSxbMywwLCJlIl0sWzIsMSwiZCJdLFs0LDAsImYiXSxbMCwxLCIiLDAseyJvZmZzZXQiOjF9XSxbMSwwLCIiLDAseyJvZmZzZXQiOjF9XSxbMiwzXSxbMiw0XSxbNCwzXSxbMyw1XSxbMCwyLCIiLDEseyJvZmZzZXQiOi0xfV1d
%
% \begin{diagram}
% a         & \rTo & c &       & \rTo  &       & e & \rTo & f \\
% \dTo \uTo &      &   & \rdTo &       & \ruTo     \\
% b         &      &   &       &  d \\
% \end{diagram}
\caption{Example graph with edges $E = \{(a,b), (a,c), (b,a), (c, d)$,
  $(c,e), (d,e), (e,f)\}$, used for the win-move game.}
\Description{Example graph with edges $E = \{(a,b), (a,c), (b,a), (c, d)$,
  $(c,e), (d,e), (e,f)\}$, used for the win-move game.}
  \label{fig:win-move}
\end{figure}

The well-founded model circumvents this problem.  It admits several,
equivalent definitions; we follow here the alternating fixpoint
semantics from~\cite{DBLP:conf/pods/Gelder89}.  Consider the grounding
of the win-move program based on the graph shown in
Fig.~\ref{fig:win-move}.  To reduce clutter, we write $W(X)$ for
$Win(X)$, and $\overline W(X)$ for $\texttt{not} (Win(X))$.  We also
write ``='' for $\cd$.  Since $E(X,Y)=1$ whenever the edge $(X,Y)$ is
present and $E(X,Y)=0$ otherwise, the grounded program is:

{\footnotesize
$$
\left.
\begin{array}{lll}
 W(a) &= & (E(a,a)\wedge \overline W(a)) \vee   (E(a,b)\wedge \overline W(b)) \vee  (E(a,c)\wedge \overline W(c)) \vee   (E(a,d)\wedge \overline W(d)) \vee  (E(a,e)\wedge \overline W(e)) \vee  (E(a,f)\wedge \overline W(f))  \\
& = &  \overline W(b) \vee \overline W(c)\\
  W(b) &= & (E(b,a)\wedge \overline W(a)) \vee  (E(b,b)\wedge \overline W(b)) \vee (E(b,c)\wedge \overline W(c)) \vee (E(b,d)\wedge \overline W(d)) \vee  (E(b,e)\wedge \overline W(e)) \vee (E(b,f)\wedge \overline W(f)) \\
 &=& \overline W(a)  \\
 W(c) &= &   \overline W(d) \vee \overline W(e)\\
 W(d) &= &  \overline W(e)\\
 W(e) &= &  \overline W(f)\\
 W(f) &= &  \bm 0
\end{array}
\right.
$$
}

The rule for $W(a)$ was obtained by substituting the existential
variable $Y$ with each node $a,b,c,d,e$, then we simplified it noting
that $E(a,b)=E(a,c)=1$ and $E(a,a)=E(a,d)=E(a,e)=0$; similarly for
$W(b)$.  For the other grounded rules, we show only the final,
simplified form.

The alternating fixpoint semantics consists of computing a sequence
$J^{(t)}$, $t \geq 0$ of standard, 2-valued IDB instances, where
$J^{(0)} = \emptyset$, and $J^{(t+1)}$ is obtained by first replacing
each negative atoms in every ground rule by its Boolean value
according to $J^{(t)}$, then $J^{(t+1)}$ is defined as the least
fixpoint of the resulting, monotone program.  In the above example, we
thus get the following sequence of truth assignments:
\begin{align*}
  &
    \begin{array}{lllllll}
           & W(a)    & W(b)    & W(c)    & W(d)    & W(e)    & W(f) \\
\hline
      J^{(0)} \ = & \bm 0 & \bm 0 & \bm 0 & \bm 0 & \bm 0 & \bm 0  \\
      J^{(1)} \ = & \bm 1 & \bm 1 & \bm 1 & \bm 1 & \bm 1  &  \bm 0  \\
      J^{(2)} \ = & \bm 0  & \bm 0  & \bm 0 & \bm 0 &  \bm 1 &   \bm 0  \\
      J^{(3)} \ = & \bm 1 & \bm 1 & \bm 1 & \bm 0 & \bm 1 & \bm 0  \\
      J^{(4)} \ = & \bm 0 & \bm 0 & \bm 1 & \bm 0 & \bm 1 &  \bm 0  \\
      J^{(5)} \ = & \bm 1 &  \bm 1 & \bm 1 & \bm 0 & \bm 1 &  \bm 0  \ = \ J^{(3)}\\
      J^{(6)} \ = & \bm 0 & \bm 0 & \bm 1 & \bm 0 & \bm 1 &  \bm 0  \ = \ J^{(4)} \\
    \end{array}
\end{align*}
The following holds:
$J^{(0)}\subseteq J^{(2)}\subseteq J^{(4)}\subseteq \ldots \subseteq
J^{(5)} \subseteq J^{(3)} \subseteq J^{(1)}$.  In other words, the
even-numbered instances form an increasing chain, while the
odd-numbered instances form a decreasing chain.  Their limits, denote
them by $L = \bigcup_t J^{(2t)}$ and $G = \bigcap_t J^{(2t+1)}$, are
the least fixpoint and the greatest fixpoint respectively, of a
certain monotone program, see~\cite{DBLP:conf/pods/Gelder89}.  The
well-founded model is defined as consisting of all positive literals
in $L$, and all negative literals missing from $G$.  In our example,
the well-founded model is
$\set{W(c),W(e), \overline W(d),\overline W(f)}$, because
$L=J^{(4)} = \set{W(c),W(e)}$ and
$G=J^{(3)}=\set{W(a),W(b),W(c),W(e)}$, and its complement is
$\set{W(d),W(f)}$.  Alternatively, the well-founded model can be
described as assigning the value 1 to the atoms $W(c), W(e)$, the
value 0 to the atoms $W(d), W(f)$, and the value $\bot$ (undefined) to
$W(a),W(b)$.

% form an
% increasing sequence, while $J^{(1)}, J^{(3)}, J^{(5)}, \ldots$ form a
% decreasing sequence.  Each can be seen as the steps of a fixpoint
% computation: $\bigcup_t J^{(2t)}$ is a least fixpoint, while
% $\bigcap_t J^{(2t+1)}$ is the greatest fixpoint of a certain program.
% Then, the well founded model is defined as follows.  If a grounded
% atom has the same value in both fixpoints, then so is its value in the
% well-founded model; otherwise its value is $\bot$, i.e. undefined.  In
% our example, the two alternating fixpoints are $J^{(4)}$ and $J^{(3)}$
% respectively, which means that the well-founded model is
% $W = (\bot,\bot, \bm 1,\bm 0,\bm 1,\bm 0)$.

\subsection{The POPS THREE}

Building on earlier work by
Przymusinski~\cite{DBLP:journals/fuin/Przymusinski90},
Fitting~\cite{DBLP:journals/jlp/Fitting85a} describes the following
three-valued semantics of logic programs with negation.  Starting from
Kleene's 3-valued logic $\set{0,0.5,0}$, where $\vee, \wedge, \neg$
are $\max, \min, 1-x$
respectively~\cite{DBLP:journals/logcom/Fitting91}, he defined two
order relations on the set THREE $=\set{\bot, 0, 1}$:
\begin{align*}
  &\mbox{The  {\em truth order:}} & 0 \leq_t & \bot \leq_t 1 \\
  &\mbox{The  {\em knowledge  order:}} & \bot \leq_k & 0 \mbox{\hspace{1cm}} \bot \leq_k  1
\end{align*}
The truth order is the same order as in Kleene's 3-valued logic, and,
for this purpose, the value $\bot$ can be interpreted as $0.5$.  This
is also the standard 3-valued logic used in SQL.  Fitting argued that,
instead of interpreting logic programs using the truth order, where
negation is not monotone, one should interpret them using the
knowledge order, where negation is monotone.

Fitting's semantics coincides, sometimes, with the well-founded model.
It also coincides with the least fixpoint of $\name$ interpreted over
the POPS THREE, and we illustrate next.  Recall from
Sec.~\ref{subsec:three:pops} that
$\texttt{THREE} \defeq (\set{\bot, 0, 1}, \vee, \wedge, 0, 1,
\leq_k)$, is a POPS where the semiring operations are the lub $\vee$
and the glb $\wedge$ of the truth order, and where the order relation
is the knowledge order $\leq_k$.  In particular, we have
$x \wedge \bm 0 = \bm 0$ and $x \vee \bm 1 = \bm 1$ for every $x$
including $x = \bot$. Hence, $\wedge$ is absorbing, i.e., THREE
constitutes a semiring and should not to be confused with the POPS
$\B_\bot \defeq \B \cup \set{\bot}$ that we get by lifting the
Booleans analogously to lifting the integers and reals to $\N_\bot$
and $\R_\bot$, respectively.

%
% In this POPS, negation can be introduced as a function
% $\mathtt{not}$ with $\mathtt{not}(\bm 0) = \bm 1$,
% $\mathtt{not}(\bm 1) = \bm 0$, and $\mathtt{not}(\bot) = \bot$. This
% function is {\em monotone\/} w.r.t.\ $\leq_k$.
%

The following $\name$ program is the win-move game~\eqref{eq:win-move}
over the POPS THREE:
\begin{align*}
  Win(Y) \cd & \bigoplus_Y  E(X,Y) \wedge \mathtt{not} ( Win(Y) )
\end{align*}

\noindent
Here,
$\mathtt{not}$
is a monotone function  w.r.t.\ $\leq_k$
defined as
$\mathtt{not}(\bm 0) = \bm 1$, $\mathtt{not}(\bm 1) = \bm 0$, and
$\mathtt{not}(\bot) = \bot$.
The fixpoint semantics computes a sequence of IDB instances
$W^{(0)} \sqsubseteq W^{(1)} \sqsubseteq W^{(2)} \sqsubseteq \cdots$,
shown here:
\begin{align*}
  &
    \begin{array}{lllllll}
           & W(a)    & W(b)    & W(c)    & W(d)    & W(e)    & W(f) \\
\hline
      W^{(0)} \ = & \bot & \bot & \bot & \bot& \bot& \bot \\
      W^{(1)} \ = & \bot & \bot & \bot& \bot& \bot &  \bm 0  \\
      W^{(2)} \ = & \bot & \bot & \bot& \bot&  \bm 1 &   \bm 0  \\
      W^{(3)} \ = & \bot & \bot & \bot& \bm 0 & \bm 1 & \bm 0  \\
      W^{(4)} \ = & \bot & \bot & \bm 1 & \bm 0 & \bm 1 &  \bm 0  \ = \ W^{(5)}
    \end{array}
\end{align*}

The least fixpoint $W^{(4)}$, is the same as the well-founded model of
the win-move program~\eqref{eq:win-move}.

\subsection{Discussion}

It should be noted that Fitting's 3-valued interpretation of logic
programs (and, likewise, $\name$ over the POPS THREE) does not always
yield the well-founded semantics. In fact, in general, it does not
even coincide with the minimal model semantics of datalog without
negation.  To see this, consider the following datalog program (or,
equivalently, $\name$ program).
\begin{align*}
  P(a) \cd & P(a)
\end{align*}
In the minimal model of this datalog program (and, equivalently, when
considering this $\name$ program over the POPS $\B$), we have
$P(a) = \bm 0$.  In contrast, Fitting's 3-valued semantics (and,
equivalently, $\name$ over the POPS THREE) yields $P(a) = \bot$. In
\cite{DBLP:journals/tcs/Fitting02}, Fitting asks ``Which is the
`right' choice?'' and he observes that a case can be made both for
setting $P(a) = \bm 0$ and for setting $P(a) = \bot$. He thus
concludes that the ``intended applications probably should decide.''

\begin{figure}[t]
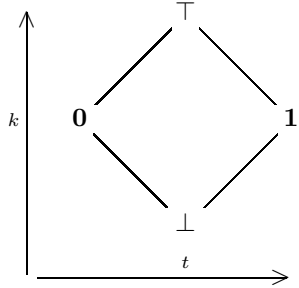

  \begin{diagram}
       &    &         & \top &         & \\
       &    & \ruLine &      & \rdLine & \\
\uTo^k & \bm 0  &         &      &         & \bm 1 \\
       &    & \rdLine &      & \ruLine & \\
       &    &         & \bot &         &\\
       &    &         & \rTo^t &   &
  \end{diagram}
\caption{Orders $\leq_k$ and $\leq_t$ in the bilattice FOUR
\cite{DBLP:journals/jlp/Fitting91}}
\Description{Orders $\leq_k$ and $\leq_t$ in the bilattice FOUR
\cite{DBLP:journals/jlp/Fitting91}}
\label{fig:FOUR}
\end{figure}

In \cite{DBLP:journals/jlp/Fitting91,DBLP:journals/jlp/Fitting93},
Fitting further extended his approach to Belnap's four-valued logic
FOUR and, more generally, to arbitrary bilattices.  FOUR
$=(\set{\bot,\bm 0,\bm 1,\top}, \leq_t, \leq_k)$ constitutes the
simplest non-trivial, complete bilattice, where the additional truth
value $\top$ denotes ``both false and true''.  We thus have a complete
lattice both w.r.t.\ the truth order
$\bm 0 \leq_t \bot, \top \leq_t \bm 1$ and w.r.t.\ the knowledge order
$\bot \leq_k \bm 0, \bm 1 \leq_k \top$, which are visualized in
Fig.~\ref{fig:FOUR} (cf.~\cite{DBLP:journals/jlp/Fitting91}, Fig.~1).
  % Fitting uses the usual Boolean connectives $\wedge,\vee$ to denote
  % the meet $\wedge$ and join $\vee$ w.r.t.\ $\leq_t$ and the
  % notation $\otimes, \oplus$ to denote the meet $\otimes$ and join
  % $\oplus$ w.r.t.\ $\leq_k$.
The \texttt{not} function is readily extended to $\top$ by setting
$\mathtt{not}(\top) = \top$. It can be shown that the additional truth
value $\top$ has no effect on the fixpoint iteration w.r.t.\ $\leq_k$
that we considered above for the POPS THREE. In fact, Fitting showed
that $\top$ can never occur as truth value in the least fixpoint
w.r.t.\ $\leq_k$ (cf.~\cite{DBLP:journals/jlp/Fitting91}, Proposition
7.1). We note that FOUR is also a POPS, where the semiring operations
$\oplus, \otimes$ are the glb and lub of the truth order, and the
partial order of the POPS is the knowledge order.

%
% The motivation in
% \cite{DBLP:journals/jlp/Fitting91,DBLP:journals/jlp/Fitting93} for
% introducing $\top$ (or, more, generally, for considering complete
% bilattices) is twofold: on the one hand, the use of $\top$ allows one
% to handle conflicting information in a meaningful way. According to
% Fitting, such conflicting information could arise from considering
% logic programs in a distributed setting.  On the other hand, $\top$
% guarantees the existence also of a {\em greatest\/} fixpoint w.r.t.\
% $\leq_k$. Fitting uses this property to establish precise lower and
% upper bounds of the set of stable models in terms of the orders
% $\leq_t$ and $\leq_k$, with the well-founded model
% \cite{DBLP:journals/jacm/GelderRS91} as the smallest stable model
% w.r.t.\ $\leq_k$.
%

\section{Related Work}\label{sec:related}

To empower Datalog, researchers have proposed amendments to make datalog capable of
expressing some problems with greedy solutions such as APSP and MST (Minimum Spanning Tree).
Most notably, the non-deterministic {\em choice} construct was extensively studied early
on~\cite{MR1842100,DBLP:conf/pods/GrecoZG92,DBLP:conf/icdt/GrecoSZ95}.
While datalog+choice is powerful, its expression and semantics are somewhat clunky,
geared towards answering optimization
questions (greedily). In particular, it was not designed to deal with general aggregations.

% Hung: I'm not completely sure what we say below about Souffle is correct. Let's revisit
% when we can confirm. This talk https://souffle-lang.github.io/pdf/SouffleLanguage.pdf
% seems to have a good summary
%
%For example, Souffl\'{e}~\cite{DBLP:conf/cav/JordanSS16}, a modern datalog
%implementation that supports both semi-naive evaluation and magic set
%optimization, restricts the program to stratified datalog, meaning
%that every recursive block is restricted to a monotone program.
%Problems like centrality in a social network cannot be expressed in
%Souffl\'{e}.  TO CHECK.

To evaluate recursive $\name$ program is to solve fixpoint equations
over semirings, which was studied in the automata
theory~\cite{MR1470001}, program
analysis~\cite{DBLP:conf/popl/CousotC77,MR1728440}, and graph
algorithms~\cite{MR556411,MR584516,MR526496} communities since the
1970s. (See~\cite{MR1059930, DBLP:conf/lics/HopkinsK99,
  DBLP:journals/tcs/Lehmann77, semiring_book,MR609751} and references
therein).  The problem took slightly different forms in these domains,
but at its core, it is to find a solution to the equation
$\bm x = \bm f(\bm x)$, where $\bm x \in \bm S^n$ is a vector over the
domain $\bm S$ of a semiring, and $\bm f : \bm S^n \to \bm S^n$ has
multivariate polynomial component functions.

When $\bm f$ is affine (what we called {\em linear} in this paper),
researchers realized that many problems in different domains are
instances of the same problem, with the same underlying algebraic
structure: transitive closure~\cite{DBLP:journals/jacm/Warshall62},
shortest paths~\cite{DBLP:journals/cacm/Floyd62a}, Kleene's theorem on
finite automata and regular languages~\cite{MR0077478}, continuous
dataflow~\cite{DBLP:conf/popl/CousotC77,DBLP:journals/jacm/KamU76},
etc.  Furthermore, these problems share the same characteristic as the
problem of computing matrix inverse~\cite{MR427338,tarjan1,MR371724}.
The problem is called the {\em algebraic path
    problem}~\cite{MR1059930}, among other names, and the main task is
to solve the matrix fixpoint equation $X = AX + I$ over a semiring.

There are several classes of solutions to the algebraic path problem, which have pros and
cons depending on what we can assume about the underlying semiring (whether or not there is
a closure operator, idempotency, natural orderability, etc.). We refer the reader
to~\cite{semiring_book,MR1059930} for more detailed discussions. Here, we briefly summarize
the main approaches.

The first approach is to keep iterating until a fixpoint is reached;
in different contexts, this has different names:
the na\"ive algorithm, Jacobi iteration, Gauss-Seidel iteration, or Kleene iteration.
The main advantage of this approach is that it assumes less about the underlying algebraic
structure: we do not need both left and right distributive law, and do not need to assume
a closure operator.

The second approach is based on Gaussian elimination (also,
Gauss-Jordan elimination), which, assuming we have oracle access to
the solution $x^*$ of the 1D problem $x= 1 + ax$, can solve the
algebraic path problem in
$O(n^3)$-time~\cite{MR1059930,DBLP:journals/tcs/Lehmann77}.

The third approach is based on specifying the solutions based on the free semiring generated
when viewing $\bm A$ as the adjacency matrix of a graph~\cite{DBLP:journals/jacm/Tarjan81a}.
The underlying graph structure (such as planarity) may sometimes be exploited for very
efficient algorithms~\cite{MR584516,MR526496}.

Beyond the affine case, since the 1960s researchers in formal languages have
been studying
the structure of the fixpoint solution to $x=f(x)$ when $f$'s component functions are
multivariate polynomials over the Kleene algebra~\cite{MR339568,MR209093,MR912710}.
It is known, for example, that Kleene iteration does not always converge (in a finite number
of steps), and thus methods based on Galois connection or on widening/narrowing
approaches~\cite{MR1252194} were studied. These approaches are (discrete) lattice-theoretic.
More recently, a completely different approach drawing inspiration from
Newton's method
for solving a system of (real) equations was
proposed~\cite{DBLP:conf/lics/HopkinsK99,DBLP:journals/jacm/EsparzaKL10}.

Recall that Newton's method for solving a system of equations $g(x) = 0$ over reals is
to start from some point $x_0$, and at time $t$ we take the first order approximation
$g(x) \approx g_t(x) := g(x_{t})+g'(x_{t})(x-x_{t})$, and set $x_{t+1}$ to be the solution of
$g_t(x)=0$, i.e. $x_{t+1}=x_t-[g'(x_t)]^{-1}g(x_t)$.
Note that in the multivariate case $g'$ is the Jacobian, and $[g'(x_t)]^{-1}$ is to compute
matrix inverse.
In {\em beautiful} papers, Esparza et al.~\cite{DBLP:journals/jacm/EsparzaKL10} and
Hopkins and Kozen~\cite{DBLP:conf/lics/HopkinsK99} were
able to generalize this idea to the case when $g(x) = f(x)-x$ is defined over
$\omega$-continuous semirings. They were able to define an appropriate {\em
    minus}
operator, derivatives of power series over semirings, matrix inverse, and prove that
the method converges at least as fast as Kleene iteration, and there are examples where
Kleene iteration does not converge, while Newton's method does.
Furthermore, if the semiring is commutative and idempotent (in addition to being
$\omega$-continuous), then  Newton's method always converges in $n$ Newton steps.
Each Newton step involves computing the Jacobian $g'$ and computing its inverse, which
is {\em exactly} the algebraic path problem!

Recently, semi-na\"ive evaluation has been extended for a higher-order functional language
called Datafun~\cite{DBLP:journals/pacmpl/ArntzeniusK20}.
The book~\cite{semiring_book} contains many fundamental results on algebraic structures
related to semirings and computational problems on such structures.

\section{Conclusions}
\label{sec:conclusions}

A massive number of application domains demand us to move beyond the confine of the Boolean
world: from program analysis~\cite{DBLP:conf/popl/CousotC77,MR1728440},
graph algorithms~\cite{MR556411,MR584516,MR526496},
provenance~\cite{DBLP:conf/pods/GreenKT07},
formal language theory~\cite{MR1470001}, to machine learning and linear
algebra~\cite{einsum:rocktaschel,DBLP:conf/osdi/AbadiBCCDDDGIIK16}.
Semiring and poset theory -- of which POPS is an instance -- is the natural
bridge connecting the Boolean island to these applications.

The bridge helps enlarge the set of problems $\name$ can express in a
very natural way.  The possibilities are endless. For example,
amending $\name$ with an interpreted function such as {\sf sigmoid}
will allow it to express typical neural network computations.  Adding
another semiring to the query language (in the sense of
FAQ~\cite{DBLP:conf/pods/KhamisNR16}) helps express rectilinear units
in modern deep learning.  At the same time, the bridge facilitates the
porting of analytical ideas from datalog to analyze convergence
properties of the application problems, and to carry over optimization
techniques such as semi-na\"ive evaluation.

This paper established part of the bridge. There are many interesting problems left
open; we mention a few here.

The question of whether a $\name$ program over $p$-stable POPS
converges in polynomial time in $p$ and the size of the input database
is open. This is open even for linear programs.  Our result on
$\trop_p$ in Sec.~\ref{sec:complexity} indicates that the linear case
is likely in PTIME.

We have discussed in Sec.~\ref{sec:datalogo} several extensions to
$\name$ that we believe are necessary in practice.  It remains open
whether our convergence results continue to hold for those extensions.

Two of the most important extensions of datalog / logic programming are
concerned with negation and aggregation. Various semantics have been proposed in the literature for these extensions,
see \cite{DBLP:journals/logcom/LiuS20}  and \cite{DBLP:journals/logcom/LiuS22} for overviews of the various approaches. 
As discussed in Sec.~\ref{sec:fitting}, negation can be added to $\name$ as an interpreted predicate. 
In particular, Fitting's three-valued semantics~\cite{DBLP:journals/jlp/Fitting85a}
of datalog with negation can thus be naturally captured in $\name$ 
by an appropriate choice of the POPS. However, the
question remains if we can also extend results for other semantics 
(above all well-founded semantics~\cite{DBLP:journals/jacm/GelderRS91} and stable model
semantics~\cite{DBLP:conf/iclp/GelfondL88}) from general datalog / logic programming to $\name$ with
negation.

The extension of datalog with aggregation was one of the original motivations for considering datalog 
over semirings \cite{DBLP:conf/pods/RossS92}. Analogously to datalog with negation, 
also for datalog with aggregation, several semantics have been proposed \cite{DBLP:journals/logcom/LiuS22}.
Again, it remains a question for future work if we can extend our results on $\name$ to 
the various semantics of datalog with aggregation. 

More generally, a systematic study of the expressive power of $\name$ is left for future work. 
While we have powerful tools
(see, e.g., \cite{DBLP:books/sp/Libkin04})
for analysing the expressive power of various logics (or, equivalently, of various query languages over the Boolean semiring) 
and pinpointing their limits, the formal study of the expressive of query languages over semirings has barely started.
In their recent work~\cite{DBLP:journals/corr/abs-2308-04910}, Brinke et al.\ have extended 
Ehrenfeucht-Fra{\"{\i}}ss{\'{e}} games to first-order logic over semirings. 
The extension of such tools to other query languages over semirings, including $\name$, 
would be a natural next step.

Beyond exact solution and finite convergence, as mentioned in the
introduction, it is natural in some domain applications to have
approximate fixpoint solutions, which will allow us to trade off
convergence time and solution quality. A theoretical framework along
this line will go a long way towards making $\name$ deal with real
machine learning, linear algebra, and optimization problems.

\begin{acks}
Suciu and Wang were partially supported by NSF IIS 1907997 and NSF IIS 1954222.
Pichler was supported by the Austrian Science Fund (FWF):P30930 and
by the
Vienna Science and Technology Fund (WWTF) [10.47379/ICT2201].
\end{acks}

% \clearpage
\bibliographystyle{acm}
\bibliography{main}

\begin{thebibliography}{10}

\bibitem{DBLP:conf/osdi/AbadiBCCDDDGIIK16}
{\sc Abadi, M., Barham, P., Chen, J., Chen, Z., Davis, A., Dean, J., Devin, M.,
  Ghemawat, S., Irving, G., Isard, M., Kudlur, M., Levenberg, J., Monga, R.,
  Moore, S., Murray, D.~G., Steiner, B., Tucker, P.~A., Vasudevan, V., Warden,
  P., Wicke, M., Yu, Y., and Zheng, X.}
\newblock Tensorflow: {A} system for large-scale machine learning.
\newblock In {\em 12th {USENIX} Symposium on Operating Systems Design and
  Implementation, {OSDI} 2016, Savannah, GA, USA, November 2-4, 2016\/} (2016),
  K.~Keeton and T.~Roscoe, Eds., {USENIX} Association, pp.~265--283.

\bibitem{DBLP:books/aw/AbiteboulHV95}
{\sc Abiteboul, S., Hull, R., and Vianu, V.}
\newblock {\em Foundations of Databases}.
\newblock Addison-Wesley, 1995.

\bibitem{DBLP:conf/pods/Khamis0PSW22}
{\sc {Abo Khamis}, M., Ngo, H.~Q., Pichler, R., Suciu, D., and Wang, Y.~R.}
\newblock Convergence of datalog over (pre-) semirings.
\newblock In {\em {PODS} '22: International Conference on Management of Data,
  Philadelphia, PA, USA, June 12 - 17, 2022\/} (2022), L.~Libkin and
  P.~Barcel{\'{o}}, Eds., {ACM}, pp.~105--117.

\bibitem{DBLP:conf/pods/KhamisNR16}
{\sc {Abo Khamis}, M., Ngo, H.~Q., and Rudra, A.}
\newblock {FAQ:} questions asked frequently.
\newblock In {\em Proceedings of the 35th {ACM} {SIGMOD-SIGACT-SIGAI} Symposium
  on Principles of Database Systems, {PODS} 2016, San Francisco, CA, USA, June
  26 - July 01, 2016\/} (2016), T.~Milo and W.~Tan, Eds., {ACM}, pp.~13--28.

\bibitem{DBLP:journals/tit/AjiM00}
{\sc Aji, S.~M., and McEliece, R.~J.}
\newblock The generalized distributive law.
\newblock {\em {IEEE} Trans. Inf. Theory 46}, 2 (2000), 325--343.

\bibitem{DBLP:journals/pacmpl/ArntzeniusK20}
{\sc Arntzenius, M., and Krishnaswami, N.}
\newblock Semina{\"{\i}}ve evaluation for a higher-order functional language.
\newblock {\em Proc. {ACM} Program. Lang. 4}, {POPL} (2020), 22:1--22:28.

\bibitem{MR427338}
{\sc Backhouse, R.~C., and Carr\'{e}, B.~A.}
\newblock Regular algebra applied to path-finding problems.
\newblock {\em J. Inst. Math. Appl. 15\/} (1975), 161--186.

\bibitem{DBLP:journals/corr/abs-2308-04910}
{\sc Brinke, S., Gr{\"{a}}del, E., and Mrkonjic, L.}
\newblock Ehrenfeucht-fra{\"{\i}}ss{\'{e}} games in semiring semantics.
\newblock {\em CoRR abs/2308.04910\/} (2023).

\bibitem{MR556411}
{\sc Carr\'{e}, B.}
\newblock {\em Graphs and networks}.
\newblock The Clarendon Press, Oxford University Press, New York, 1979.
\newblock Oxford Applied Mathematics and Computing Science Series.

\bibitem{DBLP:journals/cacm/Codd70}
{\sc Codd, E.~F.}
\newblock A relational model of data for large shared data banks.
\newblock {\em Commun. {ACM} 13}, 6 (1970), 377--387.

\bibitem{DBLP:journals/tplp/CondieDISYZ18}
{\sc Condie, T., Das, A., Interlandi, M., Shkapsky, A., Yang, M., and Zaniolo,
  C.}
\newblock Scaling-up reasoning and advanced analytics on bigdata.
\newblock {\em Theory Pract. Log. Program. 18}, 5-6 (2018), 806--845.

\bibitem{DBLP:conf/popl/CousotC77}
{\sc Cousot, P., and Cousot, R.}
\newblock Abstract interpretation: {A} unified lattice model for static
  analysis of programs by construction or approximation of fixpoints.
\newblock In {\em Conference Record of the Fourth {ACM} Symposium on Principles
  of Programming Languages, Los Angeles, California, USA, January 1977\/}
  (1977), R.~M. Graham, M.~A. Harrison, and R.~Sethi, Eds., {ACM},
  pp.~238--252.

\bibitem{MR1252194}
{\sc Cousot, P., and Cousot, R.}
\newblock Comparing the {G}alois connection and widening/narrowing approaches
  to abstract interpretation.
\newblock In {\em Programming language implementation and logic programming
  ({L}euven, 1992)}, vol.~631 of {\em Lecture Notes in Comput. Sci.} Springer,
  Berlin, 1992, pp.~269--295.

\bibitem{DBLP:conf/dagstuhl/2008coc}
{\sc Creignou, N., Kolaitis, P.~G., and Vollmer, H.}, Eds.
\newblock {\em Complexity of Constraints - An Overview of Current Research
  Themes [Result of a Dagstuhl Seminar]\/} (2008), vol.~5250 of {\em Lecture
  Notes in Computer Science}, Springer.

\bibitem{DBLP:conf/csl/DannertGNT21}
{\sc Dannert, K.~M., Gr{\"{a}}del, E., Naaf, M., and Tannen, V.}
\newblock Semiring provenance for fixed-point logic.
\newblock In {\em 29th {EACSL} Annual Conference on Computer Science Logic,
  {CSL} 2021, January 25-28, 2021, Ljubljana, Slovenia (Virtual Conference)\/}
  (2021), C.~Baier and J.~Goubault{-}Larrecq, Eds., vol.~183 of {\em LIPIcs},
  Schloss Dagstuhl - Leibniz-Zentrum f{\"{u}}r Informatik, pp.~17:1--17:22.

\bibitem{DBLP:journals/corr/abs-1909-08249}
{\sc Das, A., Li, Y., Wang, J., Li, M., and Zaniolo, C.}
\newblock Bigdata applications from graph analytics to machine learning by
  aggregates in recursion.
\newblock In {\em Proceedings 35th International Conference on Logic
  Programming (Technical Communications), {ICLP} 2019 Technical Communications,
  Las Cruces, NM, USA, September 20-25, 2019\/} (2019), B.~Bogaerts, E.~Erdem,
  P.~Fodor, A.~Formisano, G.~Ianni, D.~Inclezan, G.~Vidal, A.~Villanueva, M.~D.
  Vos, and F.~Yang, Eds., vol.~306 of {\em {EPTCS}}, pp.~273--279.

\bibitem{davey1990introduction}
{\sc Davey, B.~A., and Priestley, H.~A.}
\newblock {\em Introduction to lattices and order}.
\newblock Cambridge University Press, Cambridge, 1990.

\bibitem{DBLP:journals/constraints/Dechter97}
{\sc Dechter, R.}
\newblock Bucket elimination: a unifying framework for processing hard and soft
  constraints.
\newblock {\em Constraints An Int. J. 2}, 1 (1997), 51--55.

\bibitem{DBLP:journals/jacm/EsparzaKL10}
{\sc Esparza, J., Kiefer, S., and Luttenberger, M.}
\newblock Newtonian program analysis.
\newblock {\em J. {ACM} 57}, 6 (2010), 33:1--33:47.

\bibitem{DBLP:journals/jlp/Fitting85a}
{\sc Fitting, M.}
\newblock A kripke-kleene semantics for logic programs.
\newblock {\em J. Log. Program. 2}, 4 (1985), 295--312.

\bibitem{DBLP:journals/jlp/Fitting91}
{\sc Fitting, M.}
\newblock Bilattices and the semantics of logic programming.
\newblock {\em J. Log. Program. 11}, 1{\&}2 (1991), 91--116.

\bibitem{DBLP:journals/logcom/Fitting91}
{\sc Fitting, M.}
\newblock Kleene's logic, generalized.
\newblock {\em J. Log. Comput. 1}, 6 (1991), 797--810.

\bibitem{DBLP:journals/jlp/Fitting93}
{\sc Fitting, M.}
\newblock The family of stable models.
\newblock {\em J. Log. Program. 17}, 2/3{\&}4 (1993), 197--225.

\bibitem{DBLP:journals/tcs/Fitting02}
{\sc Fitting, M.}
\newblock Fixpoint semantics for logic programming a survey.
\newblock {\em Theor. Comput. Sci. 278}, 1-2 (2002), 25--51.

\bibitem{DBLP:journals/cacm/Floyd62a}
{\sc Floyd, R.~W.}
\newblock Algorithm 97: Shortest path.
\newblock {\em Commun. {ACM} 5}, 6 (1962), 345.

\bibitem{DBLP:conf/pods/GangulyGZ91}
{\sc Ganguly, S., Greco, S., and Zaniolo, C.}
\newblock Minimum and maximum predicates in logic programming.
\newblock In {\em Proceedings of the Tenth {ACM} {SIGACT-SIGMOD-SIGART}
  Symposium on Principles of Database Systems, May 29-31, 1991, Denver,
  Colorado, {USA}\/} (1991), D.~J. Rosenkrantz, Ed., {ACM} Press, pp.~154--163.

\bibitem{DBLP:journals/jcss/GangulyGZ95}
{\sc Ganguly, S., Greco, S., and Zaniolo, C.}
\newblock Extrema predicates in deductive databases.
\newblock {\em J. Comput. Syst. Sci. 51}, 2 (1995), 244--259.

\bibitem{DBLP:conf/pods/Gelder89}
{\sc Gelder, A.~V.}
\newblock The alternating fixpoint of logic programs with negation.
\newblock In {\em Proceedings of the Eighth {ACM} {SIGACT-SIGMOD-SIGART}
  Symposium on Principles of Database Systems, March 29-31, 1989, Philadelphia,
  Pennsylvania, {USA}\/} (1989), A.~Silberschatz, Ed., {ACM} Press, pp.~1--10.

\bibitem{DBLP:journals/jacm/GelderRS91}
{\sc Gelder, A.~V., Ross, K.~A., and Schlipf, J.~S.}
\newblock The well-founded semantics for general logic programs.
\newblock {\em J. {ACM} 38}, 3 (1991), 620--650.

\bibitem{DBLP:conf/iclp/GelfondL88}
{\sc Gelfond, M., and Lifschitz, V.}
\newblock The stable model semantics for logic programming.
\newblock In {\em Logic Programming, Proceedings of the Fifth International
  Conference and Symposium, Seattle, Washington, USA, August 15-19, 1988 {(2}
  Volumes)\/} (1988), R.~A. Kowalski and K.~A. Bowen, Eds., {MIT} Press,
  pp.~1070--1080.

\bibitem{MR371724}
{\sc Gondran, M.}
\newblock Alg\`ebre lin\'{e}aire et cheminement dans un graphe.
\newblock {\em Rev. Fran\c{c}aise Automat. Informat. Recherche
  Op\'{e}rationnelle S\'{e}r. Verte 9}, {\rm V}-1 (1975), 77--99.

\bibitem{DBLP:journals/dm/Gondran79}
{\sc Gondran, M.}
\newblock Les elements p-reguliers dans les dio{\"{\i}}des.
\newblock {\em Discret. Math. 25}, 1 (1979), 33--39.

\bibitem{semiring_book}
{\sc Gondran, M., and Minoux, M.}
\newblock {\em Graphs, dioids and semirings}, vol.~41 of {\em Operations
  Research/Computer Science Interfaces Series}.
\newblock Springer, New York, 2008.
\newblock New models and algorithms.

\bibitem{DBLP:conf/icdt/GrecoSZ95}
{\sc Greco, S., Sacc{\`{a}}, D., and Zaniolo, C.}
\newblock {DATALOG} queries with stratified negation and choice: from {P} to
  d\({}^{\mbox{p}}\).
\newblock In {\em Database Theory - ICDT'95, 5th International Conference,
  Prague, Czech Republic, January 11-13, 1995, Proceedings\/} (1995),
  G.~Gottlob and M.~Y. Vardi, Eds., vol.~893 of {\em Lecture Notes in Computer
  Science}, Springer, pp.~82--96.

\bibitem{MR1842100}
{\sc Greco, S., and Zaniolo, C.}
\newblock Greedy algorithms in {D}atalog.
\newblock {\em Theory Pract. Log. Program. 1}, 4 (2001), 381--407.

\bibitem{DBLP:conf/pods/GrecoZG92}
{\sc Greco, S., Zaniolo, C., and Ganguly, S.}
\newblock Greedy by choice.
\newblock In {\em Proceedings of the Eleventh {ACM} {SIGACT-SIGMOD-SIGART}
  Symposium on Principles of Database Systems, June 2-4, 1992, San Diego,
  California, {USA}\/} (1992), M.~Y. Vardi and P.~C. Kanellakis, Eds., {ACM}
  Press, pp.~105--113.

\bibitem{DBLP:journals/ftdb/GreenHLZ13}
{\sc Green, T.~J., Huang, S.~S., Loo, B.~T., and Zhou, W.}
\newblock Datalog and recursive query processing.
\newblock {\em Found. Trends Databases 5}, 2 (2013), 105--195.

\bibitem{DBLP:conf/pods/GreenKT07}
{\sc Green, T.~J., Karvounarakis, G., and Tannen, V.}
\newblock Provenance semirings.
\newblock In {\em Proceedings of the Twenty-Sixth {ACM} {SIGACT-SIGMOD-SIGART}
  Symposium on Principles of Database Systems, June 11-13, 2007, Beijing,
  China\/} (2007), L.~Libkin, Ed., {ACM}, pp.~31--40.

\bibitem{DBLP:conf/sigmod/0001WMSYDZ19}
{\sc Gu, J., Watanabe, Y.~H., Mazza, W.~A., Shkapsky, A., Yang, M., Ding, L.,
  and Zaniolo, C.}
\newblock Rasql: Greater power and performance for big data analytics with
  recursive-aggregate-sql on spark.
\newblock In {\em Proceedings of the 2019 International Conference on
  Management of Data, {SIGMOD} Conference 2019, Amsterdam, The Netherlands,
  June 30 - July 5, 2019\/} (2019), P.~A. Boncz, S.~Manegold, A.~Ailamaki,
  A.~Deshpande, and T.~Kraska, Eds., {ACM}, pp.~467--484.

\bibitem{MR1608370}
{\sc Gunawardena, J.}
\newblock An introduction to idempotency.
\newblock In {\em Idempotency ({B}ristol, 1994)}, vol.~11 of {\em Publ. Newton
  Inst.} Cambridge Univ. Press, Cambridge, 1998, pp.~1--49.

\bibitem{DBLP:conf/lics/HopkinsK99}
{\sc Hopkins, M.~W., and Kozen, D.}
\newblock Parikh's theorem in commutative kleene algebra.
\newblock In {\em 14th Annual {IEEE} Symposium on Logic in Computer Science,
  Trento, Italy, July 2-5, 1999\/} (1999), {IEEE} Computer Society,
  pp.~394--401.

\bibitem{DBLP:conf/cav/JordanSS16}
{\sc Jordan, H., Scholz, B., and Subotic, P.}
\newblock Souffl{\'{e}}: On synthesis of program analyzers.
\newblock In {\em Computer Aided Verification - 28th International Conference,
  {CAV} 2016, Toronto, ON, Canada, July 17-23, 2016, Proceedings, Part {II}\/}
  (2016), S.~Chaudhuri and A.~Farzan, Eds., vol.~9780 of {\em Lecture Notes in
  Computer Science}, Springer, pp.~422--430.

\bibitem{DBLP:journals/jacm/KamU76}
{\sc Kam, J.~B., and Ullman, J.~D.}
\newblock Global data flow analysis and iterative algorithms.
\newblock {\em J. {ACM} 23}, 1 (1976), 158--171.

\bibitem{MR0077478}
{\sc Kleene, S.~C.}
\newblock Representation of events in nerve nets and finite automata.
\newblock In {\em Automata studies}, Annals of mathematics studies, no. 34.
  Princeton University Press, Princeton, N. J., 1956, pp.~3--41.

\bibitem{DBLP:journals/vldb/KochAKNNLS14}
{\sc Koch, C., Ahmad, Y., Kennedy, O., Nikolic, M., N{\"{o}}tzli, A., Lupei,
  D., and Shaikhha, A.}
\newblock Dbtoaster: higher-order delta processing for dynamic, frequently
  fresh views.
\newblock {\em {VLDB} J. 23}, 2 (2014), 253--278.

\bibitem{DBLP:books/daglib/0008195}
{\sc Kohlas, J.}
\newblock {\em Information algebras - generic structures for inference}.
\newblock Discrete mathematics and theoretical computer science. Springer,
  2003.

\bibitem{DBLP:journals/ai/KohlasW08}
{\sc Kohlas, J., and Wilson, N.}
\newblock Semiring induced valuation algebras: Exact and approximate local
  computation algorithms.
\newblock {\em Artif. Intell. 172}, 11 (2008), 1360--1399.

\bibitem{DBLP:journals/iandc/Kolaitis91}
{\sc Kolaitis, P.~G.}
\newblock The expressive power of stratified programs.
\newblock {\em Inf. Comput. 90}, 1 (1991), 50--66.

\bibitem{DBLP:journals/jcss/KolaitisP91}
{\sc Kolaitis, P.~G., and Papadimitriou, C.~H.}
\newblock Why not negation by fixpoint?
\newblock {\em J. Comput. Syst. Sci. 43}, 1 (1991), 125--144.

\bibitem{MR912710}
{\sc Kuich, W.}
\newblock The {K}leene and the {P}arikh theorem in complete semirings.
\newblock In {\em Automata, languages and programming ({K}arlsruhe, 1987)},
  vol.~267 of {\em Lecture Notes in Comput. Sci.} Springer, Berlin, 1987,
  pp.~212--225.

\bibitem{MR1470001}
{\sc Kuich, W.}
\newblock Semirings and formal power series: their relevance to formal
  languages and automata.
\newblock In {\em Handbook of formal languages, {V}ol. 1}. Springer, Berlin,
  1997, pp.~609--677.

\bibitem{DBLP:journals/tcs/Lehmann77}
{\sc Lehmann, D.~J.}
\newblock Algebraic structures for transitive closure.
\newblock {\em Theor. Comput. Sci. 4}, 1 (1977), 59--76.

\bibitem{DBLP:journals/tocl/LeonePFEGPS06}
{\sc Leone, N., Pfeifer, G., Faber, W., Eiter, T., Gottlob, G., Perri, S., and
  Scarcello, F.}
\newblock The {DLV} system for knowledge representation and reasoning.
\newblock {\em {ACM} Trans. Comput. Log. 7}, 3 (2006), 499--562.

\bibitem{DBLP:books/sp/Libkin04}
{\sc Libkin, L.}
\newblock {\em Elements of Finite Model Theory}.
\newblock Texts in Theoretical Computer Science. An {EATCS} Series. Springer,
  2004.

\bibitem{MR526496}
{\sc Lipton, R.~J., Rose, D.~J., and Tarjan, R.~E.}
\newblock Generalized nested dissection.
\newblock {\em SIAM J. Numer. Anal. 16}, 2 (1979), 346--358.

\bibitem{MR584516}
{\sc Lipton, R.~J., and Tarjan, R.~E.}
\newblock Applications of a planar separator theorem.
\newblock {\em SIAM J. Comput. 9}, 3 (1980), 615--627.

\bibitem{DBLP:journals/logcom/LiuS20}
{\sc Liu, Y.~A., and Stoller, S.~D.}
\newblock Founded semantics and constraint semantics of logic rules.
\newblock {\em J. Log. Comput. 30}, 8 (2020), 1609--1668.

\bibitem{DBLP:journals/logcom/LiuS22}
{\sc Liu, Y.~A., and Stoller, S.~D.}
\newblock Recursive rules with aggregation: a simple unified semantics.
\newblock {\em J. Log. Comput. 32}, 8 (2022), 1659--1693.

\bibitem{DBLP:journals/iandc/LuttenbergerS16}
{\sc Luttenberger, M., and Schlund, M.}
\newblock Convergence of newton's method over commutative semirings.
\newblock {\em Inf. Comput. 246\/} (2016), 43--61.

\bibitem{DBLP:books/mc/18/MaierTKW18}
{\sc Maier, D., Tekle, K.~T., Kifer, M., and Warren, D.~S.}
\newblock Datalog: concepts, history, and outlook.
\newblock In {\em Declarative Logic Programming: Theory, Systems, and
  Applications}, M.~Kifer and Y.~A. Liu, Eds. {ACM} / Morgan {\&} Claypool,
  2018, pp.~3--100.

\bibitem{DBLP:journals/vldb/MazuranSZ13}
{\sc Mazuran, M., Serra, E., and Zaniolo, C.}
\newblock Extending the power of datalog recursion.
\newblock {\em {VLDB} J. 22}, 4 (2013), 471--493.

\bibitem{frankmcsherry-2022}
{\sc McSherry, F.}
\newblock Recursion in materialize.
\newblock
  \url{https://github.com/frankmcsherry/blog/blob/master/posts/2022-12-25.md},
  2022.
\newblock Accessed: December 2022.

\bibitem{MR1728440}
{\sc Nielson, F., Nielson, H.~R., and Hankin, C.}
\newblock {\em Principles of program analysis}.
\newblock Springer-Verlag, Berlin, 1999.

\bibitem{DBLP:books/sp/NocedalW99}
{\sc Nocedal, J., and Wright, S.~J.}
\newblock {\em Numerical Optimization}.
\newblock Springer, 1999.

\bibitem{MR209093}
{\sc Parikh, R.~J.}
\newblock On context-free languages.
\newblock {\em J. Assoc. Comput. Mach. 13\/} (1966), 570--581.

\bibitem{MR339568}
{\sc Pilling, D.~L.}
\newblock Commutative regular equations and {P}arikh's theorem.
\newblock {\em J. London Math. Soc. (2) 6\/} (1973), 663--666.

\bibitem{DBLP:conf/pods/Przymusinski89}
{\sc Przymusinski, T.~C.}
\newblock Every logic program has a natural stratification and an iterated
  least fixed point model.
\newblock In {\em Proceedings of the Eighth {ACM} {SIGACT-SIGMOD-SIGART}
  Symposium on Principles of Database Systems, March 29-31, 1989, Philadelphia,
  Pennsylvania, {USA}\/} (1989), A.~Silberschatz, Ed., {ACM} Press, pp.~11--21.

\bibitem{DBLP:journals/fuin/Przymusinski90}
{\sc Przymusinski, T.~C.}
\newblock The well-founded semantics coincides with the three-valued stable
  semantics.
\newblock {\em Fundam. Inform. 13}, 4 (1990), 445--463.

\bibitem{DBLP:conf/popl/RepsTP16}
{\sc Reps, T.~W., Turetsky, E., and Prabhu, P.}
\newblock Newtonian program analysis via tensor product.
\newblock In {\em Proceedings of the 43rd Annual {ACM} {SIGPLAN-SIGACT}
  Symposium on Principles of Programming Languages, {POPL} 2016, St.
  Petersburg, FL, USA, January 20 - 22, 2016\/} (2016), R.~Bod{\'{\i}}k and
  R.~Majumdar, Eds., {ACM}, pp.~663--677.

\bibitem{einsum:rocktaschel}
{\sc Rockt\"aschel, T.}
\newblock Einsum is all you need - {Einstein} summation in deep learning.
\newblock \url{https://rockt.github.io/2018/04/30/einsum}.

\bibitem{DBLP:conf/pods/RossS92}
{\sc Ross, K.~A., and Sagiv, Y.}
\newblock Monotonic aggregation in deductive databases.
\newblock In {\em Proceedings of the Eleventh {ACM} {SIGACT-SIGMOD-SIGART}
  Symposium on Principles of Database Systems, June 2-4, 1992, San Diego,
  California, {USA}\/} (1992), M.~Y. Vardi and P.~C. Kanellakis, Eds., {ACM}
  Press, pp.~114--126.

\bibitem{MR1059930}
{\sc Rote, G.}
\newblock Path problems in graphs.
\newblock In {\em Computational graph theory}, vol.~7 of {\em Comput. Suppl.}
  Springer, Vienna, 1990, pp.~155--189.

\bibitem{DBLP:conf/uai/ShenoyS88}
{\sc Shenoy, P.~P., and Shafer, G.}
\newblock Axioms for probability and belief-function proagation.
\newblock In {\em {UAI} '88: Proceedings of the Fourth Annual Conference on
  Uncertainty in Artificial Intelligence, Minneapolis, MN, USA, July 10-12,
  1988\/} (1988), R.~D. Shachter, T.~S. Levitt, L.~N. Kanal, and J.~F. Lemmer,
  Eds., North-Holland, pp.~169--198.

\bibitem{DBLP:conf/sigmod/ShkapskyYICCZ16}
{\sc Shkapsky, A., Yang, M., Interlandi, M., Chiu, H., Condie, T., and Zaniolo,
  C.}
\newblock Big data analytics with datalog queries on spark.
\newblock In {\em Proceedings of the 2016 International Conference on
  Management of Data, {SIGMOD} Conference 2016, San Francisco, CA, USA, June 26
  - July 01, 2016\/} (2016), pp.~1135--1149.

\bibitem{DBLP:conf/icde/ShkapskyYZ15}
{\sc Shkapsky, A., Yang, M., and Zaniolo, C.}
\newblock Optimizing recursive queries with monotonic aggregates in deals.
\newblock In {\em 31st {IEEE} International Conference on Data Engineering,
  {ICDE} 2015, Seoul, South Korea, April 13-17, 2015\/} (2015), J.~Gehrke,
  W.~Lehner, K.~Shim, S.~K. Cha, and G.~M. Lohman, Eds., {IEEE} Computer
  Society, pp.~867--878.

\bibitem{MR1676282}
{\sc Stanley, R.~P.}
\newblock {\em Enumerative combinatorics. {V}ol. 2}, vol.~62 of {\em Cambridge
  Studies in Advanced Mathematics}.
\newblock Cambridge University Press, Cambridge, 1999.
\newblock With a foreword by Gian-Carlo Rota and appendix 1 by Sergey Fomin.

\bibitem{tarjan1}
{\sc Tarjan, R.~E.}
\newblock Graph theory and gaussian elimination, 1976.
\newblock J.R. Bunch and D.J. Rose, eds.

\bibitem{DBLP:journals/jacm/Tarjan81a}
{\sc Tarjan, R.~E.}
\newblock A unified approach to path problems.
\newblock {\em J. {ACM} 28}, 3 (1981), 577--593.

\bibitem{DBLP:conf/stoc/Vardi82}
{\sc Vardi, M.~Y.}
\newblock The complexity of relational query languages (extended abstract).
\newblock In {\em Proceedings of the 14th Annual {ACM} Symposium on Theory of
  Computing, May 5-7, 1982, San Francisco, California, {USA}\/} (1982), H.~R.
  Lewis, B.~B. Simons, W.~A. Burkhard, and L.~H. Landweber, Eds., {ACM},
  pp.~137--146.

\bibitem{DBLP:conf/pods/Vianu21}
{\sc Vianu, V.}
\newblock Datalog unchained.
\newblock In {\em PODS'21: Proceedings of the 40th {ACM} {SIGMOD-SIGACT-SIGAI}
  Symposium on Principles of Database Systems, Virtual Event, China, June
  20-25, 2021\/} (2021), L.~Libkin, R.~Pichler, and P.~Guagliardo, Eds., {ACM},
  pp.~57--69.

\bibitem{DBLP:conf/sigmod/WangK0PS22}
{\sc Wang, Y.~R., {Abo Khamis}, M., Ngo, H.~Q., Pichler, R., and Suciu, D.}
\newblock Optimizing recursive queries with progam synthesis.
\newblock In {\em {SIGMOD} '22: International Conference on Management of Data,
  Philadelphia, PA, USA, June 12 - 17, 2022\/} (2022), Z.~Ives, A.~Bonifati,
  and A.~E. Abbadi, Eds., {ACM}, pp.~79--93.

\bibitem{DBLP:journals/jacm/Warshall62}
{\sc Warshall, S.}
\newblock A theorem on boolean matrices.
\newblock {\em J. {ACM} 9}, 1 (1962), 11--12.

\bibitem{DBLP:journals/corr/abs-1910-08888}
{\sc Zaniolo, C., Das, A., Gu, J., Li, Y., Li, M., and Wang, J.}
\newblock Monotonic properties of completed aggregates in recursive queries.
\newblock {\em CoRR abs/1910.08888\/} (2019).

\bibitem{DBLP:journals/debu/ZanioloD0LL021}
{\sc Zaniolo, C., Das, A., Gu, J., Li, Y., Li, M., and Wang, J.}
\newblock Developing big-data application as queries: an aggregate-based
  approach.
\newblock {\em {IEEE} Data Eng. Bull. 44}, 2 (2021), 3--13.

\bibitem{DBLP:conf/amw/ZanioloYDI16}
{\sc Zaniolo, C., Yang, M., Das, A., and Interlandi, M.}
\newblock The magic of pushing extrema into recursion: Simple, powerful datalog
  programs.
\newblock In {\em Proceedings of the 10th Alberto Mendelzon International
  Workshop on Foundations of Data Management, Panama City, Panama, May 8-10,
  2016\/} (2016), R.~Pichler and A.~S. da~Silva, Eds., vol.~1644 of {\em {CEUR}
  Workshop Proceedings}, CEUR-WS.org.

\bibitem{DBLP:journals/tplp/ZanioloYDSCI17}
{\sc Zaniolo, C., Yang, M., Das, A., Shkapsky, A., Condie, T., and Interlandi,
  M.}
\newblock Fixpoint semantics and optimization of recursive datalog programs
  with aggregates.
\newblock {\em Theory Pract. Log. Program. 17}, 5-6 (2017), 1048--1065.

\bibitem{DBLP:conf/amw/ZanioloYIDSC18}
{\sc Zaniolo, C., Yang, M., Interlandi, M., Das, A., Shkapsky, A., and Condie,
  T.}
\newblock Declarative bigdata algorithms via aggregates and relational database
  dependencies.
\newblock In {\em Proceedings of the 12th Alberto Mendelzon International
  Workshop on Foundations of Data Management, Cali, Colombia, May 21-25,
  2018\/} (2018), D.~Olteanu and B.~Poblete, Eds., vol.~2100 of {\em {CEUR}
  Workshop Proceedings}, CEUR-WS.org.

\bibitem{MR609751}
{\sc Zimmermann, U.}
\newblock Linear and combinatorial optimization in ordered algebraic
  structures.
\newblock {\em Ann. Discrete Math. 10\/} (1981), viii+380.

\end{thebibliography}

\appendix

\clearpage
\noindent
{\bf \LARGE APPENDIX}

\section{Lower Bound for Theorem~\ref{th:fixpoints:general}}

\label{app:fixpoint}

%%%%%%%%%%%%%%%%%%%%%%%%%%%%%%%%%%%%%%%%%%%%%%%%%%%%%%%%%%%%%%%%

\begin{proof}[Proof of the lower bound]
To keep the notation simple,
we will choose $p_1 = p_2 =  \dots = p_n = p$ below.
It is straightforward to extend this case to  arbitrary values $p_1 , p_2,  \dots, p_n$.
Moreover, we may restrict ourselves to the case $p \geq 1$, since the lower bound for
$p = 0$ is trivial.
We define $n$ posets $\bm L_1, \ldots, \bm L_n$,
and functions $f_1, \dots, f_n$
of types
$f_i : \prod_{j=1,n}\bm L_j \rightarrow \bm L_i$,
as follows:
For every  $i \in [n]$,
let $\bm L_i \defeq \{a^{(i)}_j \mid j \in \N\}$
with the linear order $a^{(i)}_0 <  a^{(i)}_1 < a^{(i)}_2 < \dots$.
By slight abuse, we will simply write $\bm L_i = \N$
with the understanding that, in $\bm L_i$,  integer $j$ actually stands for $a^{(i)}_j$.
Only when it comes to
the definition of the c-clone and we have to take care of type-compatibilities,
it will be  important to keep in mind that the $\bm L_i$'s are pairwise distinct.

For every $i \in [n]$, we define the function $f_i \colon   \prod_{j=1,n}\bm L_j \rightarrow \bm L_i$
by setting $f_i(c_1, \dots, c_n) = d_i$ via the following case distinction:

\begin{enumerate}
\item  Case $i < n$:
\begin{enumerate}
\item if $c_{i+1} \geq p^{i+1}$ then $d_i = p^i$;
\item else $d_i  = \lfloor c_{i+1} /  p \rfloor$.
\end{enumerate}

\item  Case $i =n$:
\begin{enumerate}
\item if there exists $j \in [n]$ with $c_j \geq p^j$ then $d_n = p^n$;
\item else if  $c_n \geq (c_{n-1} + 1) \cdot p$ then $d_n = c_n$;
\item else if   $c_n > c_{n-1} \cdot p $ then $d_n = c_n +1$;
\item else  if  there exists $j < n-1$ with  $c_j \cdot p^{n-j} < c_{n-1} \cdot p $ then
$d_n = c_{n-1} \cdot p$;
\item else
$d_n = c_{n-1} \cdot p + 1$;
\end{enumerate}
\end{enumerate}

\noindent
Let $h = (f_1, \dots, f_n)$ and $m \geq 0$.
It is easy to verify that the
sequence $h^{(m)}(0, \dots, 0)$ with $h = (f_1, \dots, f_n)$ and $m \geq 0$
consists of  all $n$-tuples $(c_1, \dots, c_n)$ with $0 \leq c_i \leq p^i$ in ascending lexicographical order.
The sequence is depicted in Fig.~\ref{fig:hsequence}. For every $i$, the $i$-th component is incremented
until the limit $p^i$ is reached. Moreover, with each application of $h$, we only increment one component. Hence,
function $h$ indeed has
stability index  $p + p^2 + p^3 + \dots + p^n$.

\begin{figure}
$$
\left.
\begin{array}{ccc}
(0,\dots, 0,0),
& \dots & (0,\dots, 0,p), \\
(0,\dots, 1,p), & \dots & (0,\dots, 1,2p), \\
 \vdots &  & \vdots \\
(0,\dots, p-1,p^2-p), &  \dots & (0,\dots, p-1,p^2), \\
(0,\dots, 0, p,p^2),  \\
(0,\dots, 1, p,p^2),  &  \dots & (0,\dots, 1, p,p^2+p), \\
(0,\dots, 1, p+1,p^2+p),  &  \dots & (0,\dots, 1, p,p^2+2p), \\
 \vdots &  & \vdots \\
(0,\dots, p-1, p^2-1, p^3 - p),  &  \dots & (0,\dots, p-1, p^2-1,p^3), \\
(0,\dots, 0, p-1, p^2, p^3 ), \\
(0,\dots, 0, p, p^2, p^3 ), \\
(0,\dots, 1, p, p^2, p^3 ), & \dots & (0,\dots, 1, p, p^2, p^3 +p), \\
 \vdots &  & \vdots \\
(p-1,p^2-1, \dots, \hskip20pt &                                   & (p-1,p^2-1, \dots,  \\
\hskip30pt  p^{n-1}-1, p^n -p), &  \dots  & \hskip33pt p^{n-1}-1, p^n), \\
(p-1,p^2-1, \dots, p^{n-1}, p^n ), \\
\vdots \\
(p-1,p^2, \dots, p^{n-1}, p^n ), \\
(p,p^2, \dots, p^{n-1}, p^n ).
 \end{array} \right.
$$
\caption{Sequence of tuples produced by $h^{(m)}(0, \dots, 0)$.}
\Description{Sequence of tuples produced by $h^{(m)}(0, \dots, 0)$.}
\label{fig:hsequence}
\end{figure}

It remains to show that the functions $h, f_1, \dots, f_n$ are part of  a c-clone $\calC$, such
that every function $f \colon \bm L_i \rightarrow \bm L_i$ in $\calC$ is $p$-stable.
We choose $\calC$ as the smallest c-clone containing the functions $h, f_1, \dots, f_n$.
We have to show that every function $f \colon \bm L_i \rightarrow \bm L_i$ in $\calC$ is $p$-stable and
that all functions in $\calC$ are monotone.

\medskip
\noindent
{\em Intuition.}
Before we prove these properties, we discuss the rationale behind the above definition of  the functions $f_i$.
The main challenge here is, that the functions $f_i$ have to be defined over the entire domain, i.e., we also have to consider
value combinations $(c_1, \dots, c_n)$ that are never reached by $h^{(m)}(0, \dots, 0)$.
An inspection of the sequence $h^{(m)}(0, \dots, 0)$ shows that, for every $i < n$,
component $c_{i+1}$ may only range from
$c_i \cdot p$ to $(c_i +1) \cdot p$. Equivalently, $c_i$ is fully determined by the
value of $c_{i+1}$, namely $c_i  = \lfloor c_{i+1} /  p \rfloor$.
Hence, in Case 1.b, we set $f_i(c_1, \dots, c_n)  = \lfloor c_{i+1} /  p \rfloor$.
However, in Case 1.a, we first have to make sure that
$f_i(c_1, \dots, c_n)$ never takes a value above $p^i$.

Similarly, in Case 2.a, whenever one of $c_j$ has reached its upper bound $p^j$ (or is even greater), then
we also set $f_n(c_1, \dots, c_n)$ to its upper bound, i.e., $p^n$. Of course, in the sequence produced by
$h^{(m)}(0, \dots, 0)$, the $j$-th component can never take a value $> p^j$ and also the value $p^j$ itself
is only possible if the $n$-th component has already reached $p^n$.

After having excluded the case of an excessively big component $c_j$  in Case 2.a, we still have to distinguish 4 further cases:
we have already observed above that the $(i+1)$-st component can only take values in the interval
$[c_i \cdot p, (c_i+1) \cdot p]$. That is, $c_n$ may only take values in $[c_{n-1} \cdot p, (c_{n-1}+1) \cdot p]$.
Hence, if  $c_n \geq (c_{n-1} + 1) \cdot p$, then we must not further increase $c_n$.
This is taken care of by Case 2.b.
In fact, in the sequence
$h^{(m)}(0, \dots, 0)$, we first have to increase $c_{n-1}$ and possibly further components to the left, before we may
continue incrementing $c_n$. The incrementation of $c_n$,  as long as $c_n$ is in the allowed range,
is realized via Case 2.c.

Cases 2.d and 2e only apply if
$c_n  \leq  c_{n-1} \cdot p$ holds.
Both cases cover two kinds of situations:  $c_n  =  c_{n-1} \cdot p$, which may
occur in the sequence $h^{(m)}(0, \dots, 0)$,
and $c_n  <  c_{n-1} \cdot p$, which cannot occur in this sequence.
For the case $c_n  =  c_{n-1} \cdot p$,
recall that $c_n$ may only take values in $[c_{n-1} \cdot p, (c_{n-1}+1) \cdot p]$.
An inspection of the sequence $h^{(m)}(0, \dots, 0)$ shows that also the components $c_j$ with $j < n-1$
impose a restriction on the allowed values of $c_n$, namely:
for  all $j < n-1$, component $c_n$ must be in the interval $[c_j \cdot p^{n-j},  (c_j + 1) \cdot p^{n-j}]$.
Hence, if $c_n = c_{n-1} \cdot p$, we have to distinguish the two cases, if we must keep $c_n$ unchanged and first
increment a component $c_j$ to the left of $c_{n-1}$ (i.e., Case 2.d) or if we may continue incrementing $c_n$
(i.e.,  Case 2.e).

Cases 2.d and 2e also cover the case
$c_n  <  c_{n-1} \cdot p$,  which can never occur in $h^{(m)}(0, \dots, 0)$.
When defining $f_n(c_1, \dots, c_n)$ in this case, we have to make sure that $f_n$
is monotone and every function $f \colon \bm L_n \rightarrow \bm L_n$ in $\calC$ is $p$-stable.
This goal is achieved via Cases 2.d and 2e by treating $c_n  <  c_{n-1} \cdot p$ in the same way as
$c_n  =  c_{n-1} \cdot p$.
The following example illustrates why this makes sense:
let $p = 3$ and $n=3$, and  consider the tuple $(2, p, 0)$. The third component $c_3 = 0$ is clearly below the
interval of allowed values  $[p^2,p^2+p]$ by $c_2 = p$.
When choosing an appropriate value for $f_3(2, p, 0)$,  we must be careful so as
not to destroy monotonicity. In particular, by $f_3(0, 0, 0) = 1$, we have to choose $f_3(2, p, 0) \geq 1$.
However, when choosing a concrete value $\geq 1$, we also have to keep the $p$-stability of all functions
$f \colon \bm L_n \rightarrow \bm L_n$ in $\calC$ in mind. In particular, consider the function
$f$ obtained from $f_3$ when fixing the first two components to constants $2$ and $p$, i.e.,
$f(x) = f_3(2,p,x)$.  We cannot simply set $f_3(2, p, 0) =  1$, i.e., $f(0) = 1$ for the following reason:
consider the tuple $(2, p, 1)$. By monotonicity, we have to set  $f(1) = f_3((2, p, 1) \geq f_3(0, 0, 1) = 2$.
This argument can be repeated for $f_3(2,p,x)$ for every $x \leq p^2$.
In order to ensure $p$-stability of $f$, we have to choose $f_3(2,p, 0)$ big enough so that at most $p-1$ further
incrementations of $c_3$ are possible.
That is why we simply treat $f_3(2,p, 0)$ in the same way as $f_3(2,p, p^2)$.

We now prove that $\calC$ defined as the smallest  c-clone containing functions $h, f_1, \dots, f_n$
has the desired properties in terms of monotonicity and $p$-stability:

\medskip
\noindent
{\em Proof of Monotonicity.}
We show that every function $f_i$ is monotone. From this, it follows easily that all functions in $\calC$ are monotone.
The case $i < n$ will turn out rather simple. In contrast, the case $i = n$ will require a more detailed analysis.
The following observation will be helpful: no matter which of the Cases 2.b -- 2.e is used to define
$f_n(c_1, \dots, c_n)$, the inequations
$f_n(c_1, \dots, c_n) \leq p^n$ and
$f_n(c_1, \dots, c_n) \geq c_n$ and $f_n(c_1, \dots, c_n) \geq c_{n-1} \cdot p$
always hold. This property is easy to verify by inspecting the definition
of $f_n(c_1, \dots, c_n)$
in each of the four Cases 2.b -- 2.e.

Now consider two tuples
$(c_1, \dots, c_n)$ and $(c'_1, \dots, c'_n)$ with the property
$(c_1, \dots, c_n) \sqsubseteq (c'_1, \dots, c'_n)$, i.e.,
$c_j \leq c'_j$ for every $j$. In order to show  $f_i(c_1, \dots, c_n)$ $\leq f_i (c'_1, \dots, c'_n)$,
we distinguish, which case of the definition of $f_i$ applies to $(c_1, \dots, c_n)$.

\smallskip

\noindent
{\em Case 1: $i < n$.}
If (1.a) $c_{i+1} \geq p^{i+1}$ then also $c'_{i+1} \geq p^{i+1}$ and we have
$f_i(c_1, \dots, c_n) = p^i = f_i (c'_1, \dots, c'_n)$. On the other hand, if
(1.b) $c_{i+1} < p^{i+1}$, then $f_i(c_1, \dots, c_n)  = \lfloor c_{i+1} /  p \rfloor <  p^i$.
Hence, no matter whether $f_i (c'_1, \dots, c'_n)$ is determined via Case 1.a or 1.b,
the property $f_i(c_1, \dots, c_n)$ $\leq f_i (c'_1, \dots, c'_n)$ is always guaranteed.

\smallskip

\noindent
{\em Case 2: $i = n$.}
We analyze each of the Cases 2.a -- 2.e separately.

\smallskip

\noindent
{\em Case 2.a.}
If   there exists $j \in [n]$ with $c_j \geq p^j$ then also
$c'_j \geq p^j$ and we have
$f_n(c_1, \dots, c_n) = p^n = f_n (c'_1, \dots, c'_n)$.

\smallskip

\noindent
{\em Case 2.b.}
If $c_n \geq (c_{n-1} + 1) \cdot p$ then $f_n(c_1, \dots, c_n) = c_n \leq p^n$.
On the other hand, we either have
$f_n(c'_1, \dots, c'_n) = p^n$ (if Case 2.a applies) or
$f_n(c'_1, \dots, c'_n) \geq  c'_n $ (in all other cases) by the above observation. Hence,
$f_n(c_1, \dots, c_n) \leq f_n (c'_1, \dots, c'_n)$ clearly holds.

\smallskip

\noindent
{\em Case 2.c.}
If $c_n > c_{n-1} \cdot p$ then $f_n(c_1, \dots, c_n) = c_n +1 \leq p^n$.
Again, recall that
we either have  $f_n(c'_1, \dots, c'_n) = p^n$ (if Case 2.a applies) or
$f_n(c'_1, \dots, c'_n) \geq c'_n $ (in all other cases). Hence,  the only
interesting case is that
Case 2.a does not apply to $f_n(c'_1, \dots, c'_n)$ and
$c'_n = c_n$. We now show that, no matter which of the remaining Cases 2.b -- 2.e
applies to $f_n(c'_1, \dots, c'_n)$,
we always get the desired inequation $f_n(c_1, \dots, c_n) \leq f_n (c'_1, \dots, c'_n)$.

Note that we are dealing with the case $c_n < (c_{n-1} + 1) \cdot p$.
Moreover, we are assuming  $c'_n = c_n$ and
$c'_{n-1} \geq c_{n-1}$. Hence, we also have $c'_n < (c'_{n-1} + 1) \cdot p$, which means that
$f_n(c'_1, \dots, c'_n)$  cannot be defined via Case 2.b.

If $f_n(c'_1, \dots, c'_n)$
is defined via Case 2.c, then $f_n(c'_1, \dots, c'_n) = c'_n + 1$
holds and we are done.

If one of the Cases 2.d or 2.e applies, then $c'_{n-1} \cdot p \geq c'_n$
must hold.
In contrast, we are considering the Case 2.c for $f_n(c_1, \dots, c_n)$. That is,
$c_{n-1} \cdot p < c_n$. Since we are assuming $c'_n = c_n$, we have
$c'_{n-1} > c_{n-1}$ or, equivalently, $c'_{n-1} \geq c_{n-1} + 1$.
By the above observation, we have $f_n(c'_1, \dots, c'_n) \geq c'_{n-1} \cdot p$.
On the other hand, since Case 2.b does not apply to $f_n(c_1, \dots, c_n)$, we have
$c_n < (c_{n-1} + 1) \cdot p$ or, equivalently,
$c_n + 1 \leq  (c_{n-1} + 1) \cdot p$.
In total, this gives $f_n(c'_1, \dots, c'_n) \geq c'_{n-1} \cdot p  \geq (c_{n-1} +1)  \cdot p
\geq c_n + 1 = f_n(c_1, \dots, c_n)$.

\smallskip

\noindent
{\em Case 2.d.}
If there exists $j < n-1$ with  $c_j \cdot p^{n-j} < c_{n-1} \cdot p $ then
$f_n (c_1, \dots, c_n) = c_{n-1} \cdot p \leq p^n$.
On the other hand, we either have $f_n(c'_1, \dots, c'_n)  = p^n$ (if Case 2.a applies) or
$f_n(c'_1, \dots, c'_n) \geq c'_{n-1} \cdot p$ (in all other cases) by the above observation.
Either way,
$f_n (c_1, \dots, c_n) \leq f_n(c'_1, \dots, c'_n)$
holds.

\smallskip

\noindent
{\em Case 2.e.}
If for all $j < n-1$:  $c_j \cdot p^{n-j} \geq c_{n-1} \cdot p $ then
$f_n(c_1, \dots, c_n) =  c_{n-1} \cdot p +1 \leq p^n$.
By the above observation, either
$f_n(c'_1, \dots, c'_n) = p^n$ holds (if Case 2.a applies) or
$f_n(c'_1, \dots, c'_n) \geq c'_{n-1} \cdot p $ (in all other cases)
holds. Hence,  the only
interesting case is that
Case 2.a does not apply to $f_n(c'_1, \dots, c'_n)$ and
$c'_{n-1} = c_{n-1}$.
We now show that, no matter which of the remaining Cases 2.b -- 2.e
applies to $f_n(c'_1, \dots, c'_n)$,
we always get the desired inequation $f_n(c_1, \dots, c_n) \leq f_n (c'_1, \dots, c'_n)$.

Case 2.b applies only if $c'_n \geq (c'_{n-1} + 1) \cdot p$, thus
$f_n (c'_1, \dots, c'_n) = c'_n$. We are assuming
$p \geq 1$; hence $(c'_{n-1} + 1) \cdot p \geq c'_{n-1} \cdot p +1 $ holds.
In total, we thus have $f_n (c'_1, \dots, c'_n) = c'_n \geq c'_{n-1} \cdot p +1 \geq
c_{n-1} \cdot p +1 = f_n (c_1, \dots, c_n)$.

If Case 2.c applies, then $c'_n > c'_{n-1} \cdot p$ and we set
$f_n (c'_1, \dots, c'_n) = c'_n +1$. Hence, we have
$f_n (c'_1, \dots, c'_n) = c'_n +1 > c'_{n-1} \cdot p +1  \geq
c_{n-1} \cdot p +1 = f_n (c_1, \dots, c_n)$.

Note that Case 2.d would mean that there exists $j < n-1$ with  $c'_j \cdot p^{n-j} < c'_{n-1} \cdot p $.
However, this cannot happen by our current assumptions, namely: for every $j$: $c'_j  \geq c_j$,
$c_{n-1} = c'_{n-1}$, and
for all $j < n-1$:  $c_j \cdot p^{n-j} \geq c_{n-1} \cdot p $.

Finally, if Case 2.e applies, then
$f_n(c'_1, \dots, c'_n) = c'_{n-1} \cdot p +1$. Together with $c'_{n-1} = c_{n-1}$ and
$f_n(c_1, \dots, c_n) =  c_{n-1} \cdot p +1$, we again have
 $f_n(c_1, \dots, c_n) \leq f_n (c'_1, \dots, c'_n)$.

\medskip

\noindent
{\em Proof of Stability.}
By choosing $\calC$
as the minimal c-clone containing the functions $h, f_1, \dots, f_n$,
we know that all functions $f \colon \bm L_i \rightarrow \bm L_i$ are of the form
$f_{\bm{c}} \colon \bm L_i \rightarrow \bm L_i$ with
$\bm{c} = (c_1, \dots, c_{i-1}, c_{i+1}, \dots, c_n)$
and $f_{\bm{c}}(x) = f_i (c_1, \dots, c_{i-1}, x, c_{i+1}, \dots, c_n)$
for some constants $c_j \in \bm L_j$ for every $j \neq i$.
We have to show that $f^{(p)}_{\bm{c}}(0)   = f^{(p+1)}_{\bm{c}}(0)$ holds.
To this end, we distinguish which of the cases of the definition of
$f_i$ applies to $(c_1, \dots, c_{i-1}, 0, c_{i+1}, \dots, c_n)$.

\smallskip

\noindent
{\em Case 1: $i < n$.}
If (1.a) $c_{i+1} \geq p^{i+1}$ then
$f^{(1)}_{\bm{c}}(0)   = f_i (c_1, \dots, c_{i-1}, 0$, $c_{i+1}, \dots, c_n)
= p^i = f^{(2)}_{\bm{c}}(0)$.
Otherwise,
$f^{(1)}_{\bm{c}}(0)   = f_i (c_1, \dots, c_{i-1}, 0$, $c_{i+1}, \dots, c_n)
= \lfloor c_{i+1} /  p \rfloor = f^{(2)}_{\bm{c}}(0)$.

\smallskip

\noindent
{\em Case 2: $i = n$.}
If $f_{\bm{c}}(0)   = f_n (c_1, \dots, c_{n-1}, 0)$ is determined via Case 2.a, then
we have $f^{(1)}_{\bm{c}}(0)   =
f_n (c_1, \dots, c_{n-1}, 0) = p^n = f^{(2)}_{\bm{c}}(0)$.

Case 2.b cannot apply, since this would require $c_n \geq (c_{n-1} + 1) \cdot p$, which
is impossible in case of $c_n = 0$.

Case 2.c cannot apply either, since it would require  $c_n > c_{n-1} \cdot p$, which
is impossible in case of $c_n = 0$.

If Case 2.d applies, then we have
$f^{(1)}_{\bm{c}}(0)   = f_n (c_1, \dots, c_{n-1}, 0) =
c_{n-1} \cdot p$. If we now apply $f$ again, then we are still in Case 2.d, i.e.,
$f^{(2)}_{\bm{c}}(0) = c_{n-1} \cdot p  = f^{(1)}_{\bm{c}}(0)$.

Finally, if
Case 2.e applies, then
$f^{(1)}_{\bm{c}}(0)   = f_n (c_1, \dots, c_{n-1}, 0) =
c_{n-1} \cdot p +1$ holds. By the definition of $f_n$, the last component cannot
be incremented beyond $c_{n-1} \cdot p +p$ in this case. Hence, we indeed have
$f^{(p)}_{\bm{c}}(0)   = f^{(p+1)}_{\bm{c}}(0)$.
\end{proof}

%%%%%%%%%%%%%%%%%%%%%%%%%%%%%%%%%%%%%%%%%%%%%%%%%%%%%%%%%%%%%%%%

\section{Proof of Lemma~\ref{lmm:fundamental}}

  The proof follows the same line of reasoning
  as~\cite{DBLP:journals/jacm/EsparzaKL10}.  We use the same
  definition of the height of a parse tree $T$, namely the number of
  edges on the longest path from the root to a leaf: for example,
  trees in Fig.~\ref{fig:parse:tree} have heights 1, 2, and 2
  respectively.  As usual, the internal nodes of a parse tree are
  labeled with non-terminals, and the leaves with terminals.
  Unlike~\cite{DBLP:journals/jacm/EsparzaKL10}, we do not need to
  label the internal nodes with pairs $(x_i,j)$ where $x_i$ is the
  non-terminal (variable) and $j$ represents a monomial number in
  $f_i$, but we label them only with the variable $x_i$.  This is
  sufficient because our terminal symbols $a_{\bm v, i}$ are in
  one-to-one correspondence with the monomials in the polynomial
  $f_i$.

  Recall that each component $(f^{(q)}(0))_i$ of $\bm f^{(q)}(0)$ is a
  multivariate polynomial in the coefficients
  $\bm a = (a_1, \ldots, a_M)$ that occur in all polynomials
  $f_1, \ldots, f_N$.  When we repeatedly apply $\bm f$, each monomial
  $a_1^{v_1} \cdots a_M^{v_M}$ may appear multiple times in
  $\bm f^{(q)}(0)$; in our Example~\ref{ex:running:f} the monomial
  $a^3b^4$ appears 1 times in $f^{(3)}(0)$, then 5 times in
  $f^{(4)}(0)$.  Let $M^{(q)}_i$ be the bag of monomials occurring in
  $(f^{(q)}(0))_i$, and let $N^{(q)}_i$ be the bag of yields of all
  derivation trees of depth $\leq q$, i.e.
  $\bagof{Y(T)}{T \in \calT^i_q}$.  We prove by induction on $q$ that
  $M^{(q)}_i = N^{(q)}_i$ for all $i=1,N$.

  When $q=0$ then both bags are empty.  Assuming the statement holds
  for $q\geq 0$, we prove it for $q+1$. We prove that
  $M^{(q+1)}_i \subseteq N^{(q+1)}_i$; the other direction is similar
  and omitted.
  By definition, $(\bm f^{(q+1)}(0))_i = f_i(\bm f^{(q)}(0))$, where
  $f_i$ is a sum of monomials, see Eq~\eqref{eqn:fi}.  Each monomial
  in $f_i$ has the form $a_j x_1^{\ell_1}\cdots x_N^{\ell_N}$ for some
  $a_j\in \bm a$ and some vector of exponents
  $(\ell_1, \ldots, \ell_N)$.  Then $f_i(\bm f^{(q)}(0))$ is a sum of
  expressions of the form:
  \begin{align}
    & a_j\cdot (\bm f^{(q)}(0))_1^{\ell_1}\cdots(\bm f^{(q)}(0))_N^{\ell_N} \label{eq:monomial:to:be:expanded}
  \end{align}
  In other words, we have substituted each $x_j$ in~\eqref{eqn:fi}
  with $(\bm f^{(q)}(0))_j$, which, in turn, is a sum of the monomials
  in the bag $M^{(q)}_j$.  Thus, after expanding the
  expression~\eqref{eq:monomial:to:be:expanded} we obtain a sum of
  monomials $m$ of the following form:
  \begin{align}
    m =  & a_j \cdot m_{1,1}\cdots m_{1,\ell_1}\cdots \cdots  m_{N,1}\cdots m_{N,\ell_N} \label{eq:choice:of:monomials}
  \end{align}
  where $m_{j,k} \in M^{(q)}_j$.  To summarize, each monomial $m$
  occurring in $M^{(q+1)}_i$ can be uniquely obtained as follows: (a)
  choose one monomial $a_j x_1^{\ell_1}\cdots x_N^{\ell_N}$ in the
  polynomial $f_i$, (b) for each $j=1,N$, and each $k=1,\ell_k$,
  choose one monomial $m_{j,k}$ in the bag $M^{(q)}_j$ (a total of
  $\sum_j \ell_j$ choices).  Given these choices, we define a
  unique\footnote{``Unique'' means that different choices (a) and (b)
    will lead to different parse trees.} parse tree of depth $q+1$
  whose yield is $m$, as follows.  (a') The root node is labeled with
  the production~\eqref{eq:production} associated to the monomial that
  we used in (a).  (b') For each $j=1,N$ and each $k=1,\ell_j$, we use
  the inductive hypothesis $M^{(q)}_j = N_j^{(q)}$ and argue that the
  monomial $m_{j,k} \in M^{(q)}_j$ is in $N^{(q)}_j$, thus it is the
  yield of some parse tree $T_{j,k}$ of depth $\leq q$.  Then we
  complete our parse tree by setting $T_{j,k}$ to be the $(j,k)$-th
  child of the root node. This tree generates the following word:
  \begin{align*}
     & a_j \cdot m_{1,1}\cdots m_{1,\ell_1}\cdots\cdots m_{N,1}\cdots m_{N,\ell_N}
  \end{align*}
  This is equal to $m$ in Eq.~\eqref{eq:choice:of:monomials}, proving
  that $m \in N^{(q+1)}_i$ and, thus,
  $M^{(q+1)}_i \subseteq N^{(q+1)}_i$.  Containment in the other
  direction is similar and omitted.

\end{document}